\documentclass[aps,showpacs,nofootinbib,superscriptaddress,showkeys,twocolumn,floatfix]{revtex4}

\usepackage{epsfig}
\usepackage{graphicx}
\usepackage{dcolumn}

\newcommand{\uin}{{\bf u}_{\rm in}}
\newcommand{\uout}{{\bf u}_{\rm out}}

\begin{document}

\title{Renormalization of NN-Scattering with One Pion Exchange and
       Boundary Conditions}\author{M. Pav\'on
       Valderrama}\email{mpavon@ugr.es} \affiliation{Departamento de
       F\'{\i}sica Moderna, Universidad de Granada, E-18071 Granada,
       Spain.}  \author{E. Ruiz Arriola}\email{earriola@ugr.es}
       \affiliation{Departamento de F\'{\i}sica Moderna, Universidad
       de Granada, E-18071 Granada, Spain.}

\date{\today}

\begin{abstract} 
\rule{0ex}{3ex} A non perturbative renormalization scheme for
Nucleon-Nucleon interaction based on boundary conditions at short
distances is presented and applied to the One Pion Exchange
Potential. It is free of off-shell ambiguities and ultraviolet
divergences, provides finite results at any step of the calculation
and allows to remove the short distance cut-off in a suitable way. Low
energy constants and their non-perturbative evolution can directly be
obtained from experimental threshold parameters in a completely unique
and model independent way when the long range explicit pion effects
are eliminated. This allows to compute scattering phase shifts which
are, by construction consistent with the effective range expansion to
a given order in the C.M. momentum $p$. In the singlet $^1S_0$ and
triplet $^3S_1- ^3D_1$ channels ultraviolet fixed points and limit
cycles are obtained respectively for the threshold parameters. Data
are described satisfactorily up to CM momenta of about $p \sim m_\pi$.
\end{abstract}


 \maketitle



\section{Introduction}

Effective field theories (EFT) are a powerful tool to deal with
non-perturbative low energy physics. Over the last years, they have
provided promising results as regards a systematic and model
independent understanding of hadronic and nuclear physics. The scale
separation between long and short distance physics makes the
development of a systematic power counting possible. After the
original proposal of Weinberg's~\cite{Weinberg:rz} to design a power
counting based on applying ChPT to the potential many works have
followed implementing such a
counting~\cite{Ordonez:1995rz,Park:1997kp,Epelbaum:1998ka} with finite
cut-offs or proposing a counting in the renormalized
S-matrix~\cite{Kaplan:1998we,Geg98} which has also been pursued to
NNLO~\cite{Fleming:1999ee}. The relation of both the Weinberg (W) and
Kaplan-Savage-Wise (KSW) 
counting has been understood as perturbative expansions about infrared
fixed points in the limit of small and large scattering
lengths~\cite{Birse:1998dk} respectively (see also
Ref.~\cite{Barford:2001sx} for a discussion on long range forces in
that context). For systems with a large scattering length as it turns
out to be the case in low energy NN scattering, the Weinberg counting
may be modified to iterate the scattering length to all orders, but
then the connection to ChPT must be given up~\cite{Kaplan:1996xu}. On
the other hand the KSW counting, although systematic, does not
converge at NNLO~\cite{Fleming:1999ee}.  In Ref.~\cite{Beane:2001bc} a
new counting (BBSvK) involving also the chiral limit should be
invoked. According to these authors, one should treat
non-perturbatively the NN potential in the chiral limit and consider
finite pion mass corrections perturbatively on top of that. For a
recent and more complete review on these and related issues see
e.g. Ref.~\cite{Bedaque:2002mn} and references therein.

In order to properly define a truly EFT three essential requirements
must be met. Firstly, one needs a power counting scheme, i.e. a
dimensionless expansion parameter, which controls {\it a priori} the
accuracy of a calculation and provides an error estimate of the
neglected terms. The second requirement is the mathematical need for a
regularization method and subsequent renormalization scheme which is
consistent with the physical power counting, independent on any
ultraviolet cut-off. Finally, there is the question of practical
convergence, which can only be decided {\it a posteriori} on the light
of practical calculations.

The issue of regularization and renormalization in the present context
is not at all trivial, particularly if the power-counting scheme
involves summing up some infinite set of diagrams. There are well
known examples in the literature that shows that not all
regularization methods comply to the physical power counting, like
e.g. $\pi N $ scattering for relativistic baryons in dimensional
regularization~\cite{Gasser:1987rb}. Such an approach is not
consistent with the non-relativistic limit in the case of heavy
baryons. One has to either device a specific scheme, the so-called
infrared regularization~\cite{Becher:1999he}, or to make a
non-relativistic limit first and introduce dimensional regularization
afterwards~\cite{Jenkins:1990jv}. In the context of NN scattering it
has also been shown~\cite{Phillips:1997tn} that the renormalized
scattering amplitude for a theory without pions with a contact and a
derivative term, i.e. a truncated potential, depends on whether one
uses a cut-off regularization or dimensional regularization as an
intermediate step of the calculation. The problem was latter
understood by using other subtraction schemes than the $MS$ in
dimensional regularization, the so-called power divergence scheme
(PDS)~\cite{Kaplan:1998we} (see also
Refs.~\cite{Mehen:1998tp,Mehen:1998zz} for off-shell (OS) subtraction
schemes.)

As usual in field theory, one essential ingredient in the whole
construction is the choice of an appropriate regulator, which may
eventually be removed. In perturbation theory, there are many such
regulators, like e.g. dimensional regularization. Beyond perturbation
theory, like in the two body scattering problem one can use those
regulators within an order by order analysis aiming at finite
renormalized scattering equation.  For potentials which are purely
short range, i.e., contact terms and derivative corrections there-off,
this problem has successfully been tackled. The problem becomes subtle
when, in addition, long range forces are added. Typically, these long
range potentials develop some strong singularity at short distances.
The paradigm of the problem is probably best exemplified by the
$^1S_0$ channel with One Pion Exchange (OPE) potential. On the one
hand, the regular OPE alone produces a finite scattering amplitude. On
the other hand the short range potential is infinite but
renormalizable, and hence a finite scattering amplitude can be built
according to the standard principles of renormalization theory.  This
suggests, as it turns out to be the case, that the problem may still
be renormalizable within a scheme where the long range potential is
treated perturbatively.  Whether or not this mathematical requirement
is physically justified has been the subject of much debate in the
recent past, with a negative answer after the findings of
Ref.~\cite{Mehen:1998tp,Mehen:1998zz}; iterated OPE contributions to
scattering observables are not small in the triplet $^3S_1-^3D_1$
channel. The reason may be found in the singular and attractive
$1/r^3$ nature of the tensor contribution to the OPE potential.

The mathematical problem actually arises when both short and long
range potentials are added and none of them can be considered small;
there seems no obvious way to renormalize the scattering equation
non-perturbatively. By this we mean regularizing the equations first
and removing the regulators afterwards, or at least make the mass
scale of the regulator much larger than any other mass scale in the
problem. The main interest in making such a non-perturbative
renormalization is that any perturbative scheme can be thought of as
an approximation to renormalized equations.  The non-perturbative
renormalization of NN interaction in the singlet $^1S_0$ and triplet
$^3S_1-^3D_1$ channels has been studied several times in the
literature by different regularization methods. In
Ref.~\cite{Frederico:1999ps} a subtraction method was developed for
the Lippmann-Schwinger equation to construct a finite $T$ matrix for
contact interactions added to OPE. Renormalization is indeed achieved
by taking the subtraction scale to be much larger than any other mass
scale and checking for independence of results in this limit. The
resulting description of the $^1S_0$ phase shift is only valid to very
low energies, requiring for inclusion of derivative
terms. Unfortunately, the method has not been extended to that
case. Derivative interactions can be included within a
tree-dimensional cut-off
regularization~\cite{Gegelia:2001ev,Eiras:2001hu}. Actually, in the
cut-off regularization analysis of Ref.~\cite{Eiras:2001hu} a strong
breaking of angular momentum is observed unless either a vanishing
bare mixing coupling is chosen or a (according to the authors)
unlikely fine tunning sets in for a non-trivial fixed point. We will
show below that indeed the bare mixing vanishes but not in an uniform
way, but rather following a limit cycle pattern. Inspired by the N/D
method, the work of Ref.~\cite{Oller:2002dc} makes a re-summation of
the KSW amplitudes introducing a on-shell potential to which chiral
counting is applied. A new parameter, which is considered to be of
order zero in the chiral counting is introduced. The dynamical origin
of this parameter is unclear. The non-perturbative dimensional
regularization of the OPE potential with derivative interactions has
been studied in coordinate space~\cite{Kaplan:1996nv} as well as in
momentum space~\cite{Nieves:2003uu} in the $^1S_0$ channel. In both
cases a three-parameter fit can be achieved with no explicit two pion
exchange contribution up to CM momentum as large as $p \sim 400 {\rm
MeV}$. Ref.~\cite{Nieves:2003uu} opens up some hope as how the
triplet $^3S_1-^3D_1$ channel might also be renormalized
non-perturbatively within dimensional regularization, but so far there
are no practical calculations. The early finite sharp cut-off momentum
space treatments of Ref.~\cite{Epelbaum:1998ka} have been improved by
implementing a better regularization scheme~\cite{Epelbaum:2003xx}
which allows to make take larger cut-off values, yet finite.  More
recent works supporting the W-counting have also
appeared~\cite{Yang:2003kn,Gegelia:2004pz}.

From a diagrammatic point of view momentum space treatments based on
the Lippmann-Schwinger equation are more natural within a Lagrangian
framework and allow explicit consideration of nonlocal potentials. On
the other hand, the long range NN potentials making use of chiral
symmetry constraints are local, and for those the analysis of
non-perturbative renormalization in coordinate space becomes much
simpler, as will be shown along this work. In addition, the
Schr\"odinger equation is a second order operator and mixed boundary
conditions define a complete and unique solution of the scattering
problem in the whole space at both sides of the boundary. This sharp
boundary separation of the space is naturally formulated in coordinate
space for a local potential. Boundary conditions for NN scattering
were used many years ago (see e.g. Ref.~\cite{LF67} and references
therein), and there has been renewed interest motivated by the
developments within
EFT~\cite{Phillips:1996ae,Cohen:1998bv,vanKolck:1998bw}. Actually, the
thorough analysis of Ref.~\cite{vanKolck:1998bw} shows that in the
absence of long range forces a low momentum expansion of the potential
within EFT framework for the Lippmann-Schwinger equation is completely
equivalent to an effective range expansion (ERE) and also to an energy
expansion of a generic boundary condition at the origin in coordinate
space for the Schr\"odinger equation. Moreover, the reference partial
wave analysis of the Nijmegen group~\cite{Stoks:1993tb} uses this
method to successfully describe the a large NN scattering data base,
when long range potentials are used. While in the first works
phenomenological potentials where used, more recent studies consider
potentials deduced from ChPT theory with a rather satisfactory
description of the experimental scattering
data~\cite{Rentmeester:1999vw}. The minimal boundary radius which can
still provide an acceptable $\chi^2 /DOF $ is about $ R_S=1.4-1.8 {\rm
fm} $. Obviously, if the radius cannot be lowered without spoiling the
quality of the fit, the short distance cut-off becomes an
indispensable parameter of the theory, which cannot be removed. The
corresponding momentum space cut-off $ \Lambda = 2\pi / R_S \sim 600
{\rm MeV} $ is comparable to the one needed in early momentum space
treatments~\cite{Epelbaum:1998ka}. Within the spirit of an EFT it
would actually be more appropriate to take instead larger $\Lambda$'s
or equivalently shorter $R_S$'s and to check for insensitivity of
results in the low energy regime. Thus, there arises the natural
question whether in fact this EFT procedure can be implemented.

In our previous work~\cite{Valderrama:2003np} we showed that for the
$^1S_0 $ singlet channel with OPE the boundary radius can be
effectively removed without spoiling a good description of the
corresponding phase shift up to the a priori expected CM momentum of $
k \sim m_\pi $ where the Two Pion Exchange (TPE) effects should start
playing a role. The first order differential equation satisfied by the
boundary condition of the problem defined in the interval $ R < r <
\infty $ as a function of the boundary radius was very helpful, since
the whole problem could be mapped into a variable phase
equation~\cite{Calogero} of a truncated potential in the region $ 0 <
r \le R$ with a non-trivial initial condition at the origin, encoding
the short distance physics. In this way, the long range pions could be
eliminated and the evolution of the threshold parameters as a function
of the boundary radius could be determined
non-perturbatively. Actually, a trivial ultraviolet fixed point limit
for the scattering length was found non-perturbatively. Remarkably,
this behaviour coincides with the one found in
Ref.~\cite{Kaplan:1996xu} within a perturbative treatment. This
trivial fixed point at the origin implies a fine tuning of the short
distance physics in order to reproduce the physical scattering
length. In this paper we want to extend our results for the
interesting case of the triplet $^3S_1-^3D_1$ channel. The solution of
the boundary condition problem requires solving a coupled set of
Schr\"odinger equations. Instead of doing so, we prefer to directly
compute the change of the boundary condition by an equivalent variable
phase approach~\cite{Calogero} with non-trivial initial conditions
which encode the short distance physics~\cite{Valderrama:2003np}. This
provides, in addition, a direct and quite transparent connection to
renormalization group ideas~\cite{Birse:1998dk,Barford:2001sx}.

Related works in spirit to the present are those of
Refs.~\cite{Beane:2000wh,Beane:2001bc} and \cite{Barford:2001sx}. In
Refs.~\cite{Beane:2000wh,Beane:2001bc} a square well potential is used
to regulate the short distance behaviour simulating a smeared delta
function. The renormalization group flow for the potential strength is
not uniquely defined. This phenomenon is also found in the theory of
self-adjoint extensions of the Schr\"odinger
operators~\cite{Albeverio}. In Ref.~\cite{Barford:2001sx} a delta
shell regulator located at a finite distance is assumed as the short
distance potential whereas the long distance piece is solved exactly
using a distorted wave basis. This formalism has been so far used to
the study of renormalization of repulsive singular potentials (like
$1/r^2$). The reason may have to do with the need for a well defined
renormalization at the origin. A common feature of both regularization
schemes is that the wave function at the origin is uniquely determined
by the regularity condition, $u(0)=0$. The boundary condition
regularization that we use in this paper provides a uniquely defined
renormalization group flow~\cite{Valderrama:2003np}, to treat both
repulsive and attractive singular potentials~\cite{Pavon03}. In
addition, the boundary condition admits a simple physical
interpretation: it can be mapped into a variable phase shift
problem~\cite{Calogero} with a truncated potential. This
interpretation directly provides the non-perturbative renormalization
flow of low energy parameters and a quite transparent analysis of both
infrared as well as ultraviolet fixed points and limit
cycles~\cite{Pavon03}.

In this paper we analyze precisely how the energy dependent boundary
condition must change as we move the boundary radius for fixed energy
to achieve independence of physical observables such as scattering
phase shifts. By doing so we are effectively changing the Hilbert
space since the wave function in the outer region is defined only from
the boundary to infinity. An advantage of this procedure is that we
never need to invoke off-shellness explicitly; at any step
we are dealing with an on-shell problem. In addition, we work directly
with finite quantities and no divergences appear at any step of the
calculation when the boundary radius is taken to zero from above. 

Another advantage of our construction, as it will become clear along
the paper, is that we only need the potentials and physical threshold
parameters as input of the calculation (the cut-off dependence is
removed completely, so this is not a parameter).  This implies, in
particular, that given this information we never have to make a fit
(except perhaps for the determination of the threshold parameters);
our calculations are predictions for the phase shifts that are
consistent, by construction, with a low energy expansion up to a given
order. Thus, the potential danger of compromising the low energy fit
due to a global fit up to $300 {\rm MeV} $ may be precluded from the
start. This is at difference with the standard way of proceeding where
the low energy parameters are fitted to the phase shifts and the
threshold parameters are then recomputed. Actually, our analysis is
equivalent to making a fit only in the low energy region, where
explicit pions do not contribute, and predicting the intermediate
energy region. We believe this is a possible and practical way of
learning about the role of explicit pions in the NN
interaction. Actually, our motivation was partly to see whether OPE can
actually be seen in low partial waves in the intermediate energy
region $ m_\pi /2 \le k \le m_\pi$.

In the present paper we analyze the OPE potential ($ U = 2\mu V $ and
$ \mu = M_N/2 $ ) which reads
\begin{eqnarray}
U ( \vec x ) = U_C (r) + S_{12} U_T (r)   \, , 
\end{eqnarray} 
with ( $ \hat x = \vec x / r $ )
\begin{eqnarray} 
S_{12} = 3  \vec \sigma_1 \cdot \hat x \vec \sigma_2 \cdot \hat x  -
\vec \sigma_1 \cdot \vec \sigma_2  	\, , 
\end{eqnarray} 
and 
\begin{eqnarray}
U_C &=& -\frac{m_\pi^2 M_N g_A^2 }{16 \pi f_\pi^2 } \frac{e^{-m_\pi r 
}}{r} \\ U_T &=& -\frac{m_\pi^2 M_N g_A^2 }{16 \pi f_\pi^2 }
\frac{e^{-m_\pi r }}{r} \left( 1 + \frac3{m_\pi r}+ \frac3{(m_\pi
r)^2} \right) \, 
\end{eqnarray} 
Where $M_N$ is the nucleon mass, $m_\pi$ the pion mass, $f_\pi $ the
pion weak decay constant and $g_A $ the nucleon axial coupling
constant. In the numerical calculations below we take $M_N =938.92 \,
{\rm MeV}$, $f_\pi =93 \, {\rm MeV} $ , $m_\pi =138 \, {\rm MeV} $ and
$ g_A=1.25 $. 
Note that the singularity at the origin of the tensor potential  
\begin{eqnarray}
U_T \to  -\frac{3 M_N g_A^2 }{16 \pi f_\pi^2 r^3} \qquad r \to 0  
\end{eqnarray} 
is independent on the pion mass $m_\pi$.

The plan of the paper is as follows. In Sect.~\ref{sec:smatrix} we
present the basic object of our analysis, the variable S-matrix which
we supplement with general mixed boundary conditions in the general
case of coupled channel scattering. We also discuss the role played by
the irregular solutions for singular potentials in the spirit of an
effective field theory. After that we rewrite in
Sect.~\ref{sec:threshold} the variable S-matrix equation for the
variable $K$-matrix and $R$-matrix in a way that the low energy limit
may be taken. As a result, we find the boundary radius evolution of
threshold parameters. As a first application we apply in
Sect.~\ref{sec:lecs} the obtained equations to determine the low
energy threshold parameters from well established NN potentials. In
Sect.~\ref{sec:short} we study the short distance behaviour of the
threshold parameters. There we show that one has for the $^1S_0$ and
$^3S_1-^3D_1$ channels an UV fixed point and a UV limit cycle for the
scattering lengths. In Sect.~\ref{sec:numerics} we present our
numerical results, both for the threshold parameters as well as for
the $^1S_0$ and $^3S_1-^3D_1 $ phase shifts. Finally in
Sect.~\ref{sec:concl} we present some final remarks, conclusions and
perspectives for future work.

\section{ Variable S-matrix with Boundary Conditions}
\label{sec:smatrix} 

\def\u{{\bf u}} \def\U{{\bf U}} \def\S{{\bf S}} \def\h{{\bf h}}
\def\L{{\bf L}} \def\E{{\bf 1}} \def\j{{\bf j}} \def\y{{\bf y}}
\def\K{{\bf K}} 
\def\f{{\bf f}} 

In order to generalize to triplet states the results of
Ref.~\cite{Valderrama:2003np} for the singlet channel case, we
introduce the variable S-matrix formalism for the general coupled
channel case. For potentials which are either regular or singular
repulsive at the origin the procedure is standard~\cite{Calogero} and
it has many variants. For completeness and to make the exposition more
self contained we present here our particular derivation which also
applies to singular attractive potentials and at the same time introduce
our basic notation for the rest of the paper. Although in the case
under study we are interested in, at most, two coupled channels, the
formalism can be developed for the general case with almost no
additional effort.

The scattering amplitude for NN scattering can be written as a partial
wave expansion
\begin{eqnarray}
\f = \frac1{ 2i k} \sum_{l=0}^\infty \left( \S -1 \right) P_l ( \cos
\theta ) \, ,  
\end{eqnarray} 
where $\S$ is the S-matrix for coupled channels.  The coupled channel
Schr\"odinger equation for the relative motion reads
\begin{eqnarray}
-\u '' (r) + \left[ \U (r) + \frac{{\bf l}^2}{r^2} \right] \u (r) =
 k^2 \u (r) \, , 
\label{eq:sch_cp} 
\end{eqnarray} 
where $\U (r)$ is the coupled channel matrix potential, $ {\bf l}^2 =
{\rm diag} ( l_1 (l_1+1), \dots, l_N (l_N +1) )$ is the angular
momentum, $\u(r)$ is the reduced matrix wave function and $k$ the
C.M. momentum. We assume for $\u(r)$ the mixed boundary
condition~\footnote{This is the most general boundary condition that
makes the coupled channel Hamiltonian self-adjoint in the interval $ R
\le r < \infty$.}
\begin{eqnarray}
\u ' (R)  + \L_k (R) \u (R) = 0 \, ,  
\label{eq:bc}
\end{eqnarray}   
where $\L_k(R) $ is a real hermitean matrix in coupled channel space,
which in our framework encodes the {\it unknown} physics at distances
$r$ below the boundary radius $R$. In addition, we assume the
asymptotic normalization condition
\begin{eqnarray}
\u (r)  \to \uin (r) -  \uout (r) \S \, , 
\label{eq:asym}
\end{eqnarray} 
with $\S$ the standard coupled channel S-matrix. The corresponding
out-going and in-going free spherical waves are given by 
\begin{eqnarray}
\uout (r) &=& {\rm diag} ( \hat h^+_{l_1} ( k r) , \dots , \hat
h^+_{l_N} (k r) ) \, , \\ \uin (r) &=& {\rm diag} ( \hat h^-_{l_1} ( k r) ,
\dots , \hat h^-_{l_N} (k r) ) \, , 
\end{eqnarray} 
with $ \hat h^{\pm}_l ( x) $ the reduced Hankel functions of order
$l$, $ \hat h_l^{\pm} (x) = x H_{l+1/2}^{\pm} (x) $ ( $ \hat h_0^{\pm} =
e^{ \pm i x}$ ), and satisfy the free Schr\"odinger's equation for a free
particle, 
\begin{eqnarray}
-\uout '' (r) + \frac{{\bf l}^2}{r^2} \uout (r) &=&  
 k^2 \uout  (r) \, , \\ 
-\uin '' (r) + \frac{{\bf l}^2}{r^2}   \uin (r) &=& 
 k^2 \uin  (r) \, , 
\label{eq:sch_cp_free} 
\end{eqnarray} 
The boundary condition, Eq.~(\ref{eq:bc}), for the {\it outer}
boundary values problem, Eq.~(\ref{eq:sch_cp}) and Eq.~(\ref{eq:asym}),
 can be interpreted in simple physical terms of a complementary {\it
inner} problem where the potential ${\bf U}(r)$ acts in the interval $
R \le r < \infty$. If we switch off the potential above a given
boundary radius $R$ we have, at the boundary
\begin{eqnarray} 
 {\bf L}_k (R) &=& \u' (R) \u^{-1} (R) \nonumber \\ &=& \left[ \uin'
(R) - \uout' (R) \S (R) \right] \nonumber \\ &\times& \left[ \uin (R)
- \uout (R) \S (R) \right]^{-1} \, , 
\label{eq:bc_s}
\end{eqnarray}
where ${\bf S (R)}$ is the S-matrix associated to the potential ${\bf
U} (r) $ acting in the region $ 0 < r \le R$, which inherits the
dependence on the chosen boundary radius $R$. The equation satisfied
by the variable S-matrix can be obtained from Schr\"odinger's equation
applied to the matrix ${\bf L} (R)$ yielding
\begin{eqnarray}
{\bf L}_k (R) ' + {\bf L}_k (R)^2 = \U (R) + \frac{{\bf l}^2}{R^2} -
k^2 \, . 
\label{eq:bc_ren}  
\end{eqnarray} 
From here~\footnote{An alternative derivation of Eq.~(\ref{eq:bc_ren})
, closer in spirit to the renormalization group and the
Callan-Symanzik equation will be presented elsewhere
Ref.~\cite{Pavon03}.} it is straightforward to obtain the equation for
the variable S-matrix,
\begin{eqnarray}
2 {\rm i} k \frac{ d \S (R)}{dR} &=& \left[ \S(R) \hat \h^{(+)} (R) - \hat \h^{(-)}
(R) \right] \U(R) \nonumber \\ & \times& \left[ \hat \h^{(-)} (R) - \hat \h^{(+)} (R) \S (R)
\right] \, . 
\label{eq:vs}
\end{eqnarray} 
This is a first order non-linear matrix differential equation which
can be solved by standard means, provided the S-matrix is known at one
given scale. One of the interesting aspects of this equation is that
there is no need to invoke any off-shellness; for any value of the
boundary radius we have a different on-shell scattering problem. In
appendix~\ref{sec:app1} we show an alternative derivation based on
continuous deformations of the potential with a fixed boundary
condition. As we will discuss below, Eq.~(\ref{eq:vs}) describes the
renormalization group flow of the S-matrix as a function of the
distance scale $R$ where the long range potential is cut-off.

In the case of a regular potential, Eq.~(\ref{eq:vs}) has to be
supplemented with an initial condition at the origin, namely the
trivial one (corresponding to the absence of a potential), and its
asymptotic value yields the full $S-$matrix ;
\begin{eqnarray}
\S(0) = {\bf 1} \qquad , \qquad  \S  = \S (\infty)  \qquad ( {\rm regular
} ) 
\label{eq:ini_reg} 
\end{eqnarray} 
In this paper we are concerned with the OPE potential which has a
singular $1/r$ behaviour at the origin in the $^1S_0$ singlet channel
and singular $1/r^3$ behaviour at the origin due to the tensor force
in the $^3S_1-^3D_1$ triplet channel. While in the single channel the
singularity is a mild one in the sense that there still exists a
unique regular solution at the origin, $ u(0)=0$ (like in the Coulomb
potential) , in the triplet channel both linearly independent solutions to
Schr\"odinger's equation vanish at the origin, and the regularity
condition $ u(0)=0$ does not uniquely specify the solution.

The point of view we take in the present work is that of an EFT; low
energy physics should not depend on the detailed knowledge of the
interaction at short distances. This applies, in particular, to the
case of a singular potential as will also become clear
below. Following the lines already sketched in our previous
work~\cite{Valderrama:2003np}, we take instead the value at infinity
as the initial value for the variable $S-$matrix. Of course, for a
short range potential, this procedure corresponds to start integrating
at sufficiently large distances (where the potential may be
neglected). An advantage of this procedure is that {\it by
construction} a unique solution $\S(R)$ is obtained. Even for a
regular potential, it is clear that a generic choice of ${\bf S}
(\infty) $ cannot yield by integrating towards the origin the
result $ {\bf S} (0) = {\bf 1} $ besides the very exceptional cases
which accidentally correspond to the regular solution at the
origin. Thus, we expect in general an admixture of both the regular
and irregular solutions, which corresponds to a mixed boundary
condition close to the origin,
\begin{eqnarray}
\lim_{R \to 0^+} \left\{ \u ' (R) + \L (R) \u(R) \right\} =0  
\label{eq:bc0}
\end{eqnarray} 
In the case of a singular potential both solutions vanish and we
equally have a unique mixed boundary condition as in
Eq.~(\ref{eq:bc0}). Thus, we may define the {\it short distance}
S-matrix as the extrapolation to the origin of a given solution at
infinity,
\begin{eqnarray}
\S_s \equiv \lim_{R \to 0^+} \S (R) \qquad , \qquad \S = \S (\infty)
\qquad ( {\rm general } )
\label{eq:ini_sing} 
\end{eqnarray}
Actually, the precise meaning of the previous limit will be the main
topic of the present work. We anticipate already that we will find
ultraviolet fixed points for the singlet $^1S_0$ channel and limit
cycles for the $^3S_1-^3D_1$ triplet channel. Eqs.~(\ref{eq:vs}) and
(\ref{eq:vf}) are well known in potential scattering ( for a review
see e.g. Ref.~\cite{Calogero}), but they have always been used
assuming the trivial initial conditions $\S (0)= {\bf 1} $.

Obviously, if one would literally use the full $S-$ matrix and
integrate downward, nothing could be achieved, since that would
correspond to eliminating the full potential. A more interesting
perspective, already pursued in Ref.~\cite{Valderrama:2003np} for the
singlet $^1S_0$ channel, consists of regarding the low energy limit of
the previous equations, extracting the threshold parameters at short
distances by integrating downward from their experimental values and
integrate back upward the variable $S$-matrix equation to
infinity. Physically, this procedure corresponds to explicitly
separate the OPE contributions on top of any good low energy
approximation, like e.g. the effective range expansion.

In the case of one channel, like the $^1S_0$, the S-matrix can be
parameterized as $S_l (k,R) = \exp( 2 {\rm i} \delta_l (k,R) ) $ with
$\delta_l (k, R) $ the variable phase. Eq.~(\ref{eq:vs}) becomes
rather simple~\cite{Calogero} for $s-$ waves, yielding
\begin{equation}
\frac{ d \delta_0 (k,R) }{dR} = -\frac1k U(R) \sin^2 (k R+ \delta_0 (k,R))
\, . 
\label{eq:vf}
\end{equation}
and the obvious conditions both at the origin and at infinity must be
satisfied
\begin{equation}
\lim_{R \to 0} \delta_0 (k,R) = \delta_0^S (k) \qquad  \lim_{R \to
\infty} \delta_0 (k,R) = \delta_0 (k) \,  . 
\end{equation}
The OPE potential in the coupled $^3S_1-^3D_1 $ in the triplet channel
space is given by
\begin{eqnarray}
\U(r) &=& \left( \matrix{ U_s (r) & U_{sd} (r) \cr U_{sd} (r) & U_d
(r) } \right) \, , 
\end{eqnarray} 
where 
\begin{eqnarray}
U_s = U_C \qquad U_{sd} = 2 \sqrt{2} U_T \qquad U_d = U_C - 2 U_T \, .
\end{eqnarray} 
The two coupled channels S-matrix can be represented in the
Blatt-Biedenharn (BB or Eigen phase) parameterization
\begin{eqnarray}
S &=& \left( \matrix{ \cos \epsilon & -\sin \epsilon \cr \sin \epsilon
& \cos \epsilon } \right) \left( \matrix{ e^{2 {\rm i} \delta_1} & 0
\cr 0 & e^{2 {\rm i} \delta_2} } \right) \left( \matrix{ \cos \epsilon
& \sin \epsilon \cr -\sin \epsilon & \cos \epsilon } \right) \nonumber
\\
\end{eqnarray} 
which will be used along this paper. The relation to the standard
coupled channel K-matrix is given by
\begin{eqnarray} 
\S = \left({\bf K} + {\rm i} k \right) \left({\bf K} - {\rm
i} k \right)^{-1} \, , 
\end{eqnarray} 
where 
\begin{eqnarray} 
\tan ( 2\epsilon) &=& \frac{2 K_{12}}{K_{11}-K_{22}} \, , \\ -\tan \delta_-
&=& \frac12 \left[ K_{11} + K_{22} + \frac{K_{11} - K_{22}}{\cos 2
\epsilon } \right] \, , \\ -\tan \delta_+ &=& \frac12 \left[ K_{11} +
K_{22} -\frac{K_{11} - K_{22}}{\cos 2 \epsilon }
\right] \, . 
\end{eqnarray} 
Due to unitarity of the S-matrix in the low energy limit, $ k\to 0$ we
have
\begin{eqnarray}
\left(\S - \E \right)_{l',l}=- 2 {\rm i} \alpha_{l', l}  k^{l'+l+1} +
\dots   \, ,  
\end{eqnarray} 
with $\alpha_{l' l} $ the (hermitean) scattering length matrix. The
low energy limit acquires its simplest form in the
Stapp-Ypsilantis-Metropolis (SYM or Nuclear bar) parameterization
\begin{eqnarray}
S &=& \left( \matrix{ e^{2 {\rm i} \bar \delta_1} \cos 2 \bar \epsilon
& {\rm i} e^{ {\rm i} ( \bar \delta_1+ \bar \delta_2 )} \sin 2 \bar
\epsilon \cr {\rm i} e^{ {\rm i} ( \bar \delta_1+ \bar \delta_2) } \sin 2 \bar
\epsilon & e^{2 {\rm i} \bar \delta_2 } \cos 2 \bar \epsilon } \right) 
\end{eqnarray} 
which is related to the BB phase shifts by 
\begin{eqnarray}
\bar \delta_1 + \bar \delta_2 &=& \delta_+ + \delta_- \, , \\ 
\sin( \bar \delta_1 - \bar \delta_2  ) &=& \frac{\tan( 2\bar
\epsilon)}{\tan(2\epsilon)}  \, . 
\end{eqnarray} 
The low energy limit in the SYM representation becomes   
\begin{eqnarray} 
\bar \delta_1 \to  - \alpha_0 k \, , \qquad 
\bar \delta_2 \to  -\alpha_2 k^5  \, , \qquad 
\bar \epsilon \to  - \alpha_{02} k^3   \, .
\end{eqnarray} 
The scaled K-matrix, ${\bf \hat K} $, has a good low energy
behaviour and is defined by making an energy dependent transformation
\begin{eqnarray}
{\bf \hat K} = k {\bf D} {\bf K} {\bf D} \, , 
\end{eqnarray} 
with ${\bf D} = {\rm diag} ( k^{l_1}, \dots , k^{l_N}) $.  The scaled
K-matrix admits the coupled channel analog of the effective range
expansion
\begin{eqnarray}
{\bf \hat K} = -{\bf a}^{-1} + \frac12 {\bf r}k^2 + {\bf v} k^4 +
\dots \, , 
\label{eq:c-ere}
\end{eqnarray} 
where ${\bf a} $, ${\bf r}$ and ${\bf v} $ are the scattering length
matrix, effective range and curvature parameters respectively.

\section{Evolution of low energy parameters}
\label{sec:threshold} 

In order to take this low energy limit and corrections there-off, we
introduce the variable or running $K-$matrix
\begin{eqnarray}
\S (R) = \left({\bf K} (R)+ {\rm i} k \right) \left({\bf K} (R) - {\rm
i} k \right)^{-1} \, , 
\end{eqnarray} 
as well as the reduced Bessel functions 
\begin{eqnarray} 
\hat j_l (x) = x j_l (x) \, ,  \qquad \hat y_l (x) = x y_l (x) \, ,  
\end{eqnarray} 
i.e. $\hat j_0 (x)= \sin x $ , $\hat y_0 (x) = -\cos x $. Thus, 
\begin{eqnarray}
{\bf \hat j} &=& \frac1{2 {\rm i}} \left( \hat \h^{(+)} - \hat
\h^{(-)} \right) \, , \\ -{\bf \hat y} &=& \frac12 \left( \hat \h^{(+)} +
\hat \h^{(-)} \right) \, .
\end{eqnarray} 
Then,  we get 
\begin{eqnarray}
\K ' (k,R) &=& \left( \frac1k \K (k,R) {\bf \hat j} (kR) - {\bf \hat y}
(kR) \right) \U (R) \nonumber \\ & \times & \left( \frac1k {\bf \hat
j} (k R) \K (k,R) - {\bf \hat y} (kR) \right) \, . 
\end{eqnarray} 
The scaled K-matrix, ${\bf \hat K} (R)$, has a better low energy
behaviour and is defined by making an energy dependent transformation
\begin{eqnarray}
{\bf \hat K} (R) = k {\bf D} {\bf K}(R) {\bf D} \, , 
\end{eqnarray} 
with ${\bf D} = {\rm diag}  ( k^{l_1}, \dots , k^{l_N}) $. We get  
\begin{eqnarray}
\hat \K ' (k,R) &=& \left( \hat \K (R, k) \frac1k {\bf j} (kR) {\bf D}^{-1} -
{\bf y} (kR) {\bf D} \right) \U (R) \nonumber \\ & \times &
\left( \frac1k {\bf j} (kR) {\bf D}^{-1} \hat \K (R, k) -
{\bf y} (kR) {\bf D} \right) \, . 
\label{eq:vkhat} 
\end{eqnarray} 
The scaled K-matrix admits the analog of the effective range expansion 
\begin{eqnarray}
{\bf \hat K} (R) = -{\bf a} (R)^{-1} + \frac12 {\bf r}(R) k^2 + {\bf v}
(R) k^4 + \dots \, . 
\end{eqnarray} 
where ${\bf a} (R)$, ${\bf r}(R)$ and ${\bf v}(R)$ are the
corresponding running scattering length matrix, effective range and curvature
parameters respectively. In this form the low energy limit can be
easily taken. Defining the matrix functions and their low energy
expansion
\begin{eqnarray}
{\bf A}_k(R) &=& \frac{{\bf j} (kR)}k {\bf D}^{-1} = A_0 + k^2 A_2 +
k^4 A_4 + \dots \, , \nonumber \\ {\bf B}_k(R) &=& {\bf y} (kR) {\bf
D} \quad = B_0 + k^2 B_2 + k^4 B_4 + \dots \, , \nonumber \\ 
\end{eqnarray} 
we get the system of coupled equations~\footnote{The Equation for
${\bf v_2} $ in the coupled channel case is too long to be reproduced
here, but can be obtained in a straightforward way.}
\begin{eqnarray} 
\frac{d}{dR} [{\bf a} (R)]^{-1} &=& -\left( [{\bf a} (R)]^{-1} {\bf
A}_0 + {\bf B}_0 \right) \U (R) \nonumber \\ &\times & \left( {\bf
A}_0 [{\bf a} (R)]^{-1} + {\bf B}_0 \right) \\ \frac{d}{dR} {\bf r}
(R) &=& \left( [{\bf a} (R)]^{-1} {\bf A}_0 + {\bf B}_0 \right) \U (R)
\nonumber \\ &\times & \left( {\bf r} (R) {\bf A}_0 + 2 [{\bf a}
(R)]^{-1} {\bf A}_2 + 2 {\bf B}_2 \right) \nonumber \\ &+& \left( {\bf
r} (R) {\bf A}_0 + 2 [{\bf a} (R)]^{-1} {\bf A}_2 + 2 {\bf B}_2
\right) \U (R) \nonumber \\ & \times& \left( [{\bf a} (R)]^{-1} {\bf
A}_0 + {\bf B}_0 \right) \, . 
\label{eq:var}
\end{eqnarray} 
These equations generalize to the coupled channel case those already
found in Ref.~\cite{Valderrama:2003np} and have to be supplemented
with some initial conditions, at e.g. infinity,
\begin{eqnarray} 
{\bf a} ( \infty)= {\bf a} \, , \qquad {\bf r} ( \infty)= {\bf r}
\qquad ,   \dots 
\end{eqnarray} 
For the case of $s-$wave one channel scattering Eq.~(\ref{eq:vkhat}) 
becomes 
\begin{eqnarray}
\frac{ d K(k,R)}{dR} = U(R) \left[ K(k,R) \frac{\sin k R}k + \cos k R
\right]^2 \, . 
\label{eq:vk}
\end{eqnarray}
where 
\begin{equation} 
K(k,R) = k \cot \delta(k,R) \, , 
\end{equation} 
yielding at low energies an effective range expansion,
\begin{equation}
k \cot \delta (k,R) = -\frac1{\alpha_0 (R) } + \frac12 r_0 (R)  k^2 + 
v_2 (R) k^3 \cdots \, , 
\end{equation}
where 
\begin{eqnarray}
\frac{d \alpha_0}{dR} &=& U(R) \left( \alpha_0 -R \right)^2
 \label{eq:valpha} \\ \frac{ d r_0}{dR} &=& 2 U(R) R^2 \left( 1- \frac
 {R}{ \alpha_0 } \right) \left( \frac{r_0}R + \frac{R}{3 \alpha_0 } -1
 \right) \label{eq:vr0}\\ \frac{d v_2 }{d R} &=& \frac{U(R)}R \left\{
 \frac14 \left( \frac{r_0}R + \frac{R}{3 \alpha_0 } -1 \right)^2
 \right. 
 \\ &+& \left. 
 2\left( 1- \frac{R}{ \alpha_0} \right) \left(-\frac1{12} \frac{r_0}R
 + \frac{v_2}{R^3} - \frac1{120} \frac{R}{ \alpha_0} +
 \frac1{24}\right) \right\}  \nonumber \\   
\label{eq:vv2}
\end{eqnarray}
These equations have been studied by us in
Ref.~\cite{Valderrama:2003np} for analyzing the OPE in the singlet
$^1S_0$ channel.

In the $^3S_1-^3D_1$ coupled channel case the threshold parameters
matrices are 
\begin{eqnarray}
{\bf a} &=& \left(\matrix{ \alpha_0 & \alpha_{02} \cr \alpha_{02} &
\alpha_2 } \right) \, , \\ {\bf r} &=& \left(\matrix{ r_0 & r_{02} \cr
r_{02} & r_2 } \right) \, , \\ {\bf v} &=& \left(\matrix{ v_0 & v_{02}
\cr v_{02} & v_2 } \right) \, . 
\end{eqnarray} 
The explicit form of the equations for the $^3S_1-^3D_1$ running
scattering lengths reads
\begin{widetext} 
\begin{eqnarray} 
R^4 \alpha_0' &=& 9 U_d \alpha_{02}^2 + ( \alpha_0 - R) R^2 \left[ ( 
\alpha_0 -R ) U_s + 6 \alpha_{02} U_{sd} \right] \, , \nonumber \\ 15 R^5
\alpha_{02}' &=& -15 \alpha_{02} R^4 \left( -\alpha_0 + R \right) U_s
+ R^2 \left( 45 \alpha_{02}^2 - \left( \alpha_0 - R \right) \left( -45
\alpha_2 + R^5 \right) \right) U_{sd} - 3 \alpha_{02}\left( -45
\alpha_2 + R^5 \right) U_d \, , \nonumber \\ 225 R^4 \alpha_2' &=& 225
\alpha_{02}^2\,R^4 U_s - 30 \alpha_{02} R^2 \left( -45 \alpha_2 + R^5
\right) U_{sd} + \left( -45 \alpha_2 + R^5 \right)^2 U_d 
\label{eq:va_c}  \, . 
\end{eqnarray} 
\end{widetext} 
Note that all three running low energy parameters $\alpha_0 $,
$\alpha_{02}$ and $\alpha_2 $ (the explicit R-dependence has been
suppressed for simplicity) are coupled due to the mixing potential
$U_{sd}$.  Thus, it would be inconsistent to take any of them as a
constant; exact renormalization group invariance requires mixing
between the $S$ and $D$ channels. As we see the mixing is related both
to a non-vanishing of the mixing potential $U_{sd}$ and a non
vanishing value of $\alpha_{sd}$ at a given point. If by some accident
both vanish at a given point, the mixing will vanish. 

The evolution of the low energy parameters can be translated into the
corresponding evolution of the short distance boundary condition as a
function of the boundary radius. Defining the dimensionless quantity
\begin{eqnarray}
{\bf C}_k (R) = {\bf 1}-R {\bf L}_k (R) =
{\bf 1}- R {\bf u}'_k (R) {\bf u}_k (R)^{-1}  \, , 
\end{eqnarray}   
and using  Eq.~(\ref{eq:bc_ren}) we get 
\begin{eqnarray}
R {\bf C}_k^\prime (R) = {\bf C_k} ( {\bf 1} - {\bf C_k} ) + {\bf U}(R) R^2
+ {\bf l}^2 - k^2 R^2  \, . 
\label{eq:bc_ren_ck}
\end{eqnarray}   
Expanding into powers of the momentum $k$ one gets 
\begin{eqnarray}
{\bf C}_k (R) =  {\bf C}_0 (R) + k^2 R^2 {\bf C}_2 (R)  + \dots 
\end{eqnarray} 
For the singlet $^1S_0$ channel we have, in particular, the following
relation 
\begin{eqnarray}
C_0 = \frac{\alpha_0 (R)}{R-\alpha_0(R)}  \, . 
\label{eq:bc_alpha} 
\end{eqnarray} 
Note that for $R \to \infty $ we have a fixed point behaviour $C_0 \to
0 $ unless $ \alpha = \infty $ in which case $ C_0 \to 1 $.  The
evolution of the boundary condition with the short distance boundary
radius for the $^3S_1-^3D_1$ in terms of the running scattering
lengths is given by
\begin{eqnarray} 
C^0_s &=& 1+\frac{R(R^5 - 45 \alpha_{22})}{45 \alpha_{02}^2 + (
\alpha_{00} -R) (R^5 - 45 \alpha_{22})} \, , \\ C^0_{sd} &=& \frac{15
\alpha_{02} R^3 }{45 \alpha_{02}^2 + ( \alpha_{00} -R) (R^5 - 45
\alpha_{22})} \, , \\ C^0_d &=& 3 - \frac{5 ( R-\alpha_{00})R^5 }{45
\alpha_{02}^2 + ( \alpha_{00} -R) (R^5 - 45 \alpha_{22})} \, . 
\label{eq:bc_alpha_coupled} 
\end{eqnarray}
(R-dependence has been suppressed for simplicity).  Again, for $R \to
\infty $ we have for non exceptional values of the parameters $ C^0_s
\to 0 $ , $ C^0_{sd} \to 0 $ and $ C^0_{d} \to -2 $. In
Ref.~\cite{Pavon03} a more detailed study on these issues will be
carried out.

\section{Determination of low energy parameters and the theory without
explicit pions}
\label{sec:lecs} 

\begin{figure*}[tbc]
\begin{center}
\epsfig{figure=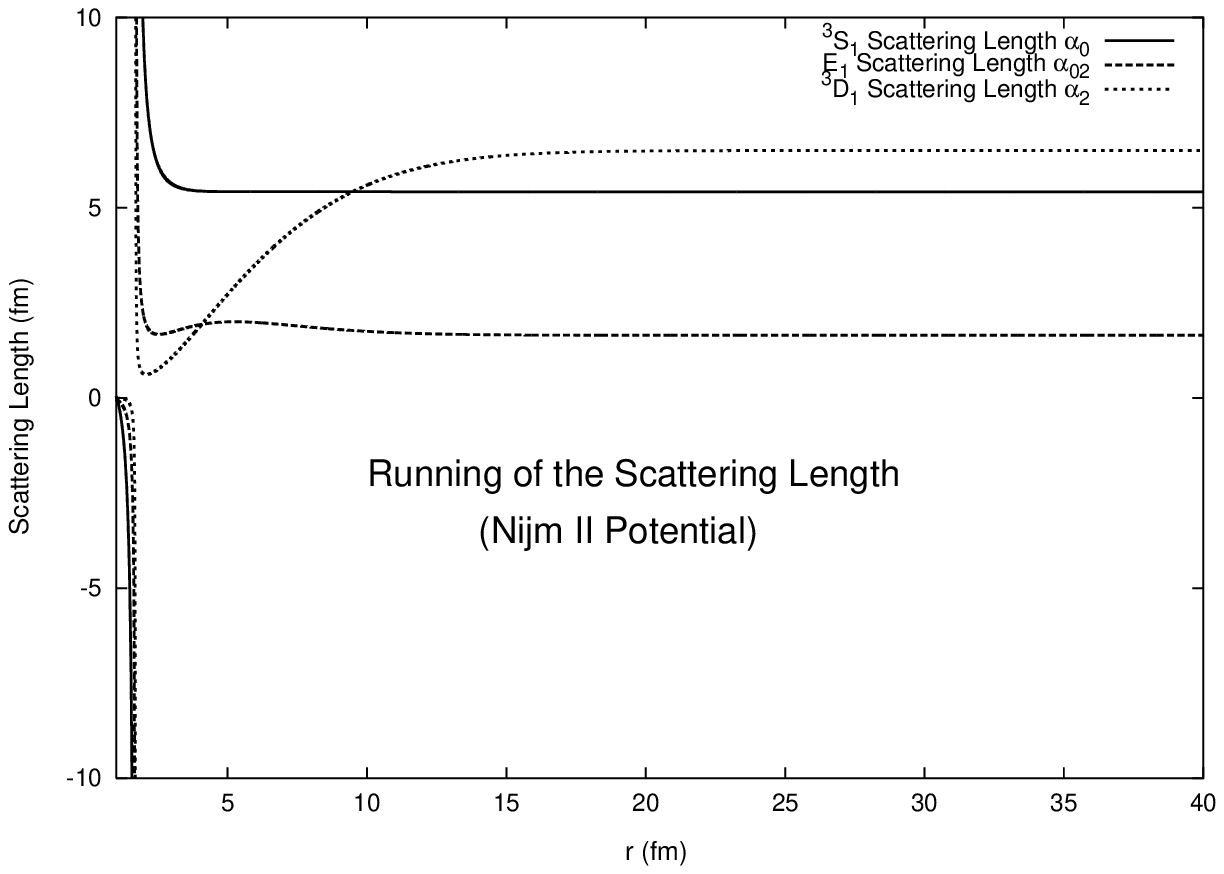,height=6cm,width=6cm}
\epsfig{figure=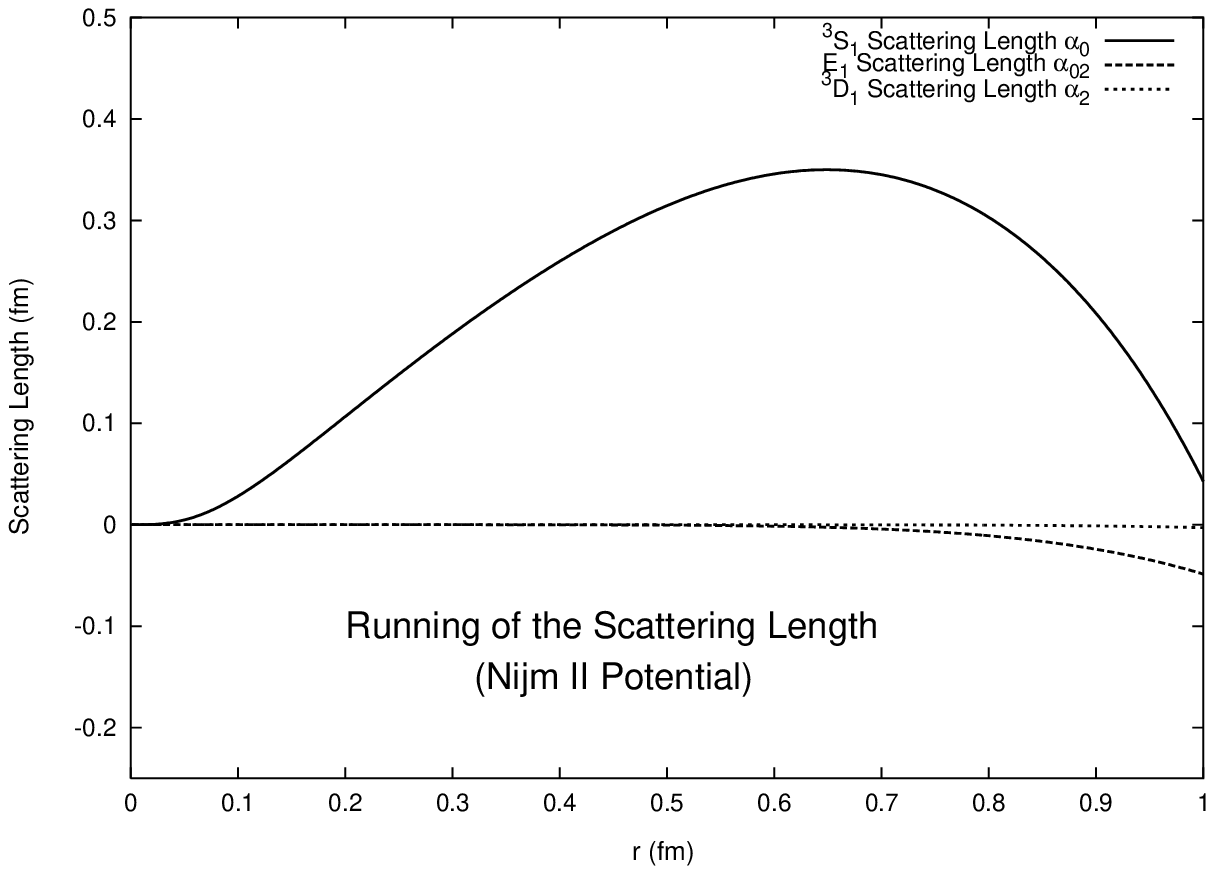,height=6cm,width=6cm}
\end{center}
\begin{center}
\epsfig{figure=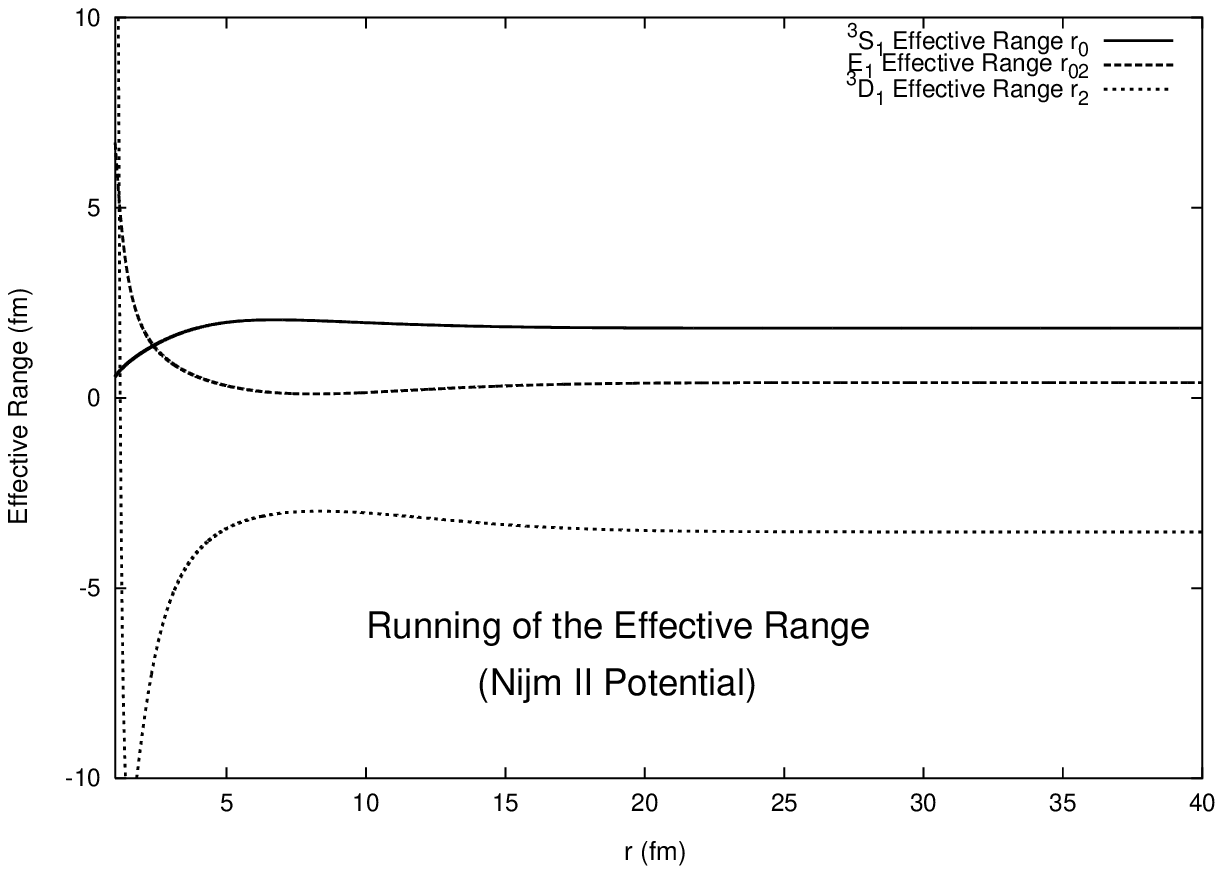,height=6cm,width=6cm}
\epsfig{figure=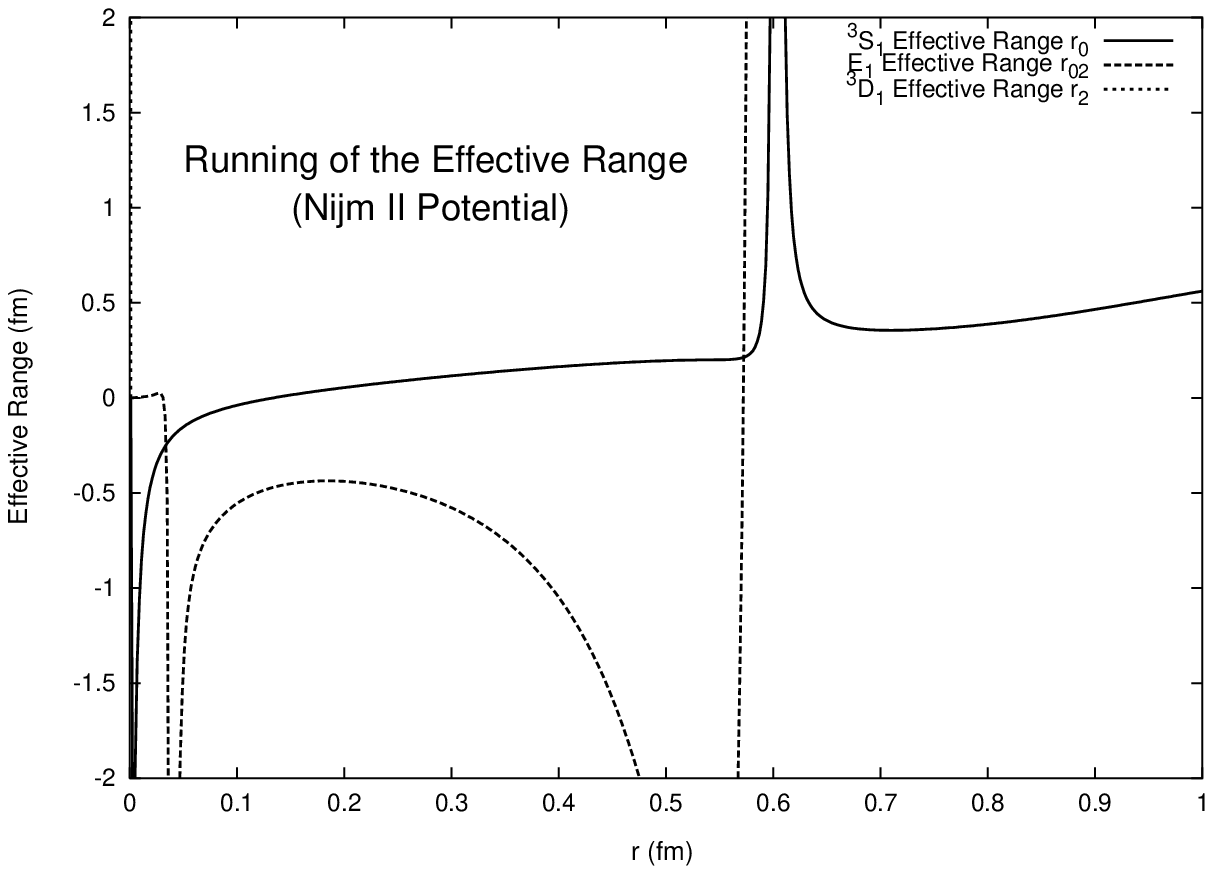,height=6cm,width=6cm}
\end{center}
\begin{center}
\epsfig{figure=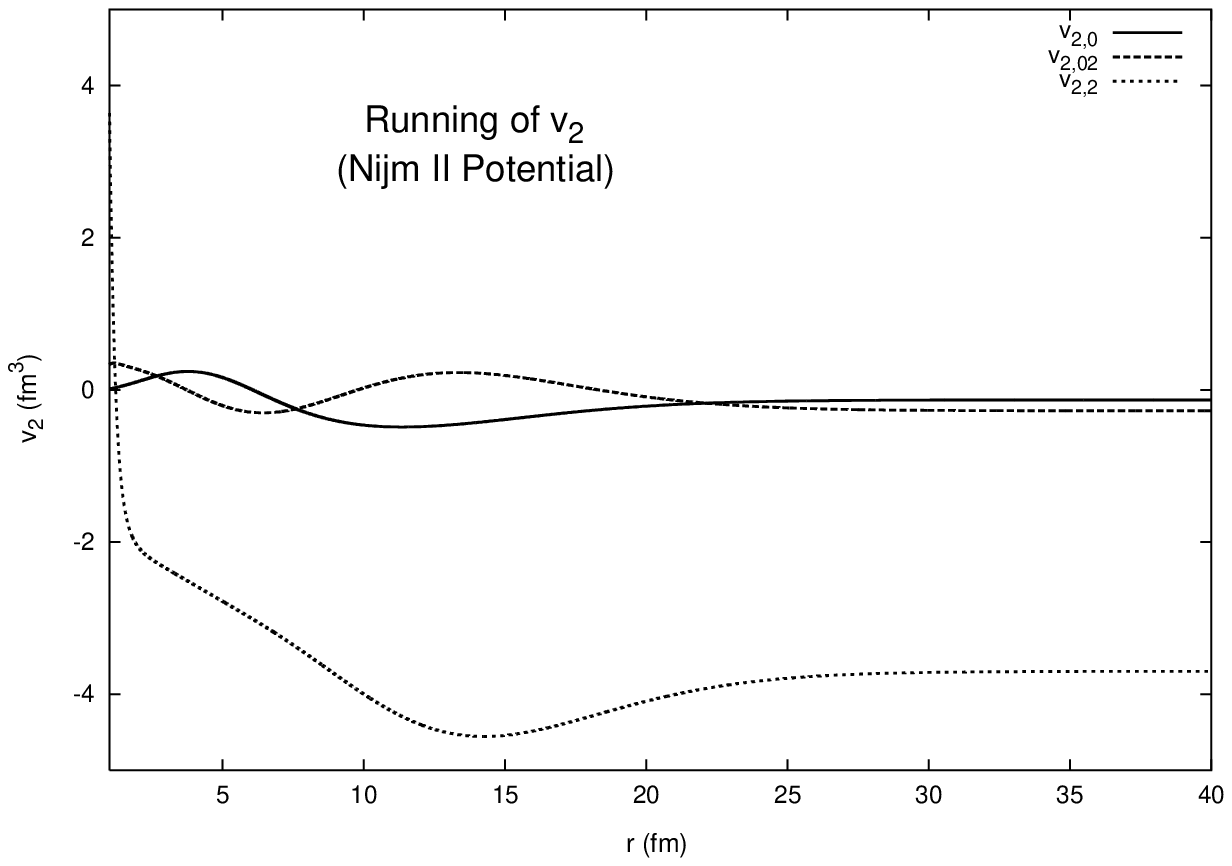,height=6cm,width=6cm}
\epsfig{figure=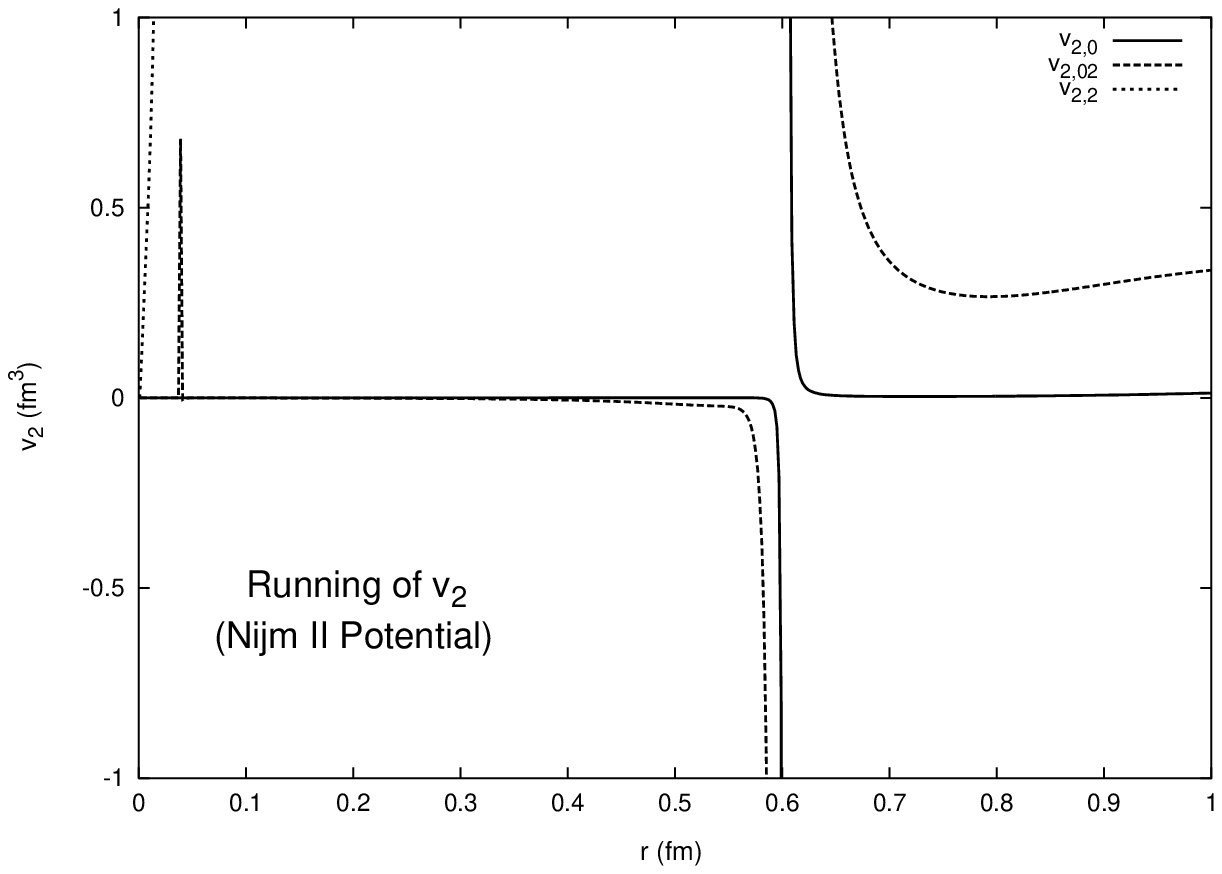,height=6cm,width=6cm}
\end{center}
\caption{Evolution of the $^3S_1$, $^3D_1$ and $E_1$ NN-threshold
parameters from the trivial values at the origin using the NijmII
potential to infinity. Top panel: scattering lengths $\alpha_0 (R)$
(in $\, {\rm fm}$),$\alpha_{02} (R)$ (in $\, {\rm fm}^3$) and
$\alpha_2 (R)$ (in ${\rm fm}^5$). Bottom panel: effective ranges $v_0
(R)$ (in $\, {\rm fm}$),$v_{02} (R)$ (in $\, {\rm fm}^3$) and $v_2
(R)$ (in ${\rm fm}^5$).  In the left panel we represent a global
picture from the 1fm to 40 fm. The lower panel is a detailed picture
in the short distance region below 1fm.}
\label{fig:threshold_evol_up}
\end{figure*}

An essential ingredient of our formalism is to parameterize the
scattering data directly in terms of low energy threshold parameters,
like $\alpha$, $r$ and $v$, defined through
Eq.~(\ref{eq:c-ere}). Unfortunately, besides $ \alpha $ and $r_0$ in
the singlet and triplet channels, the PWA data
base~\cite{Stoks:1993tb} does not provide values for them.  They could
be obtained from a fit to the NN data base in the pertinent channels,
at sufficiently low energies.  Such a procedure turns out to be
numerically unstable, particularly for the $v$ parameter, because it
depends very strongly on the energy window chosen for the fit (see
Appendix \ref{sec:app3}). On the other hand, the NN data base provides
explicit potentials, some of them local like the NijmII and Reid93
potentials, for which the variable phase approach may directly be
applied. In such a way we can uniquely and accurately determine all
the needed low energy threshold parameters by integrating
Eqs.~(\ref{eq:var}) {\it upwards} from the origin to infinity with
trivial boundary conditions. For illustration purposes the evolution
for the NijmII potential for the threshold parameters is depicted in
Fig.~(\ref{fig:threshold_evol_up}). We remind that these curves
represent the low energy threshold parameters corresponding to a
potential truncated at a given distance, $R$. The behaviour at the
origin has to do with the strong repulsive core of the potential in
the $s-$wave channel.  Also, the divergence of the scattering lengths
at about $ R \sim 2 {\rm fm }$ signals that a bound stated has
appeared. The values at infinity correspond to the physical
values. Our results, in appropriate powers of {\rm fm}, can be
summarized as follows for the NijmII and the Reid93 (in brackets)
potential
\vskip.5cm 
\begin{itemize} 
\item \underline{Singlet $^1S_0$ NijmII (Reid93)}
\begin{eqnarray}
\alpha_0 = -23.74(3) \, , \quad r_0 = 2.67(75) \, ,
\quad v_2 = -0.48(9) 
\label{eq:1S0-exp}
\end{eqnarray} 
\item \underline{Triplet $^3S_1$ without mixing NijmII (Reid93)}
\begin{eqnarray}
\alpha_0 = 5.001(3) \,, \quad r_0 = 1.833 \, , \quad v_2
= 0.131(41)  
\label{eq:3S1-exp}
\end{eqnarray} 
\item \underline{Triplet $^3S_1-^3D_1$  with mixing NijmII (Reid93)} 
\begin{eqnarray}
\\ {\bf a} &=&
\left(\matrix{ 5.419(22) & 1.647(6) \cr - & 6.504(453) } \right) \\ {\bf r} &=&
\left(\matrix{ 1.833 & 0.404(12)\cr - & -3.522(66) } \right) \\
{\bf v} &=&
\left(\matrix{ -0.131(41) & -0.274(64) \cr - & -3.70(80)} \right) 
\label{eq:3S13D1-exp} 
\end{eqnarray} 
\end{itemize}
The $^3S_1$ channel without mixing parameters have been obtained from
the $^3S_1-^3D_1$ channel for the $^3S_1$ component, and $\alpha_0 =
1/ ( {\bf a^{-1}})_{00} $, complying to the low energy expansion of the
scaled $ \hat {\bf K}$ matrix, Eq.~(\ref{eq:c-ere}). Although we will
be using the NijmII parameters, we have also presented the ones
corresponding to the Reid93 case to provide an idea on the size of
errors.

\begin{figure*}
\begin{center}
\epsfig{figure=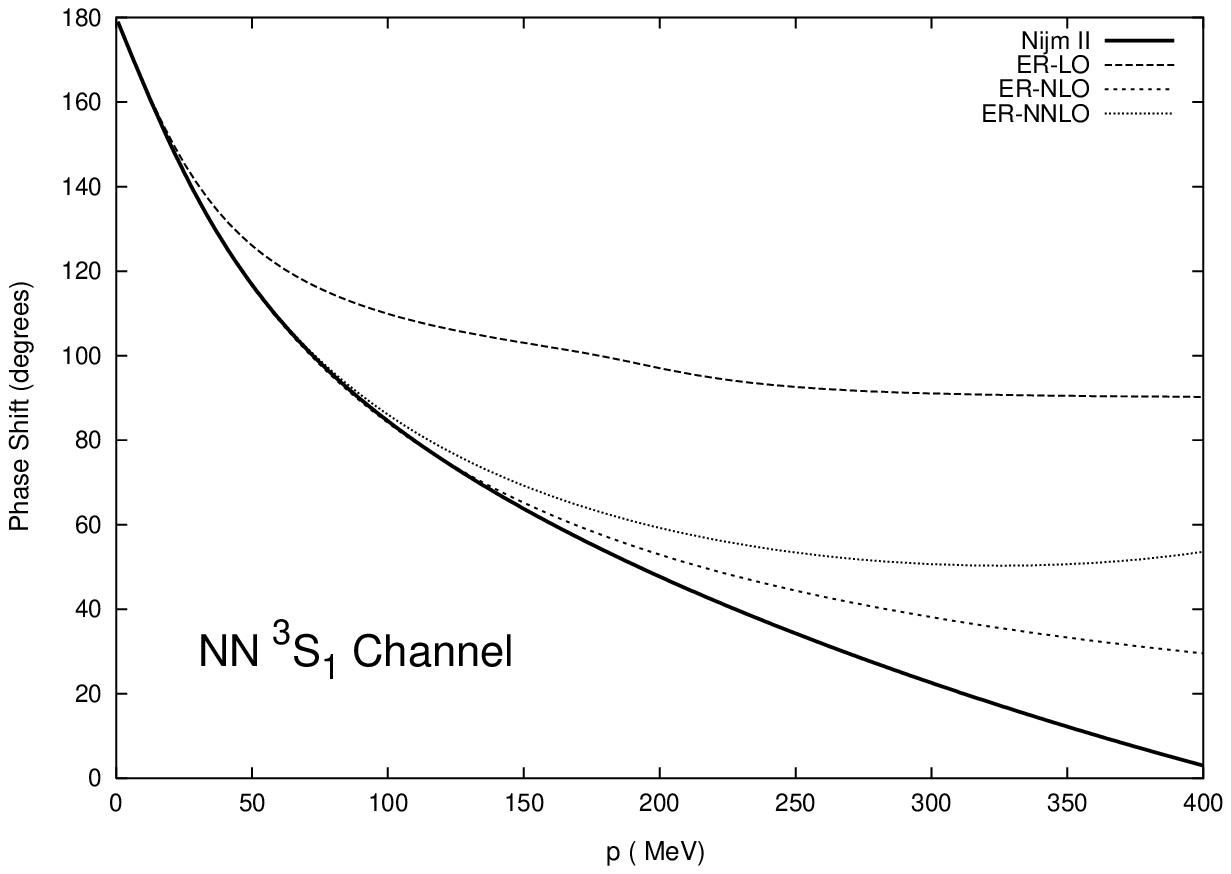,height=8cm,width=8cm} 
\epsfig{figure=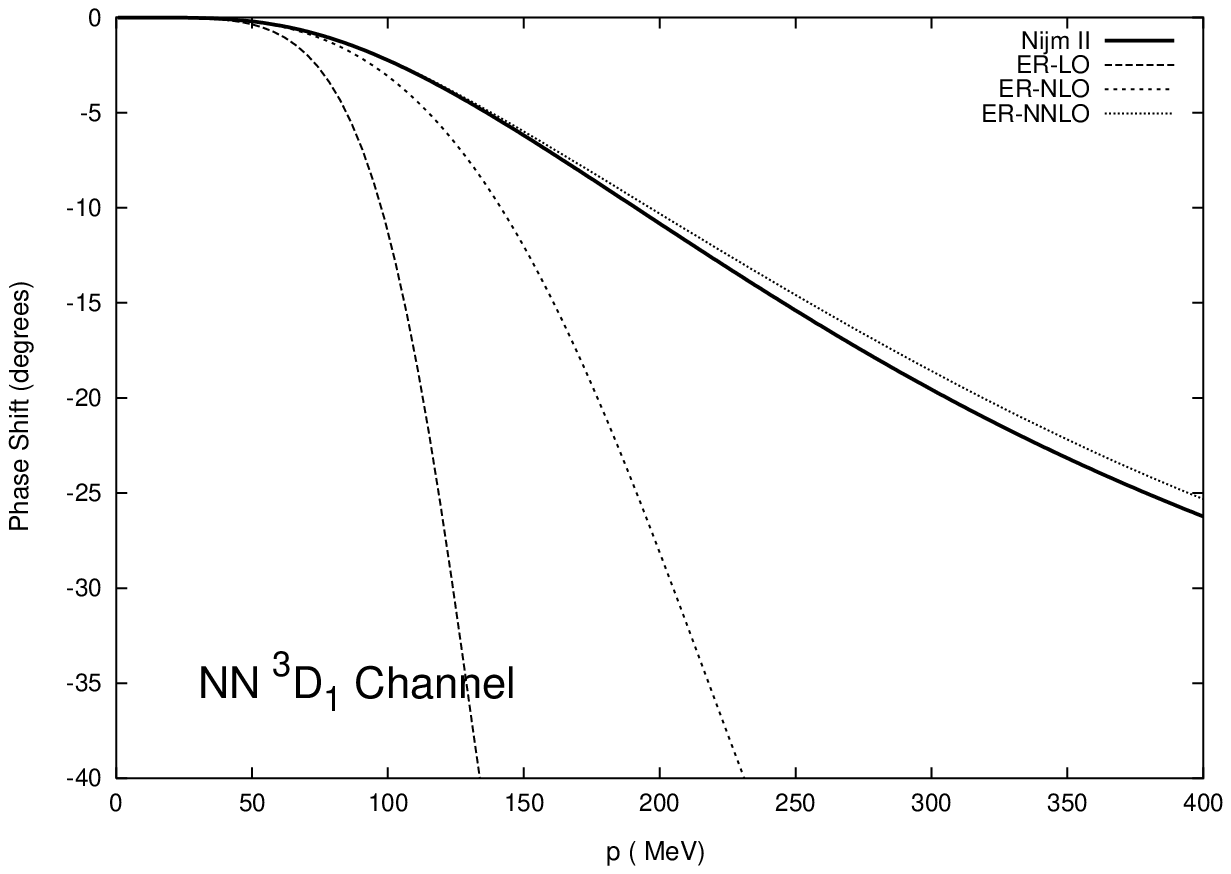,height=8cm,width=8cm} 
\epsfig{figure=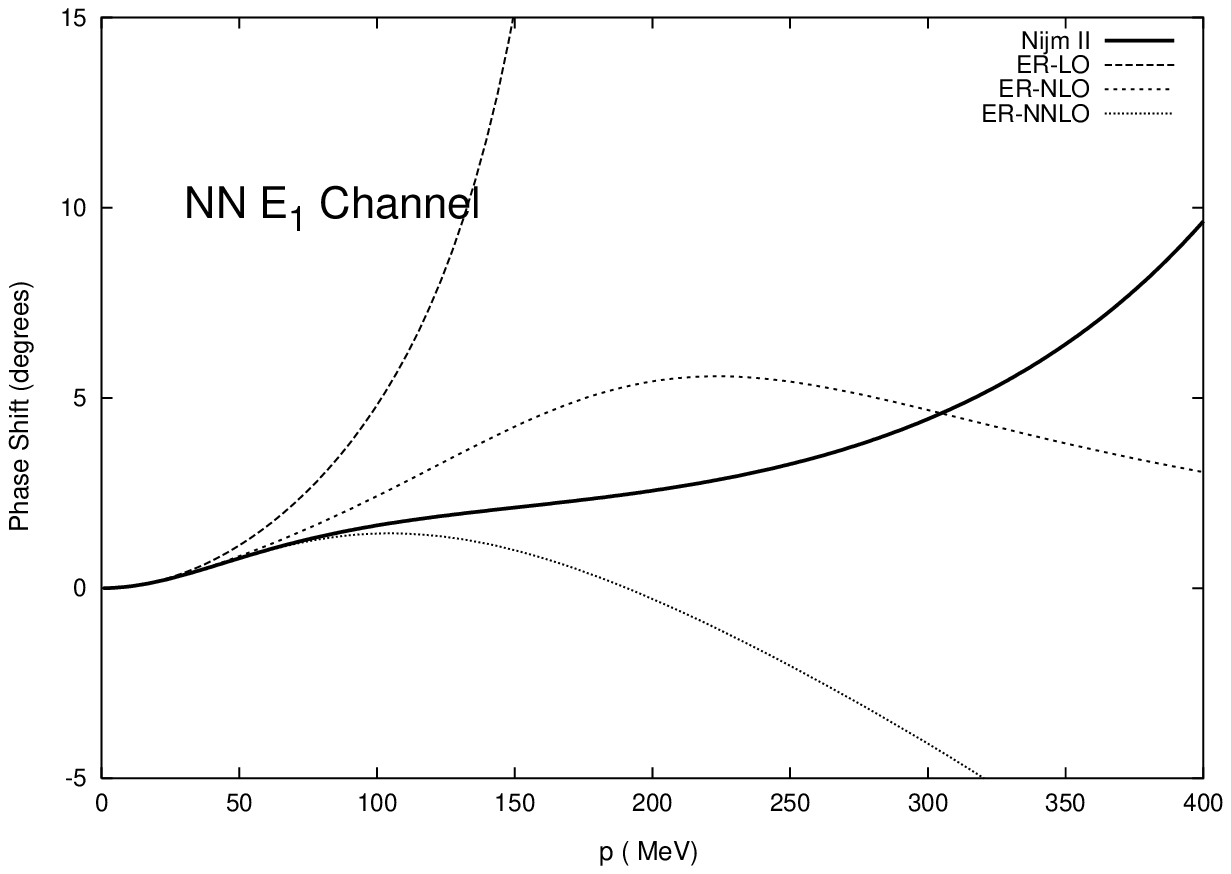,height=8cm,width=8cm} 
\end{center}
\caption{Theory without explicit pions. $^3S_1$, $^3D_1$ and $E_1$ at
LO (contact terms), NLO ($k^2$ terms) and NNLO ( $k^4$ terms)
predicted phase shifts for the triplet channel within a pure effective
range expansion approximation, with the low energy threshold
parameters obtained from solving the evolution equations for the
threshold parameters, Eq.~(\ref{eq:var}) with the Nijm II potential
and using the reduced K-matrix, Eq.~(\ref{eq:c-ere}). Data are the PWA
from Ref.~\cite{Stoks:1993tb}.}
\label{fig:phase_shifts_ere}
\end{figure*}

Once the threshold parameters have been determined we can use the
coupled channel effective range expansion, Eq.(\ref{eq:c-ere}) to find
out to what extent does this expansion apply. On theoretical grounds
we expect this expansion to converge within the region of analyticity
of the $S$-matrix, which presents a left cut at $ k= \pm {\rm i} m_\pi
/2 $. In Fig.~(\ref{fig:phase_shifts_ere}) we compare the quality of
the ERE including LO, NLO and NNLO contributions to the original data
of Ref.~\cite{Stoks:1993tb}. As we see, to describe the data within
the ERE approach up to the convergence radius $m_\pi /2 $ one has to
go at least to NLO. The description of the data below $m_\pi /2 $ is
improved, as expected, with higher orders in the ERE. Above this
region, where OPE should play a role, this is not necessarily so.
Actually, we see that in the $^3S_1$ end $E_1$ channels the NNLO is
worse than the NLO approximation. We emphasize that these curves are
not obtained from a fit to the data. 

\section{Short distance behaviour for OPE: Fixed Points and Limit cycles} 
\label{sec:short} 

In this section we analyze the short distance behaviour of the
equations for the scattering lengths for the singlet $^1S_0$,
Eq.~(\ref{eq:valpha}), and the triplet $^3S_1$, Eq.~(\ref{eq:va_c}),
channels in the short distance limit. According to Eq.~(\ref{eq:bc_s})
this is equivalent to study the mixed boundary condition at short
distances.

We study first the case of OPE in the singlet $^1S_0$ channel.  At
short distances $R << 1/m_\pi $ the OPE potential behaves like the
Coulomb potential. Eq.~(\ref{eq:valpha}) can be easily solved in two
extreme cases, $ \alpha_0 << R $ and $\alpha_0 >> R $. While in the
first case we get
\begin{eqnarray}
\alpha_0(R) & \to & - \frac{g_A^2 m_\pi^2 M_N }{32 \pi
f_\pi^2} R^2  \qquad , \quad \alpha_0 << R
\label{eq:asy_reg} 
\end{eqnarray}
in the second case one solution behaves as
\begin{eqnarray}
\alpha_0(R) & \to \frac{16 \pi f_\pi^2}{g_A^2 m_\pi^2 M_N }
\frac1{\log (R/R_0)} \qquad , \quad \alpha_0 << R
\label{eq:asy} 
\end{eqnarray}
where $R_0$ is a reference scale fulfilling $ R < R_0 << 1/m_\pi $. As
we see, $\alpha_0(R)$ goes to zero in both case but, while
Eq.~(\ref{eq:asy_reg}) goes rapidly $\alpha_0'(R) \to 0 $,
Eq.~(\ref{eq:asy}) goes very slowly and with $ \alpha_0' (R) \to
-\infty$ at short distances. In momentum space the $R \to 0$ limit
corresponds to the ultraviolet limit. Eq.~(\ref{eq:asy}) resembles a
sort of asymptotic freedom and, hence we have an ultraviolet fixed
point.  One can see that the first case, Eq.~(\ref{eq:asy_reg}),
corresponds to selecting the regular solution at the origin, whereas
Eq.~(\ref{eq:asy}) corresponds to a generic case, which always
contains an admixture of the irregular solution. The regular case at
the origin corresponds to integrate from the origin starting with the
trivial initial condition $\delta(k,0)=0$ up to infinity.  As we have
discussed in Ref.~\cite{Valderrama:2003np} the result corresponds
to a pure OPE interaction, with no short-distance interactions. The
important thing to realize is that regardless of the value of
$\alpha_0$ at infinity, removing one-pion exchange goes into the same
value at the origin, as implied by Eq.~(\ref{eq:asy}). This also
implies that any tiny deviation of the $\alpha_0(R )$ at small
distances results in huge variations at infinity. Thus, removing OPE
in the $^1S_0$ channel implies an extreme fine tuning of the
scattering length at short distances, and hence of the boundary
condition at the origin.

We turn now to the case of the $^3S_1-^3D_1$ channel, where the tensor
force plays a role. In the region close to the origin the wave
function oscillates wildly and hence a WKB approximation may be
used. The calculation is simplified by taking into account that for
the OPE interaction the potential matrix is diagonalized by an
$r$-independent unitary transformation, i.e.
\begin{eqnarray} 
{\bf M} {\bf U}(r) {\bf M}^{-1} = \left( \matrix{ U_C(r)-4 U_T (r) & 0
\cr 0  & U_C(r) + 2 U_T (r)  } \right)  \, , \nonumber \\ 
\end{eqnarray} 
with 
\begin{eqnarray}
{\bf M}= \left( \matrix{ -\frac1{\sqrt{2}} & 1 \cr \sqrt{2} & 1 }
\right) \, . 
\end{eqnarray}  
Note that this transformation {\it does not} diagonalize the full
potential $ \U + {\bf l}^2 /r^2 $ including the centrifugal barrier,
which for $r\to 0 $ may be neglected.  Thus, in the short distance
limit we may decouple all our equations into pairs, and in particular
we can apply the transformation to the boundary condition,
Eq.~(\ref{eq:bc}) at zero energy
\begin{eqnarray} 
{\bf M} {\bf L}_0 (R) {\bf M}^{-1} = {\rm diag} ( l_1(R) , l_2(R) )
\, , 
\end{eqnarray}  
where $l_1 (R) $ and $l_2(R) $ are the logarithmic derivatives at zero
energy of the decoupled problem with potentials $ U_1= U_C - 4 U_T $
and $ U_2= U_C + 2U_T $ respectively. After straightforward algebra we
get
\begin{eqnarray}
\alpha_0 (R) &=& {3 } \frac{ R l_2 (R) ( R l_1(R) +1) -2 }{4
l_2 (R) + l_1 (R) ( 3 R l_2 (R) +2 ) } \, , \\
\alpha_{02} (R) &=& -\frac{\sqrt{2} R^3 }3 \frac{ l_1
(R) - l_2 (R) }{4 l_2 (R) + l_1 (R) ( 3 R l_2 (R) +2 ) } \, , \\ 
\alpha_{2}
(R) &=& \frac{R^5 }{15} \frac{ l_1 (R) (R l_2 (R) -1)- 2 l_2 (R) }{4 l_2
(R) + l_1 (R) ( 3 R l_2 (R) +2 ) } \, . 
\end{eqnarray} 
Now, as we approach the origin the tensor potential dominates, and the
potential $U_1$ and $U_2$ behave as repulsive and attractive $ 1/r^3 $
potentials respectively, corresponding to take $ l_1 \to \infty $ and
$ l_2 (R) $ by the zero energy limit of the logarithmic derivative of
a WKB function,
\begin{eqnarray}
\alpha_0 (R) & \to & {R } \frac{ 3 R l_{\rm WKB} (R) }{ 3 R
l_{\rm WKB} (R) +2  } \, , \\
\alpha_{02} (R) &\to & -\frac{\sqrt{2} R^3 }3 \frac{1}{ 3 R l_{\rm
WKB} (R) +2  } \, , \\
\alpha_{2} (R) & \to & \frac{R^5 }{15} \frac{R l_{\rm WKB} (R) -1 }{
3 R l_{\rm WKB} (R) +2  } \, , 
\end{eqnarray} 
with 
\begin{eqnarray}
R l_{\rm WKB} (R) = \frac{3}{4} + \frac12\sqrt{\frac{R_M }{ R}} \cot \left(
\Delta + \sqrt{\frac{R_M }{R_0}} - \sqrt{\frac{R_M }{R}} 
\right) \, , \nonumber \\ 
\end{eqnarray} 
Here $\Delta $ is an energy independent phase, and $R_0$ a reference
point, given by 
\begin{eqnarray}
R_0 l_{\rm WKB} (R_0 ) = \frac{3}{4} + \frac12 \sqrt{\frac{R_M }{R_0}}
\cot \left( \Delta \right) \, ,
\end{eqnarray} 
and  
\begin{eqnarray}
R_M= \frac{3 g^2 M }{2 f^2 \pi}  = 16 {\rm fm} \, .  
\end{eqnarray} 
As we see the scattering lengths $\alpha_0 $, $\alpha_{02} $ and
$\alpha_2$ present on oscillatory behaviour as we approach the origin,
so they do not converge to a well defined value; as we approach to the
origin the $\alpha$'s take all possible values. This situation
corresponds to a limit cycle at short distances. A way of avoiding the
unbound variation of the scattering lengths consists of going to the
origin stepwise through some envelope subsequence defined by a fixed
condition for $ l_{\rm WKB} (R) $. For instance, if we define a cycle
by the condition $\alpha (R_n)=0$, we have $ R_n l_{\rm WKB} (R_n) =0
$, yielding
\begin{eqnarray}
\alpha_0 (R_n) & = & 0 \nonumber \\
\alpha_{02} (R_n) & = & -\frac{\sqrt{2} R_n^3 }6 \nonumber \\
\alpha_{2} (R_n ) & = & -\frac{R_n ^5 }{30} \nonumber 
\end{eqnarray} 
Another possibility would to take $ l_{\rm WKB} (R_n) = \infty $ in
which case one has
\begin{eqnarray}
\alpha_0 (R_n) & = & {R_n} \nonumber \\
\alpha_{02} (R_n) & = & 0 \nonumber \\
\alpha_{2} (R_n ) & = & \frac{R_n ^5 }{30} \nonumber  
\end{eqnarray} 
As we see, there are infinitely many such possibilities, although all
of them go towards the trivial values, $\alpha_0 (0^+ ) = \alpha_{02}
(0^+) = \alpha_2 (0^+) =0$. Actually, any of the choices corresponds
to a different starting condition at infinity, modulo a
cycle. Conversely, if we go to very short distances, where the
scattering lengths vary wildly; any tiny perturbation there results in
a completely different value at infinity. So, we see again that an
extreme fine tunning of the threshold parameters at short distances is
required. Finally, let us mention that close to the origin the
sequence of cycles can be determined by the solution of the equation
\begin{eqnarray} 
\frac32 + x_n \cot( \Delta + x_n ) = 0 \quad , \qquad x_n =
\sqrt{\frac{R_M}{R_n}} \, 
\end{eqnarray}   
in the limit $x_n \to 0$.

In practical numerical calculations the finite integration step
$\Delta R$ provides a given resolution scale, and these infinite limit
cycles may not observed due to the rapid oscillations. Instead, one
sees the envelope corresponding to the stationary points of the
scattering lengths. This point will become clear below,
Sect.~\ref{sec:numerics}.

\section{NON-PERTURBATIVE SOLUTIONS} 
\label{sec:numerics} 

\subsection{Evolution of the Low energy parameters} 

The exact mathematical analysis of the general set of Equations is
rather complicated since we are dealing with a non-linear system of
equations. In Ref.~\cite{Pavon03} simple cases are analyzed
analytically and the general features which can be deduced there are
consistent with the numerical results we have obtained in the present
work.

As we have said the set of equations, Eq.~(\ref{eq:valpha}) and
Eq.~(\ref{eq:va_c}) can be numerically solved. Given the
fact that as we approach the origin the tensor part of the potential
develops a singularity it is important to carefully check for
numerical accuracy at short distances. A crucial property which must
be fulfilled by any algorithm is that of exact reversibility;
i.e. evolving upwards or downwards should be inverse operations of
each other. This is an stringent test and, moreover, the only way to
make sure that when the long range piece of the potential is switched
on for the K-matrix integration we have consistency with the effective
range expansion up to the relevant order (see also below). We prefer
to impose this reversibility exactly, independently on the number of
mesh points used in the integration, so that any numerical
irreversibility is merely attributable to computer arithmetic
round-off errors. This feature will prove extremely relevant when
computing the phase shifts below since our calculation requires
upwards integration from lower distances. In all calculations
presented in this paper we have checked that the correct threshold
behaviour is obtained.

Quite generally, we find stable results when we take the long distance
cut-off to be $R_\infty = 20 {\rm fm} $. On the other hand the lowest
radius we can achieve numerically and preserving reversibility is $R_S
=0.1 {\rm fm} $, mainly due to computer arithmetic round-off errors
triggered by the singularity of the potential. One could further lower
the radius by a semi-classical approximation as outlined in
Sect.~\ref{sec:short} since as the origin is approached the wave
function undergoes an increasing number of oscillations and WKB
methods can be applied. Nevertheless, as we will see below, for our
short distance cut-off the phase shifts for CM momenta up to $k = 250
{\rm MeV}$ are rather stable numerically.

The strong dependence of the low energy threshold parameters on the
short distance cut-off provides a clue to the fact that there seems to
be a lower finite limit for the boundary radius $R_S=R_{\rm min} \sim
1.4 {\rm fm}$~\cite{Rentmeester:1999vw} with still an acceptable fit;
if the boundary radius is lowered, the parameters encoding the short
distance boundary condition which are used as fitting parameters
depend in a non-smooth way on $R_S$. In addition, the strong
singularity at the origin triggers a fine tuning in those
parameters. According to our previous discussion, this short distance
fine tuning of low energy parameters is absolutely necessary to comply
with the independence of the scattering amplitude on the short
distance boundary radius.  For such a situation, a fit based on
successive adiabatic changes of $R_S$ becomes impractical since the
fitting parameters do not change adiabatically and also because these
parameters should have to be determined to extraordinary high
precision.  In addition, the way how the limit $R_S \to 0 $ should be
taken differs from channel to channel. Our method provides a practical
way to overcome the difficulty, given the fact that the boundary
radius is taken exactly to zero along the renormalization trajectories
while keeping the low energy threshold parameters at fixed values.

\subsubsection{$^1S_0 $ and $^3S_1$-without mixing channels} 

In Fig.~(\ref{fig:threshold_evol}) we show our results for the
evolution of the threshold parameters $\alpha_0$, $r_0$ and $v_2 $ in
the singlet $^1S_0$ and triplet $^3S_1$-without mixing
(i.e. neglecting the tensor force) channels. The main difference one
can appreciate from the comparison of both channels is that while the
scattering length for the $^1S_0$ channel exhibits a monotonic trend
towards the origin, the scattering length in the $^3S_1$ channel
diverges at a distance of about $0.7 {\rm fm} $. The interpretation of
this fact in our framework is clear; the central part of the OPE
potential is purely attractive. Thus, by eliminating the pions down to
a certain distance, we are effectively building some repulsion, until
we lose a bound state. An alternative interpretation is that as we
switch on the OPE potential from the origin up to certain distance we
can accommodate a bound state above $0.7 {\rm fm} $. With this
interpretation in mind, we should add $180^0 $ to the $^3S_1$
phase shift to comply with Levinson's theorem.

\subsubsection{$^3S_1-^3D_1$ channel} 

\begin{figure*}[tbc]
\begin{center}
\epsfig{figure=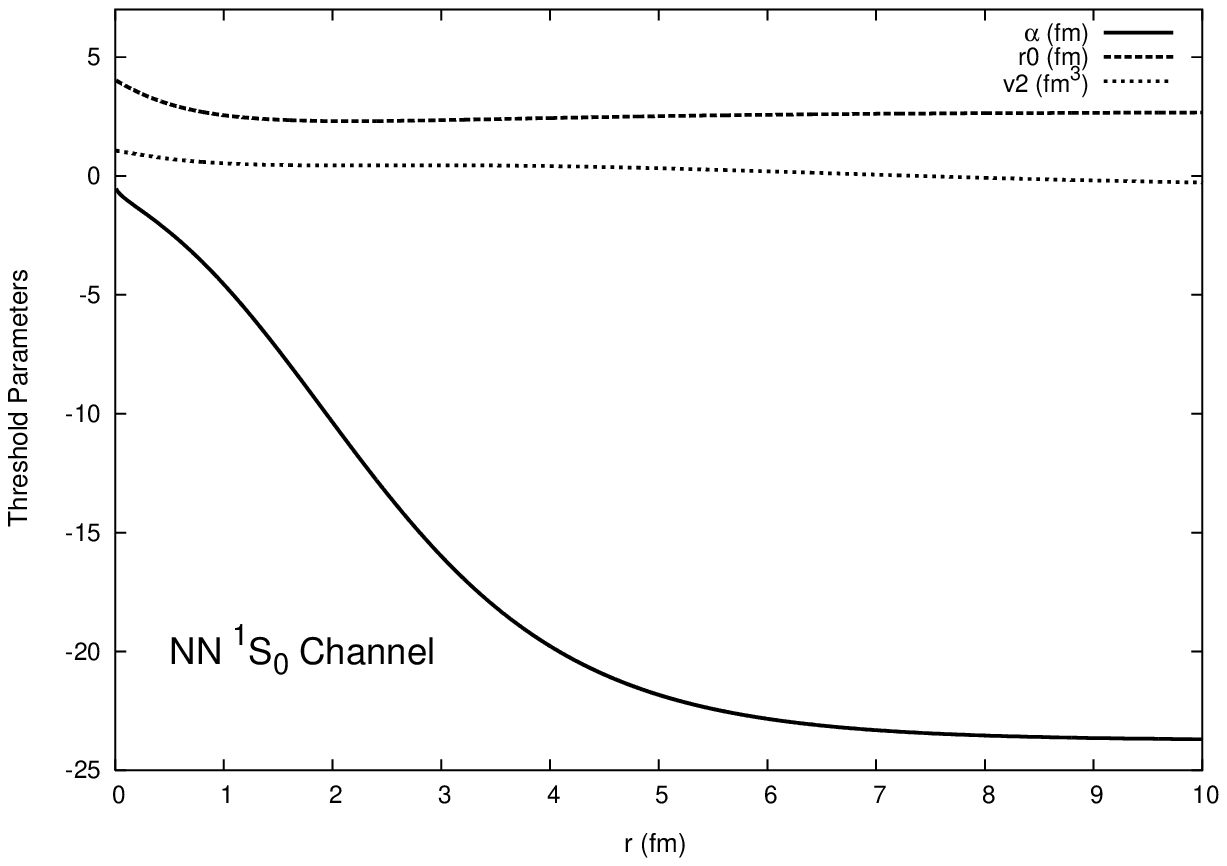,height=8cm,width=8cm}
\epsfig{figure=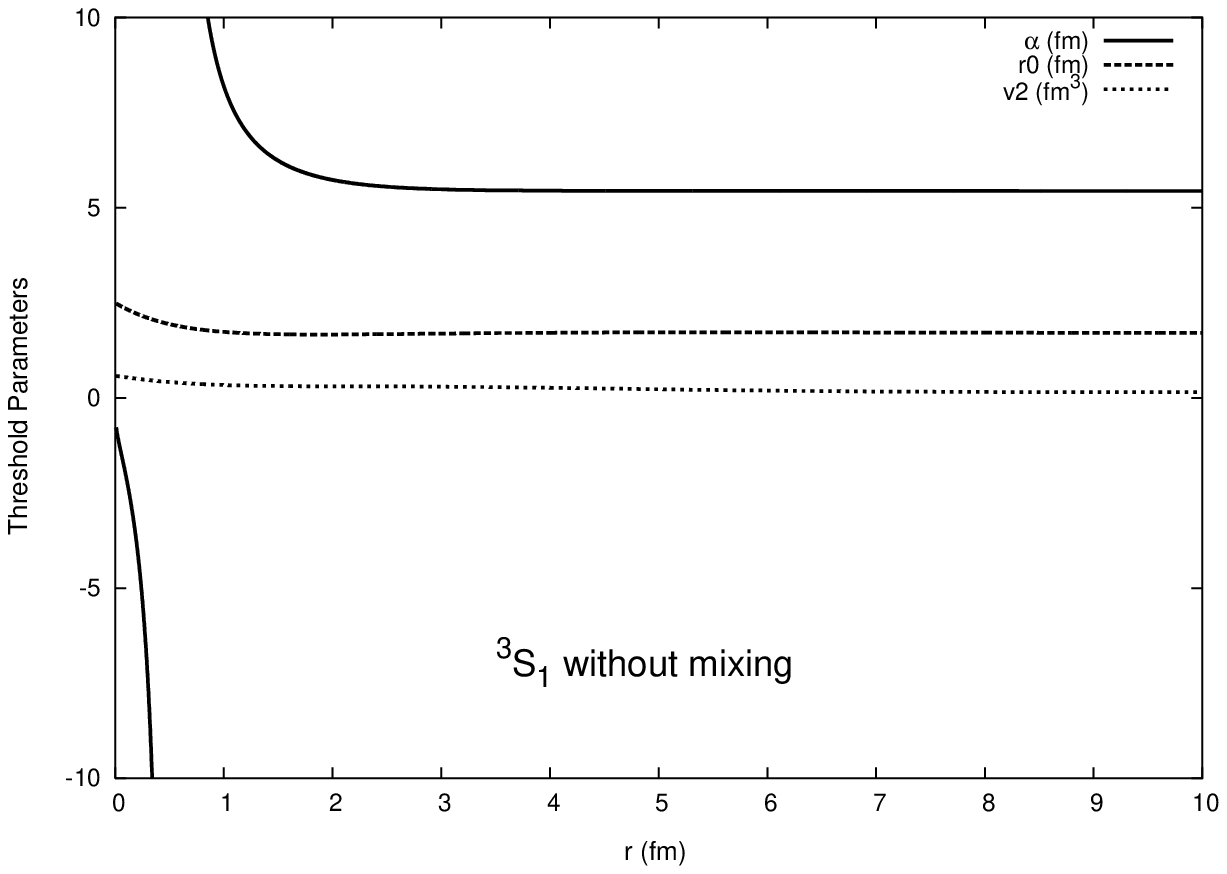,height=8cm,width=8cm}
\end{center}
\caption{Evolution of the scattering $^1S_0$ (left panel) and
$^3S_1$-without mixing (right panel ) NN-threshold
parameters $\alpha_0(R)$ (in $\, {\rm fm}$),$r_0(R)$ (in $\, {\rm
fm}$) and $v_2(R)$ (in ${\rm fm}^3$) from the asymptotic values at
infinity (which we take in practice $R_\infty= 20 \, {\rm fm}$ ) when
OPE effects are removed down to the origin.}
\label{fig:threshold_evol}
\end{figure*}

We finally analyze the triplet $^3S_1 - ^3D_1$-channel taking into
account the tensor mixing. In Fig.~(\ref{fig:threshold_evol}) we show
our numerical solutions of the set of Eqs.~(\ref{eq:var}). Starting at
sufficiently long distances (in practice $R_\infty = 20 {\rm fm}$
turns out to be adequate) and evolve downwards to the
origin. Operationally this corresponds to eliminate OPE in the triplet
channel. As can be clearly seen for distances above $ R \sim 3{\rm fm}
$ nothing dramatic happens and a monotonic trend is observed. At
smaller distances $ \sim {\rm 2 fm}$ , however, we note a rapid change
in the running scattering lengths. Again, a rather flat evolution
follows until the region below $ 1 {\rm fm}$. A zoomed picture is
plotted in the lower panel for distances shorter than $1 {\rm fm}$
where the cyclic structure of evolution becomes evident, as expected
from our analysis in Sect.~(\ref{sec:short}). The number of cycles
increases without any bound as the origin is approached. This
situation is dramatically different from that found in the case
without tensor mixing, since there OPE produced an ultraviolet fixed
point. The situation we encounter here is not new and has already been
described in the context of non-coupled channels. The limit cycle
structure naturally raises the problem of undefined values of the
short distance parameters as we take the limit $R \to 0$. The point is
that there is a way of taking the limit through equivalent points
defined by the property $ \alpha (R_n) = \alpha (R_{n+1}) $; any two
such points produce identical low energy parameters at infinity. Thus,
the limit $ R_n \to \infty$ through equivalent points produces the
same parameters at long distances. The cycles in $\alpha_{02}$ and
$\alpha_2 $ are hardly seen in the plot due to a low resolution
$\Delta R$ compared with the typical cycle spacing. 

\begin{figure*}[tbc]
\begin{center}
\epsfig{figure=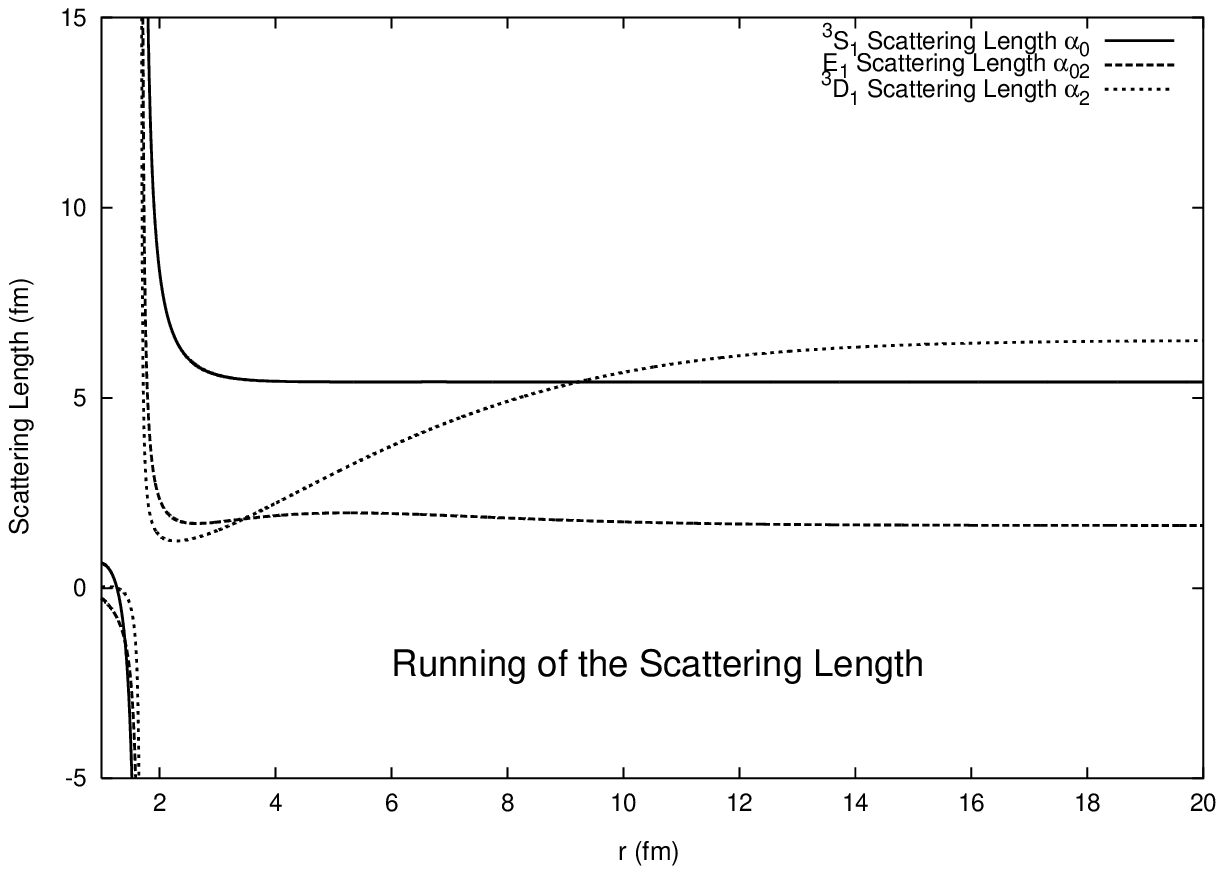,height=6cm,width=6cm}
\epsfig{figure=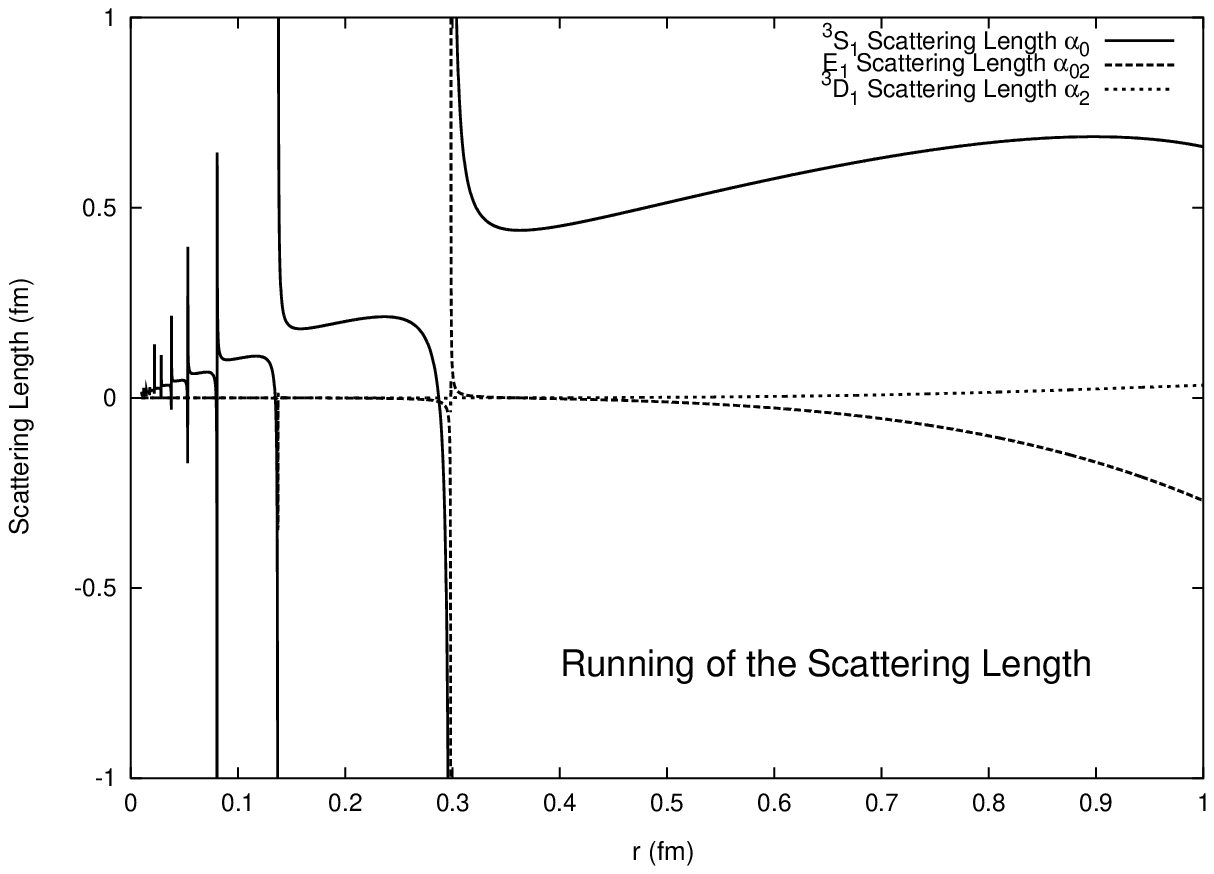,height=6cm,width=6cm}
\end{center}
\begin{center}
\epsfig{figure=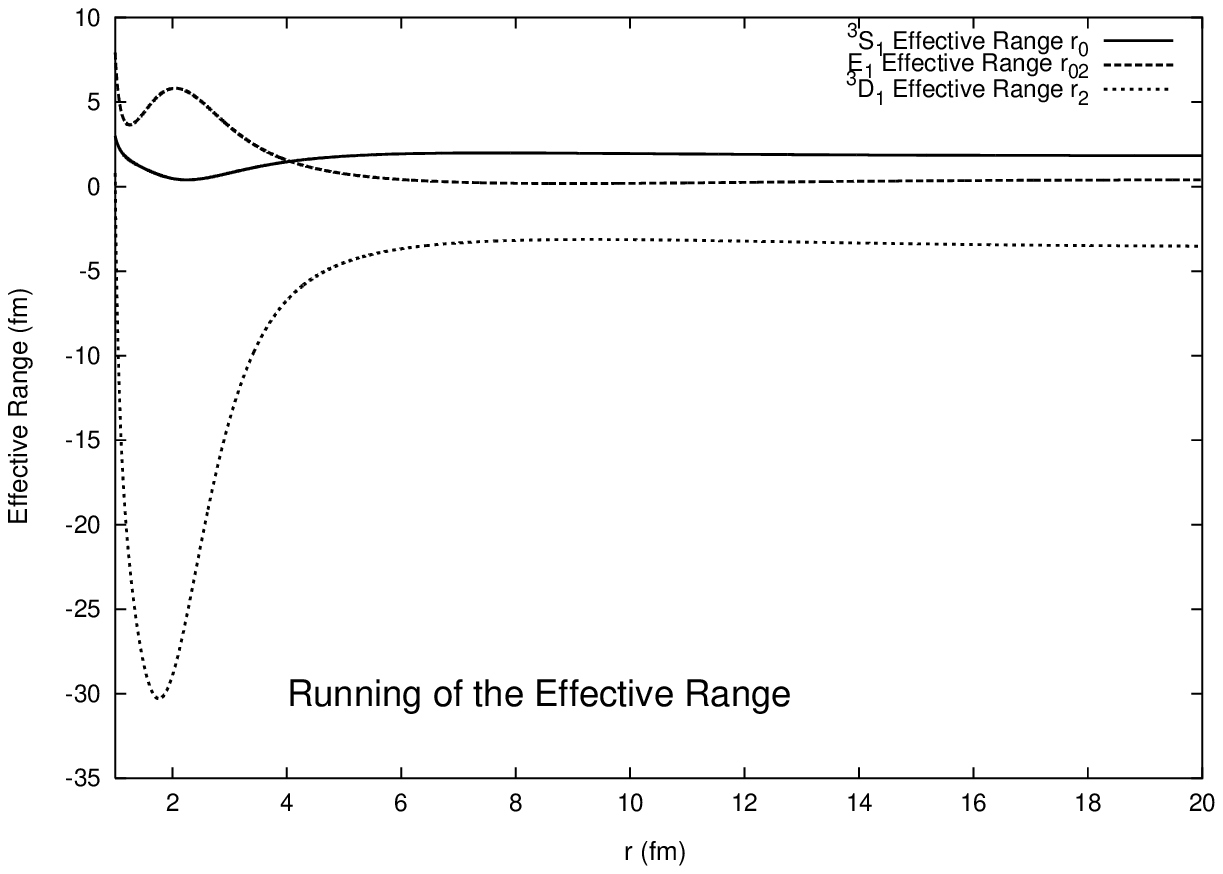,height=6cm,width=6cm}
\epsfig{figure=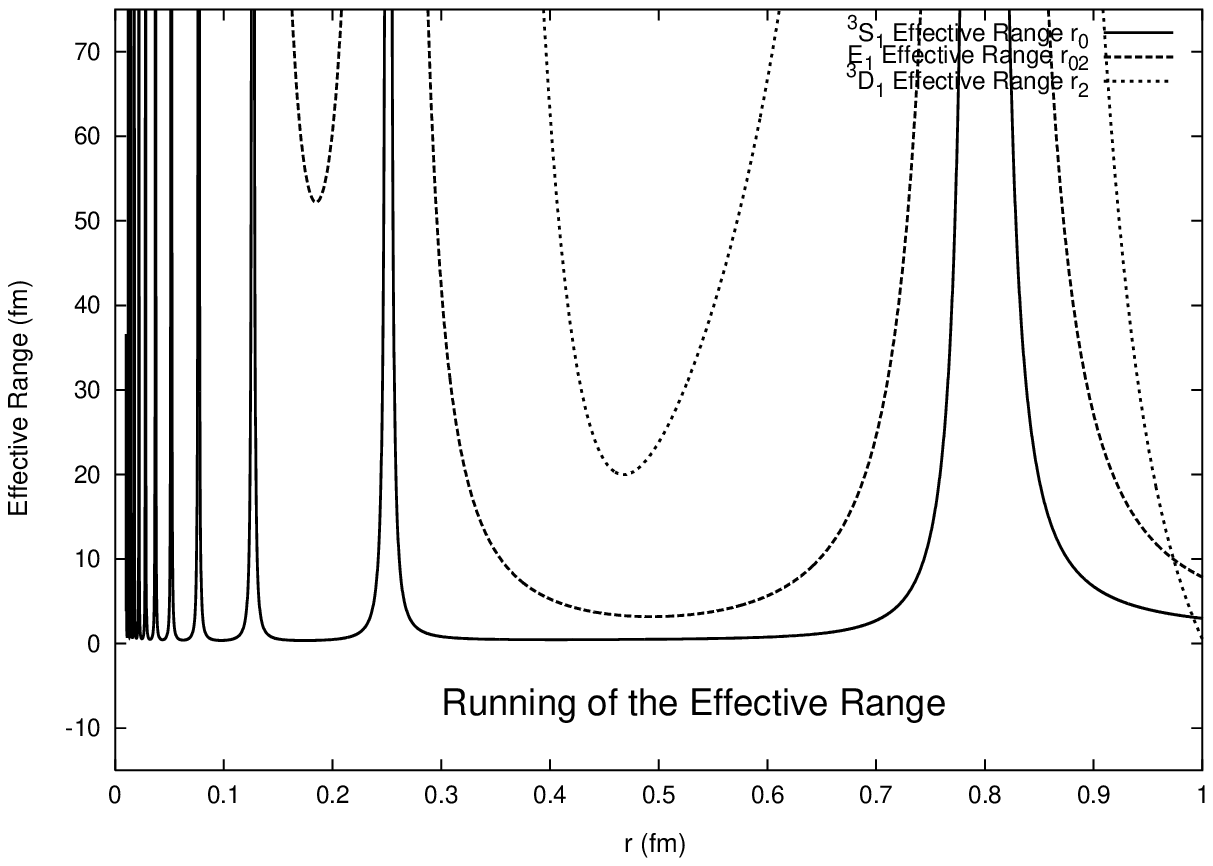,height=6cm,width=6cm}
\end{center}
\begin{center}
\epsfig{figure=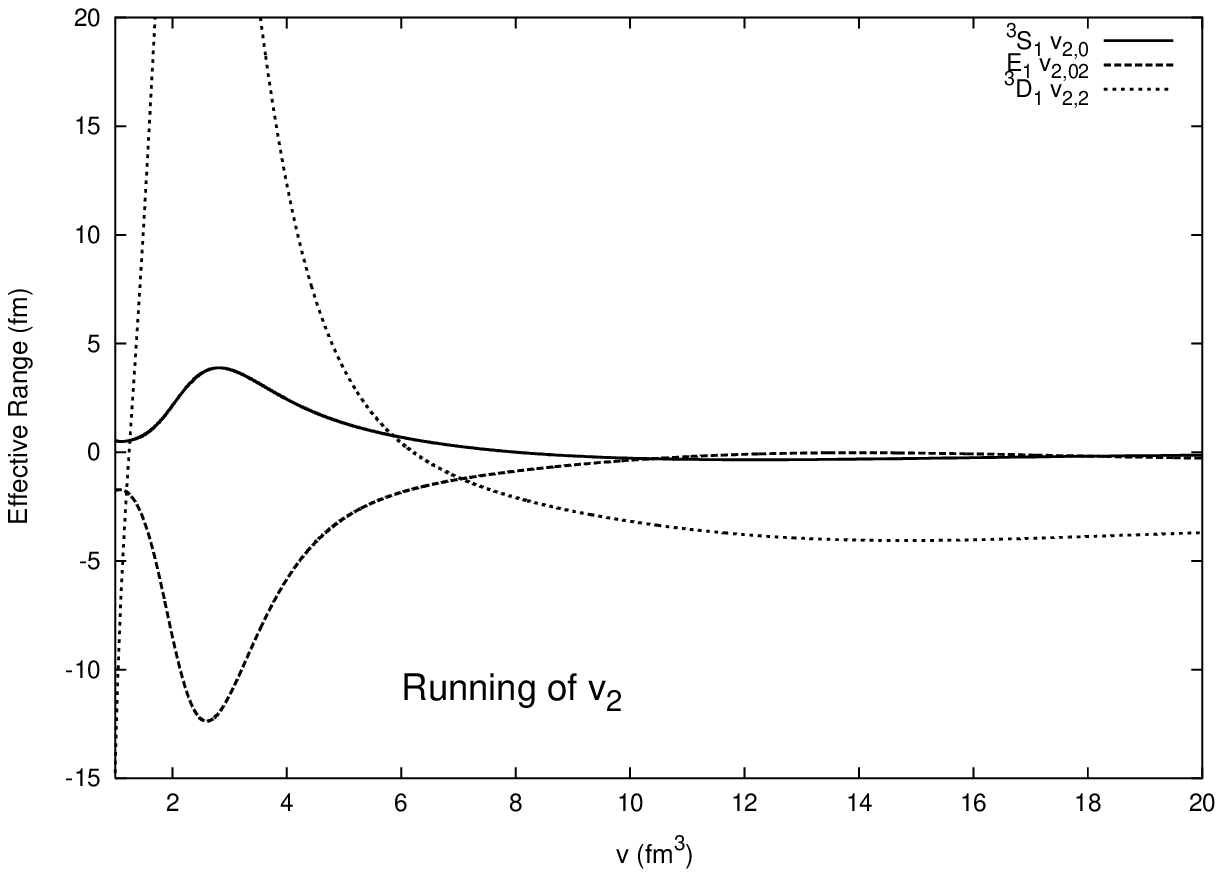,height=6cm,width=6cm}
\epsfig{figure=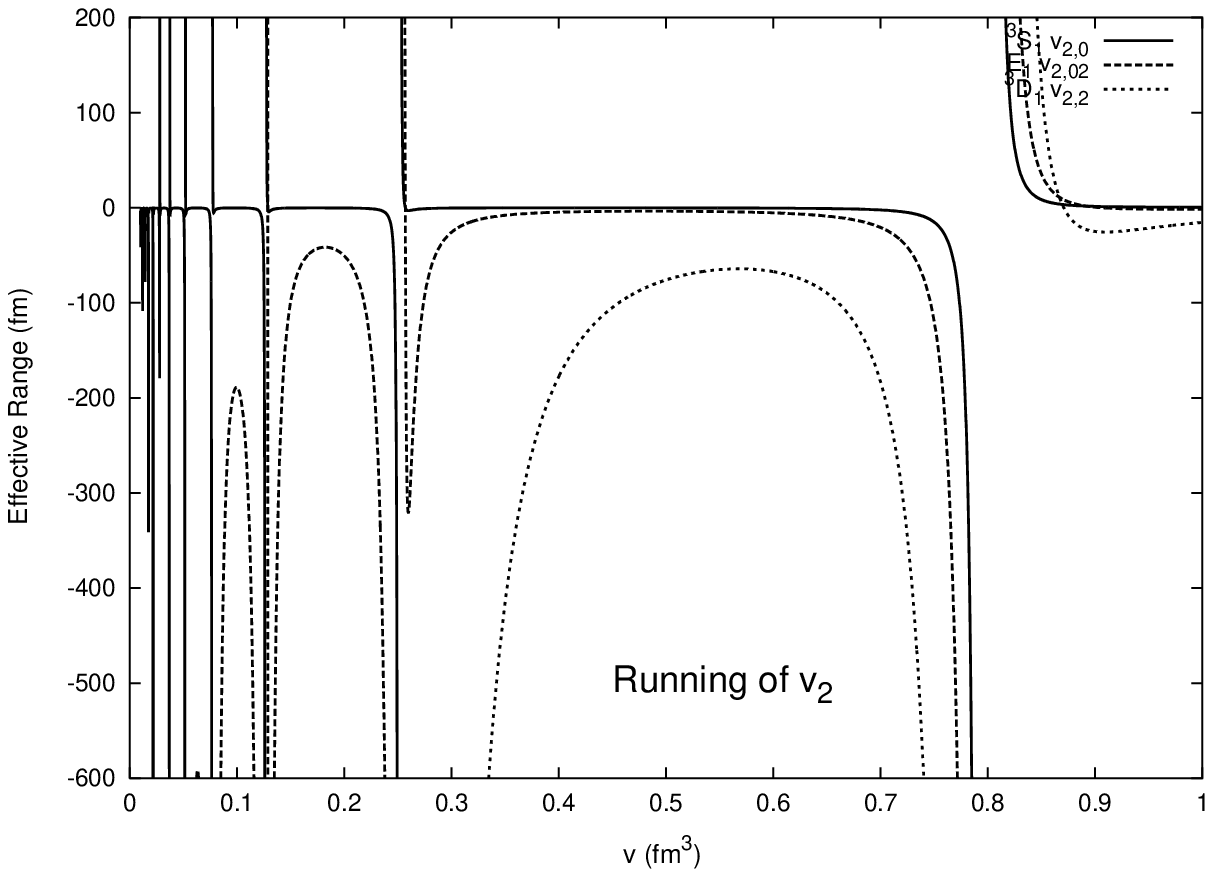,height=6cm,width=6cm}
\end{center}
\caption{Evolution of the $^3S_1$, $^3D_1$ and $E_1$ NN-threshold
parameters from the physical values at infinity down to the origin
using the OPE potential. Top panel: scattering lengths $\alpha_0 (R)$
(in $\, {\rm fm}$),$\alpha_{02} (R)$ (in $\, {\rm fm}^3$) and
$\alpha_2 (R)$ (in ${\rm fm}^5$). Bottom panel: effective ranges $v_0
(R)$ (in $\, {\rm fm}$),$v_{02} (R)$ (in $\, {\rm fm}^3$) and $v_2
(R)$ (in ${\rm fm}^5$).  In the left panel we represent a global
picture from the 1fm to 20 fm. The lower panel is a detailed picture
in the short distance region below 1fm. Limit cycles are clearly
visible in the s-wave scattering length $\alpha_0$ and effective
ranges. The $E_1$ and $^3D_1$ scattering lengths $\alpha_{02}$ and
$\alpha_2$ go quickly to zero below 0.25 fm.}
\label{fig:threshold_evol_down}
\end{figure*}

\begin{figure*}[tbc]
\begin{center}
\epsfig{figure=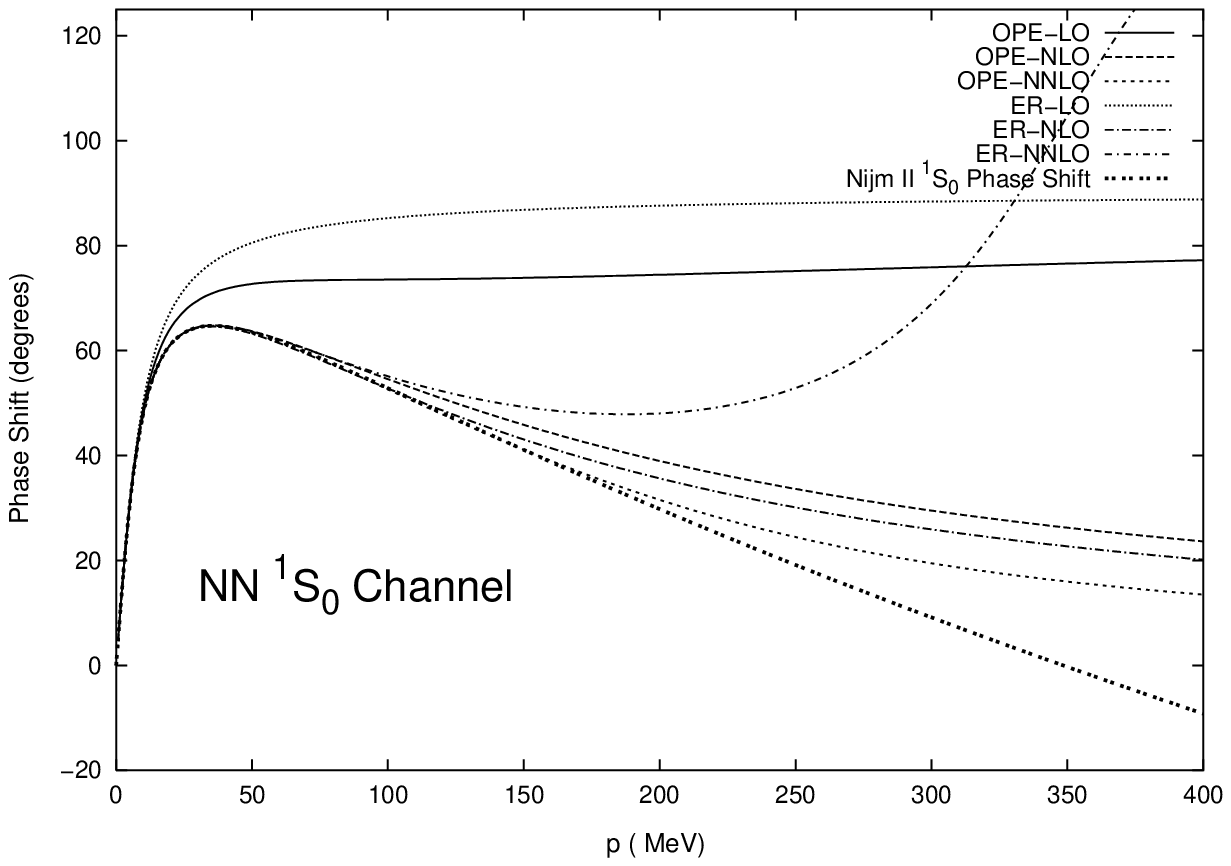,height=8cm,width=8cm}
\epsfig{figure=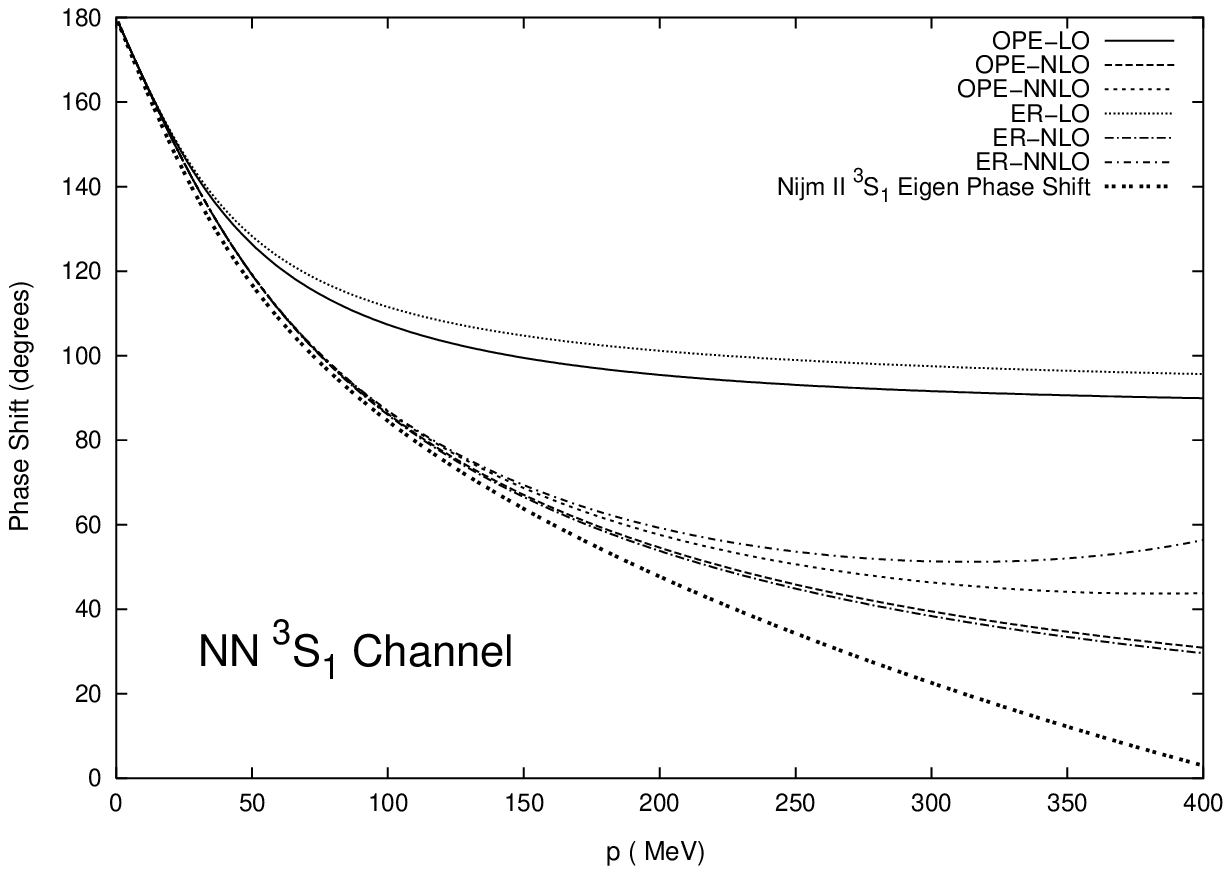,height=8cm,width=8cm}
\end{center}
\caption{Predicted phase shifts in the $^1S_0$ (left panel) and the
$^3S_1$-without mixing (right panel) channels for NN scattering as a
function of the CM momentum in {\rm MeV}. In the $^3S_1$ channel we
assume no mixing according to Eq.~(\ref{eq:vk}) when OPE potential is
switched on and the initial condition is a low energy expansion of the
$K$-matrix at short distances, (See Eq.~(\ref{eq:eff_short}) in the
main text). LO means keeping $\alpha_{S,0}$ only, NLO keeping $
\alpha_{S,0}$ and $r_{0,S}$ and NNLO keeping $ \alpha_{S,0}$,
$r_{0,S}$ and $v_{2,S}$. The short range parameters are directly
determined by evolving the low energy parameters from their
experimental values, Eq.~\ref{eq:1S0-exp}) and Eq.~\ref{eq:3S1-exp})
ER-LO, ER-NLO and ER-NNLO corresponds to a pure effective range
expansion keeping $\alpha_0$ only, $ \alpha_{S,0}$ and $r_{0}$, $
\alpha_0 $, $r_0$ and $v_2$ respectively. No further fit is
involved. Data are the PWA from Ref.~\cite{Stoks:1993tb}.}
\label{fig:one_channel_phase}
\end{figure*}

\begin{figure*}[tbc]
\begin{center}
\epsfig{figure=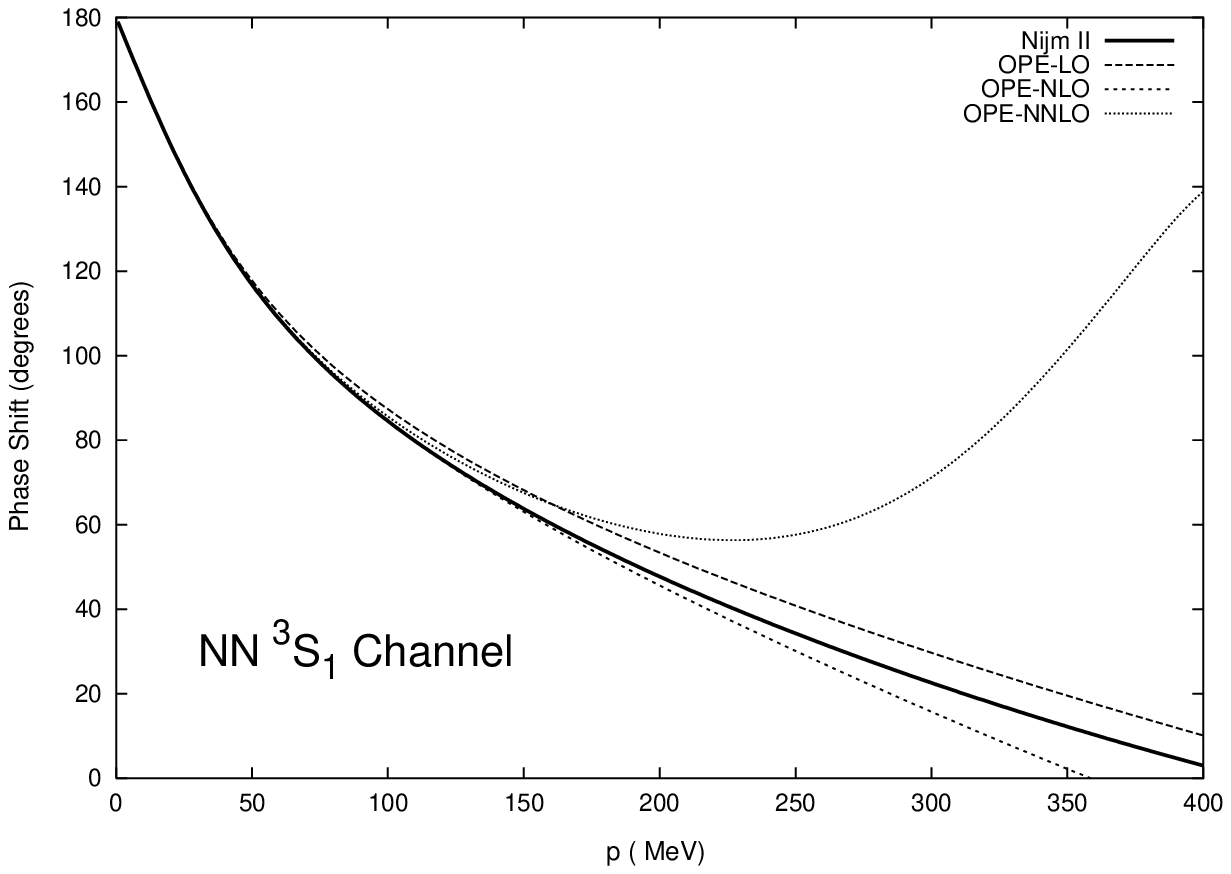,height=8cm,width=8cm} 
\epsfig{figure=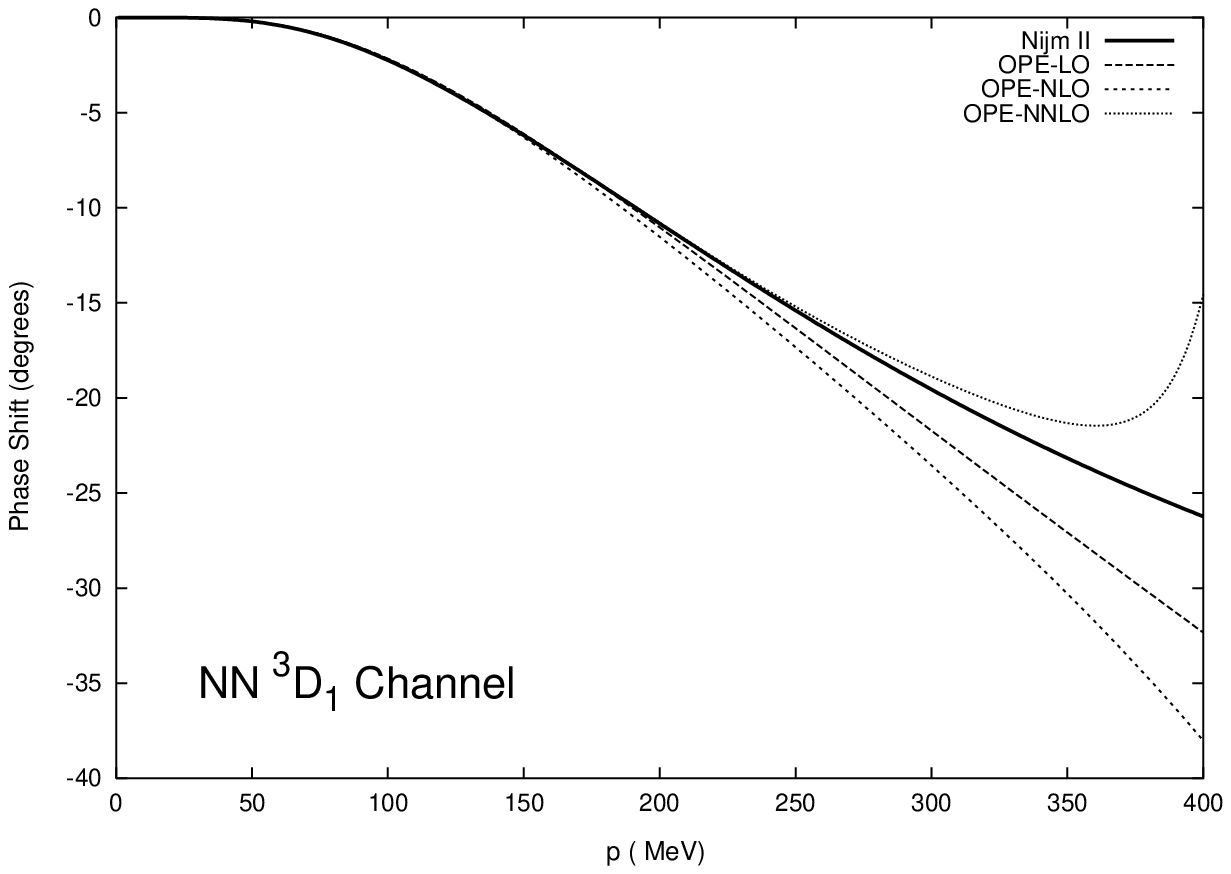,height=8cm,width=8cm} 
\epsfig{figure=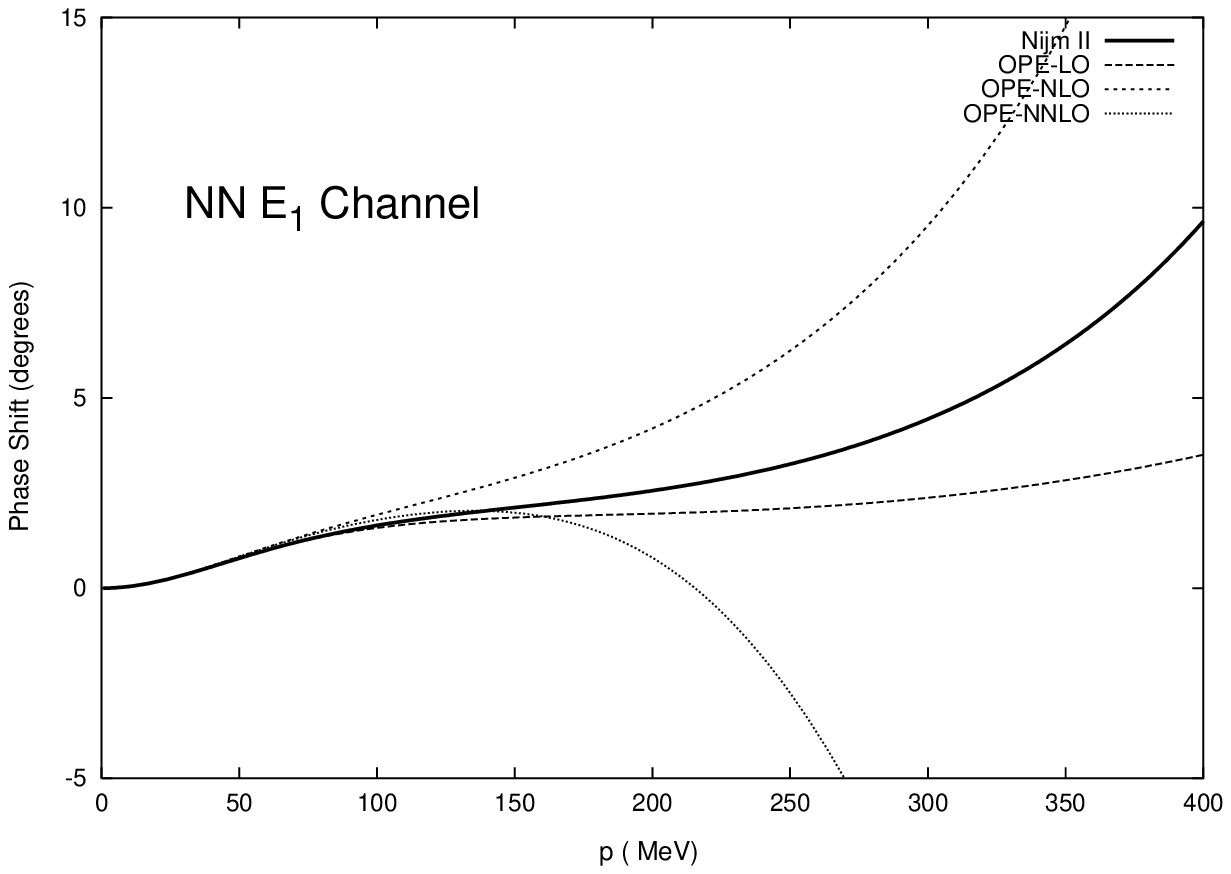,height=8cm,width=8cm} 
\end{center}
\caption{$^3S_1$, $^3D_1$ and $E_1$ at LO (contact terms), NLO ($k^2$
terms) and NNLO ( $k^4$ terms) predicted phase shifts for the triplet
channel. The initial condition is a low energy expansion of the
$K$-matrix at short distances. By construction the low energy
parameters $ \alpha$ , $r$ and $v$ coincide with those extracted from
the NijmII potential. Data are the PWA from Ref.~\cite{Stoks:1993tb}.}
\label{fig:phase_shifts}
\end{figure*}

\begin{figure*}[tbc]
\begin{center}
\epsfig{figure=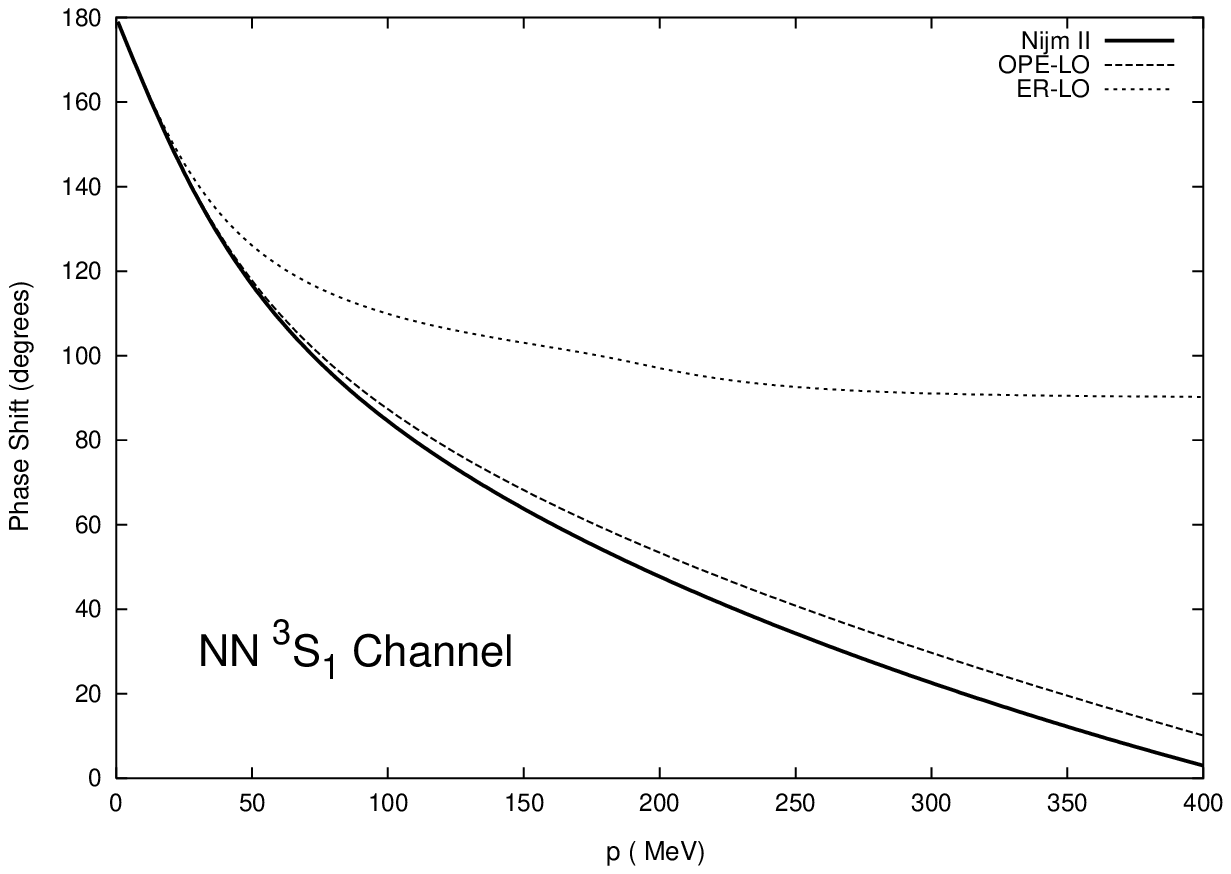,height=5.5cm,width=5.5cm} 
\epsfig{figure=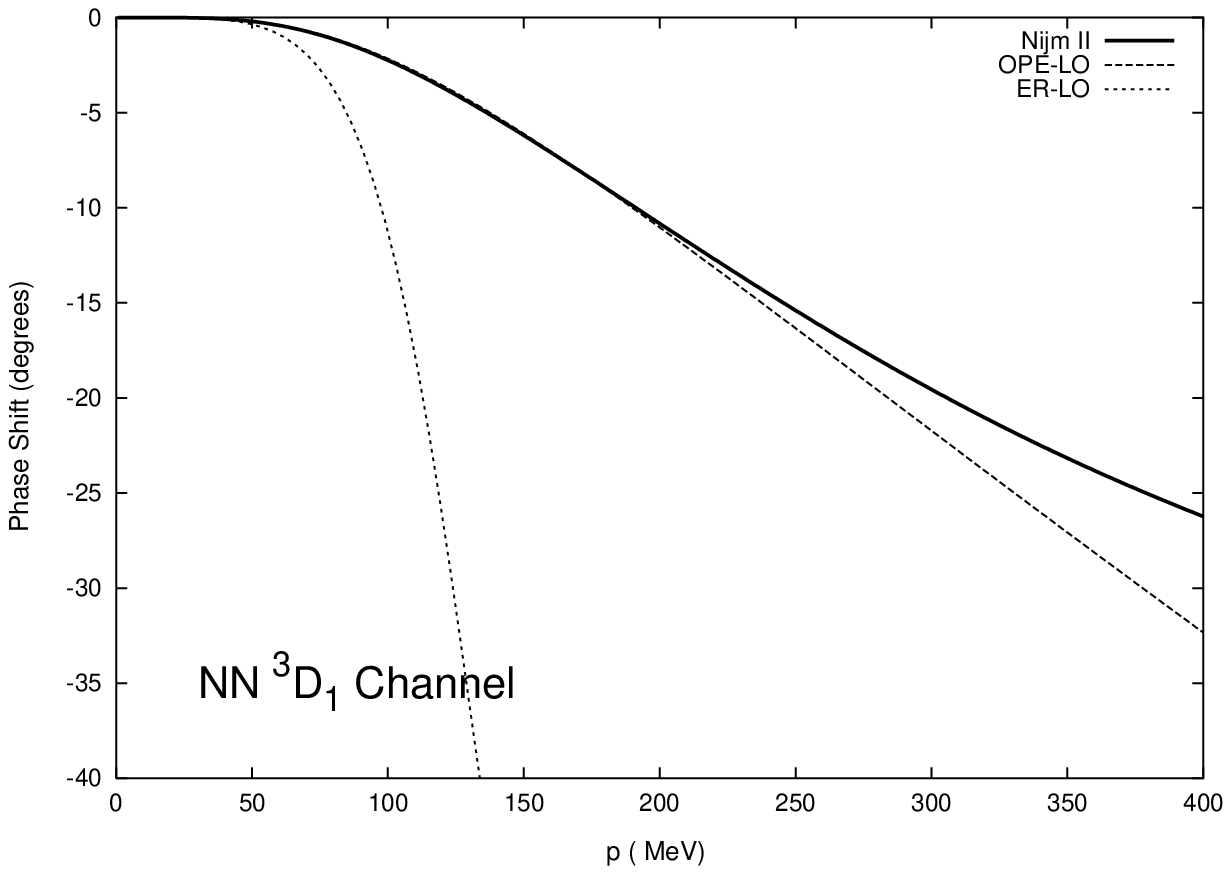,height=5.5cm,width=5.5cm} 
\epsfig{figure=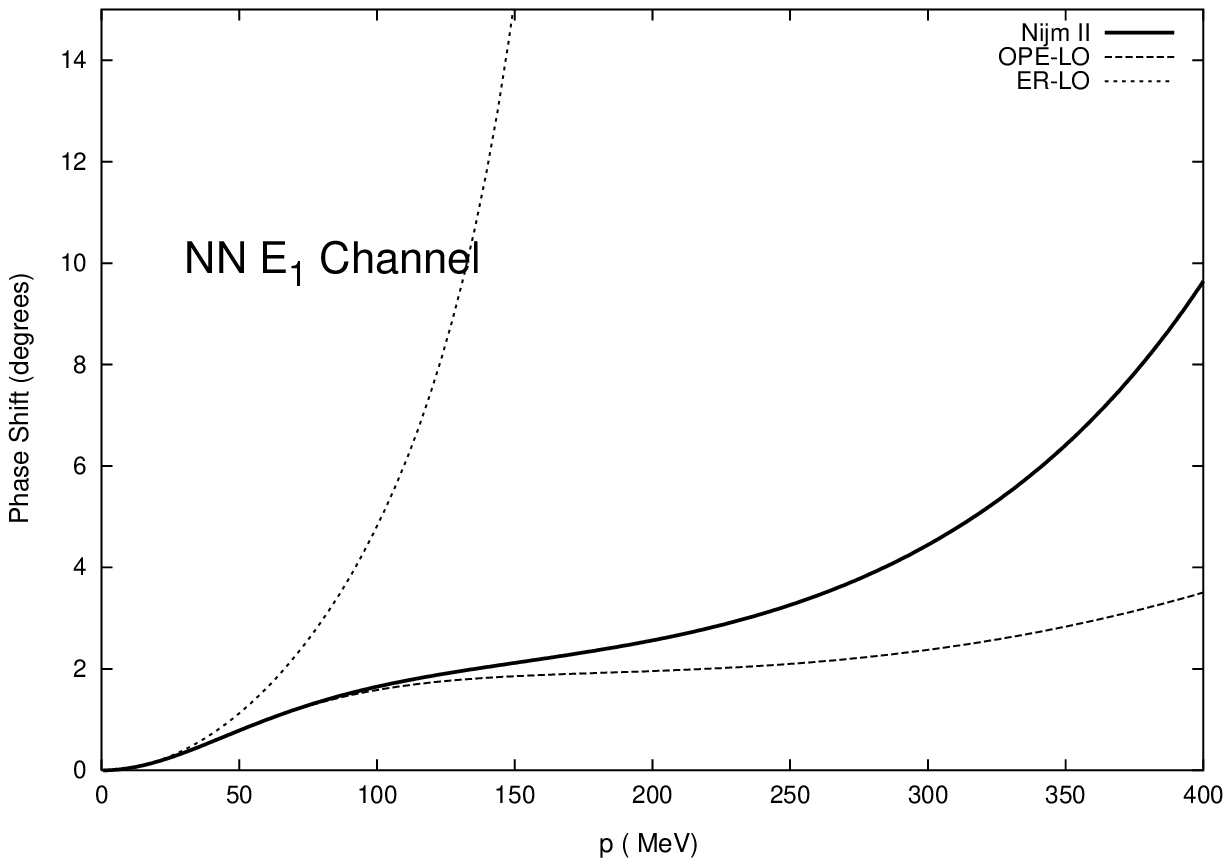,height=5.5cm,width=5.5cm} 
\end{center}
\begin{center}
\epsfig{figure=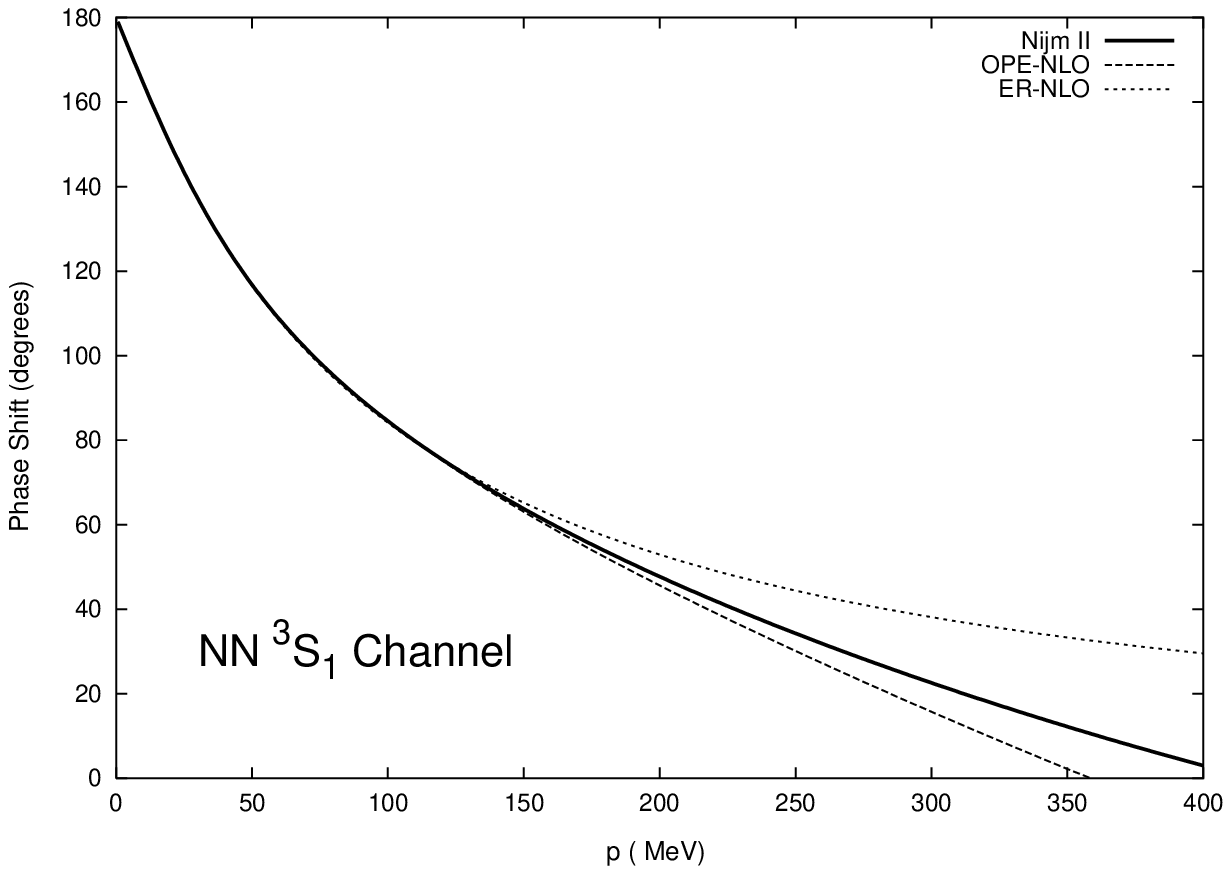,height=5.5cm,width=5.5cm} 
\epsfig{figure=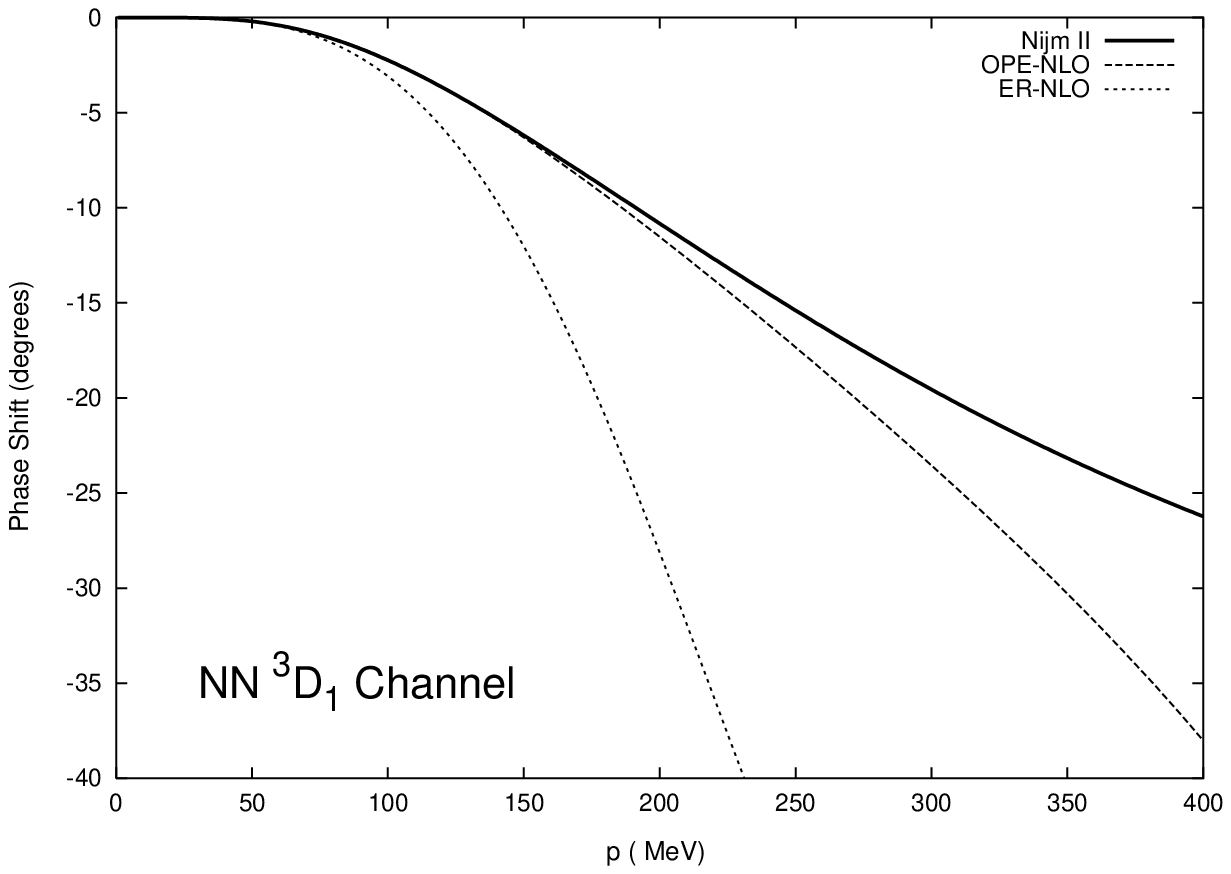,height=5.5cm,width=5.5cm} 
\epsfig{figure=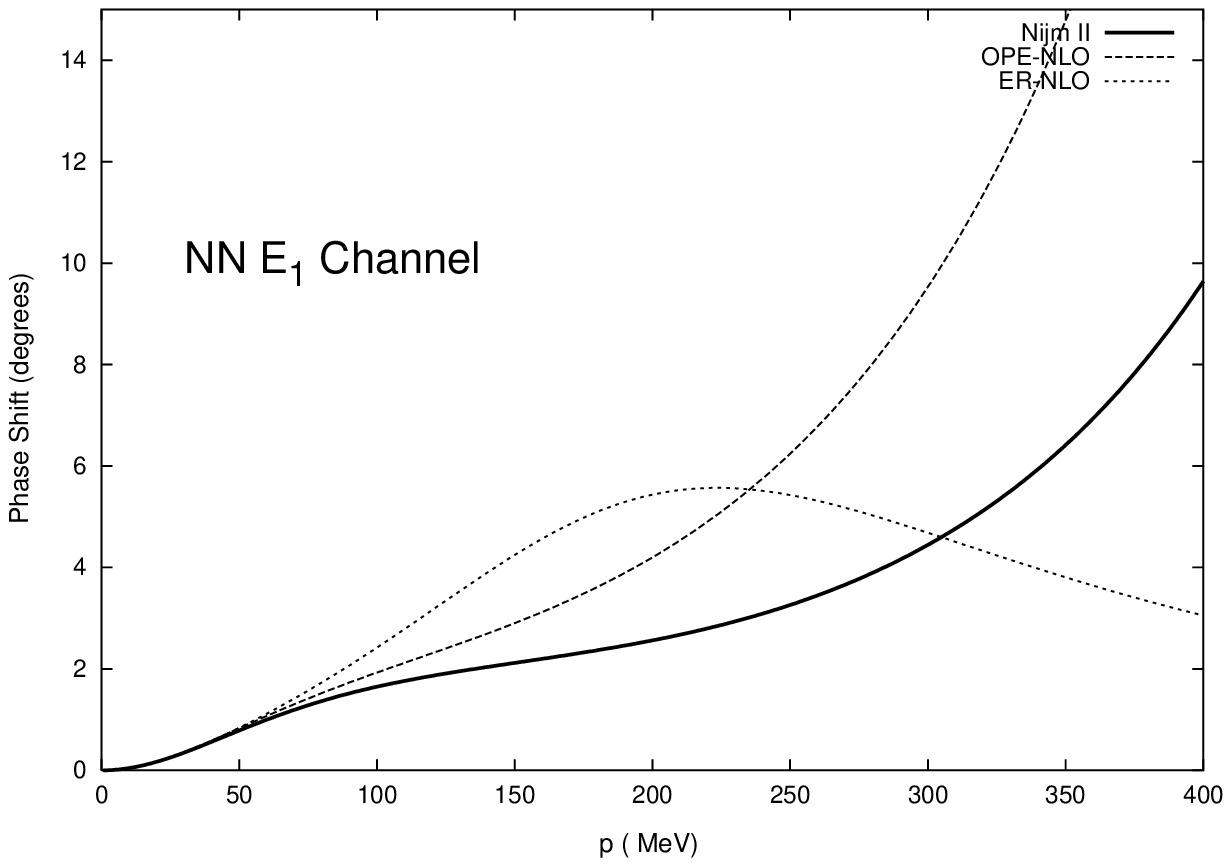,height=5.5cm,width=5.5cm} 
\end{center}
\begin{center}
\epsfig{figure=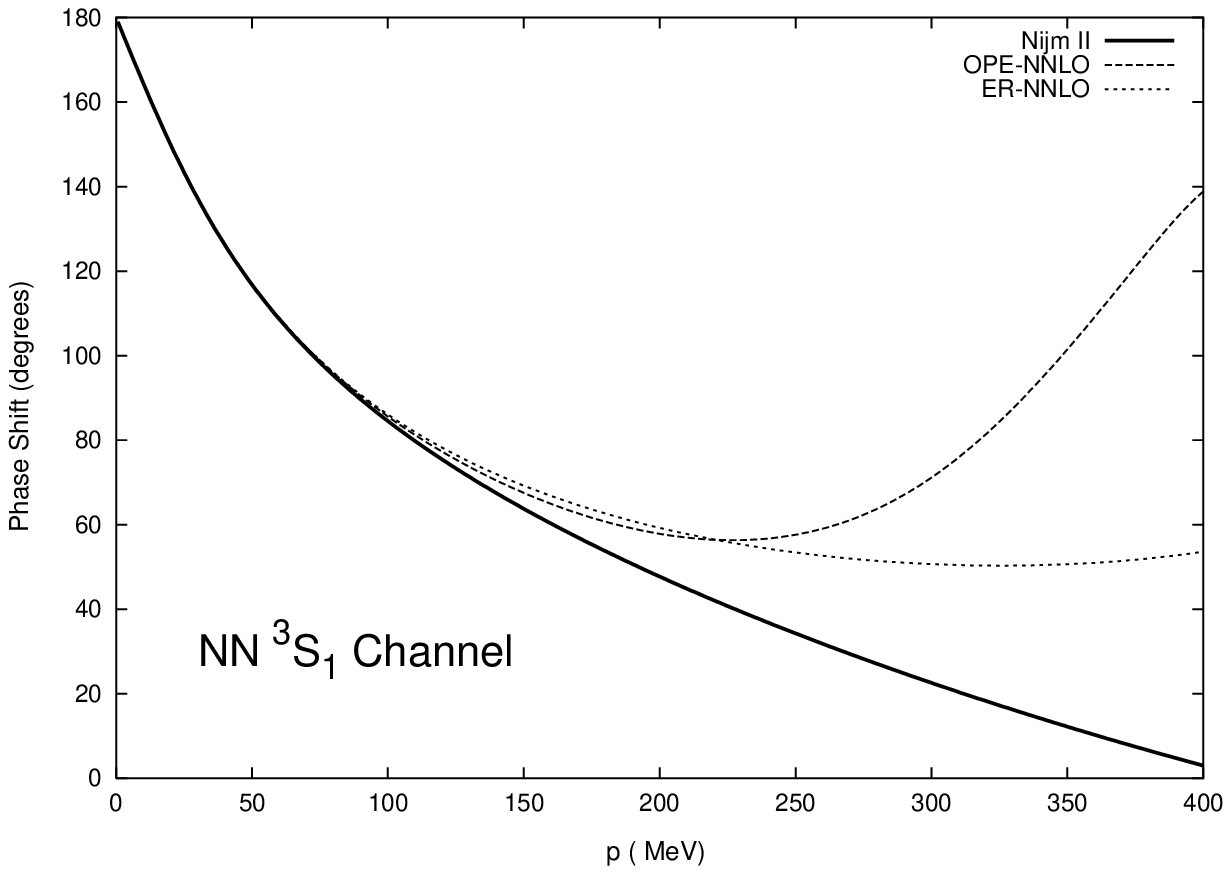,height=5.5cm,width=5.5cm} 
\epsfig{figure=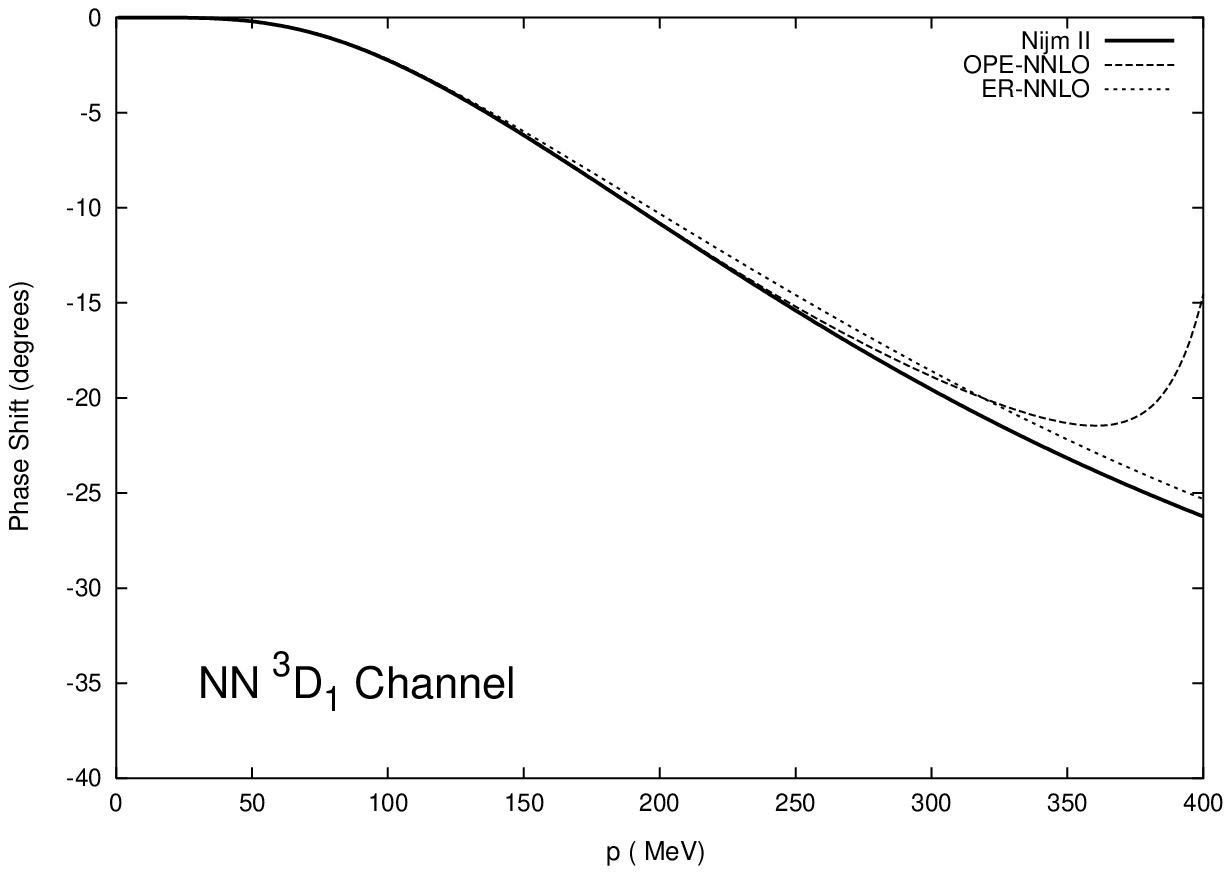,height=5.5cm,width=5.5cm} 
\epsfig{figure=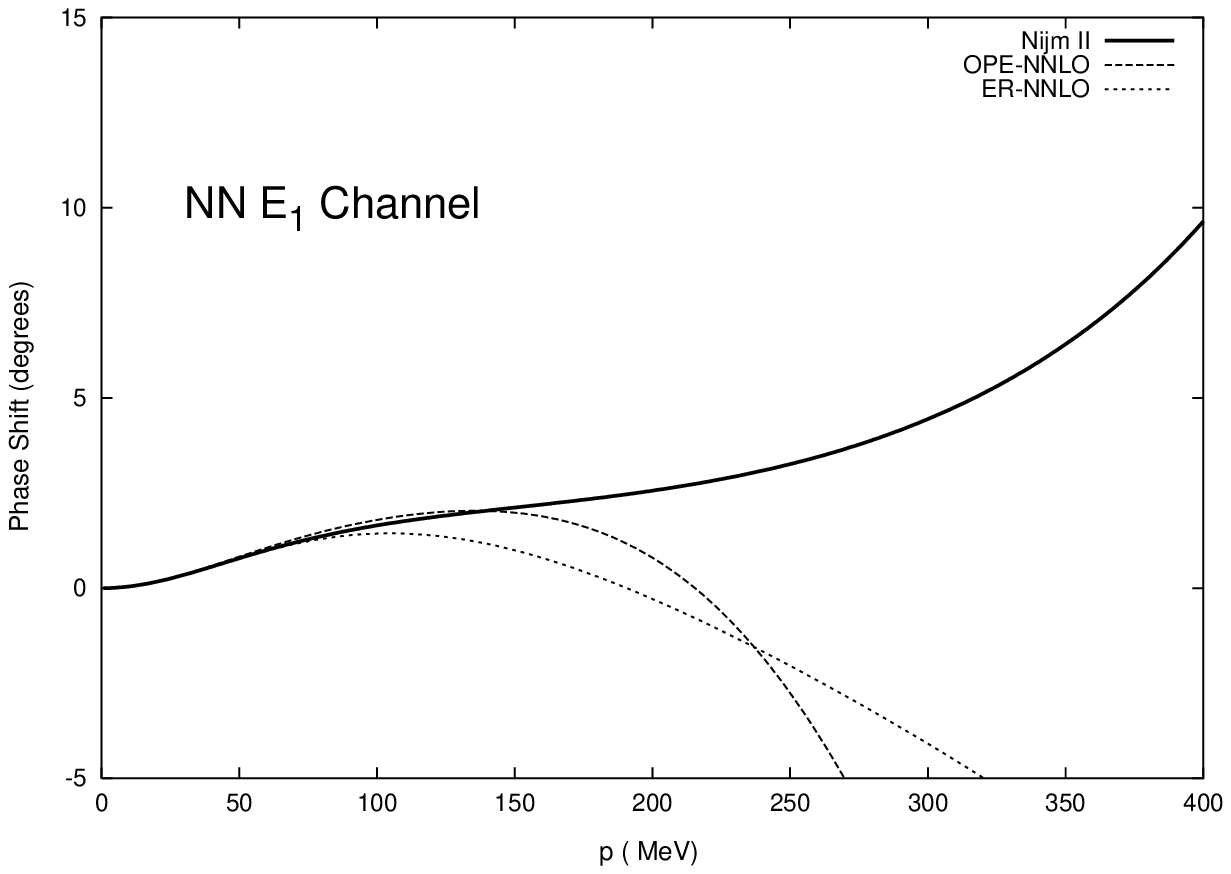,height=5.5cm,width=5.5cm} 
\end{center}
\caption{The effect of including the OPE potential on top of the
effective range expansion (ERE).  Top panel : $^3S_1$, $^3D_1$ and
$E_1$ at LO with (OPE-LO) and without (ER-LO) OPE explicit
effects. Middle panel: same but for NLO. Bottom panel: same but
NNLO.  The difference between ER and OPE indicates the size of the
explicit effects due to the OPE potential to LO (contact terms), NLO
($k^2$ terms) and NNLO ( $k^4$ terms). In both cases the low energy
threshold parameters coincide with those extracted from the NijmII
potential. Data are the PWA from Ref.~\cite{Stoks:1993tb}.}
\label{fig:OPE-ER}
\end{figure*}

\subsection{Phase-Shifts} 

The standard way of proceeding would be to determine the low energy
constants or, equivalently, the short distance parameters directly
from a fit to the data in a large energy range ( say up to $k \sim
m_\pi $ where the two pion exchange left cut opens up should start
contributing) for the theory with OPE. The low energy parameters would
have to be recomputed, and the description at lower energies ($ k <
m_\pi$ ) might become even worse than a pure effective range expansion
(see e.g. Refs.~\cite{Cohen:1998jr,Cohen:1999ia}). Obviously, this is
an undesirable situation. The effective range expansion is convergent
up to the OPE left cut, located at $k= \pm {\rm i} m_\pi /2 $ and
should be applied only there~\footnote{The fact that only two terms in
the expansion, involving the scattering length and the effective range
only, works so well at high momenta, almost up to $ k \sim m_\pi $, is
purely accidental. Actually, including the next $v_2$ term in the
expansion and fitting it in the region $ k < m_\pi /2 $ does not
reproduce the data for $ m_\pi / 2 < k < m_\pi $, but improves the fit
for $ k < m_\pi /2 $. This is obviously an indication of the breakdown
of the expansion beyond the analyticity domain.}. Our formalism can be
specifically constructed to avoid such a situation. Once the threshold
parameters are determined in the short distance limit $R_S \to 0$, our
phase shifts become pure predictions {\it without any additional
parameter fitting} obtained to a given order $k^2 $ expansion of the
initial condition by integrating Eq.~(\ref{eq:vk}) using the effective
range type of initial condition,
\begin{equation}
\hat \K_S   = \hat \K (R_S ) =   
-{\bf a}_S^{-1} + \frac12 {\bf r}_S k^2 + {\bf v}_S k^4 + \dots
\label{eq:eff_short}
\end{equation}
with with $R_S \to 0$. The solution of Eq.~(\ref{eq:vk}) at $R \to
\infty $ gives a solution which when expanded in powers of $k^2$
exactly reproduces ERE to the order imposed by the initial condition,
Eq.~(\ref{eq:c-ere}). Thus, the difference beyond the displayed terms
is merely attributable to the OPE potential. 

In what follows we use LO, NLO,NNLO, etc. to denote keeping up to the
first, second, third order terms in Eq.~(\ref{eq:eff_short})
respectively.

\subsubsection{$^1S_0 $ and $^3S_1$-without mixing channels} 

In Fig.~(\ref{fig:one_channel_phase}) we show the results for the
phase shifts for both $^1S_0 $ and $^3S_1$-without mixing channels
depending on the number of terms kept in the low energy expansion at
short distances. Our results exhibit a good convergence rate.  For
comparison we also depict the effective range expansion results
without explicit pions, which is expected to work at low energies
only. As we see, the effect of introducing pions always improves the
results. This can be fully appreciated at NNLO, where ERE does a poor
job above CM momenta $\sim 100 {\rm MeV}$, but explicit OPE effects
enlarge the energy range up to about $\sim 140 {\rm MeV} \sim m_\pi$.
where we expect explicit two pion exchange contributions to start
playing a role.

\subsubsection{$^3S_1-3D_1$ channel} 

Once the short distance evolution of the low energy parameters are
known one may compute the phase shifts to any order of the
approximation in a $k^2 $ expansion of the initial condition {\it
without any additional parameter fitting} by integrating
Eq.~(\ref{eq:vk}) upwards with a suitable initial condition at a short
distance radius. As a matter of fact the practical choice of the
radius in the numerical calculation is far from obvious, particularly
in the triplet channel case where the low energy parameters take
unbounded values in an increasingly finer scale at short distances
(see e.g. Fig.~(\ref{fig:threshold_evol_down}). It is most practical
to use the WKB approximation to match the numerical solution at a
radius $R_{\rm WKB}$ which can safely be taken in the range $ \sim 0.5
{\rm fm}$. The results for LO (contact terms), NLO ($k^2$ terms) and
NNLO ( $k^4$ terms) are presented in Fig.~(\ref{fig:phase_shifts}) and
compared to the partial wave analysis of Ref.~\cite{Stoks:1993tb}. As
we see the best scheme to take into account the OPE potential
corresponds to use the NLO initial condition. This means on the one
hand that while the scattering lengths may be considered large and
comparable to the effective ranges the curvature parameters $ v_2$ can
be considered to be small. 

\subsection{Finite cut-off effects} 

\begin{figure*}[tbc]
\begin{center}
\epsfig{figure=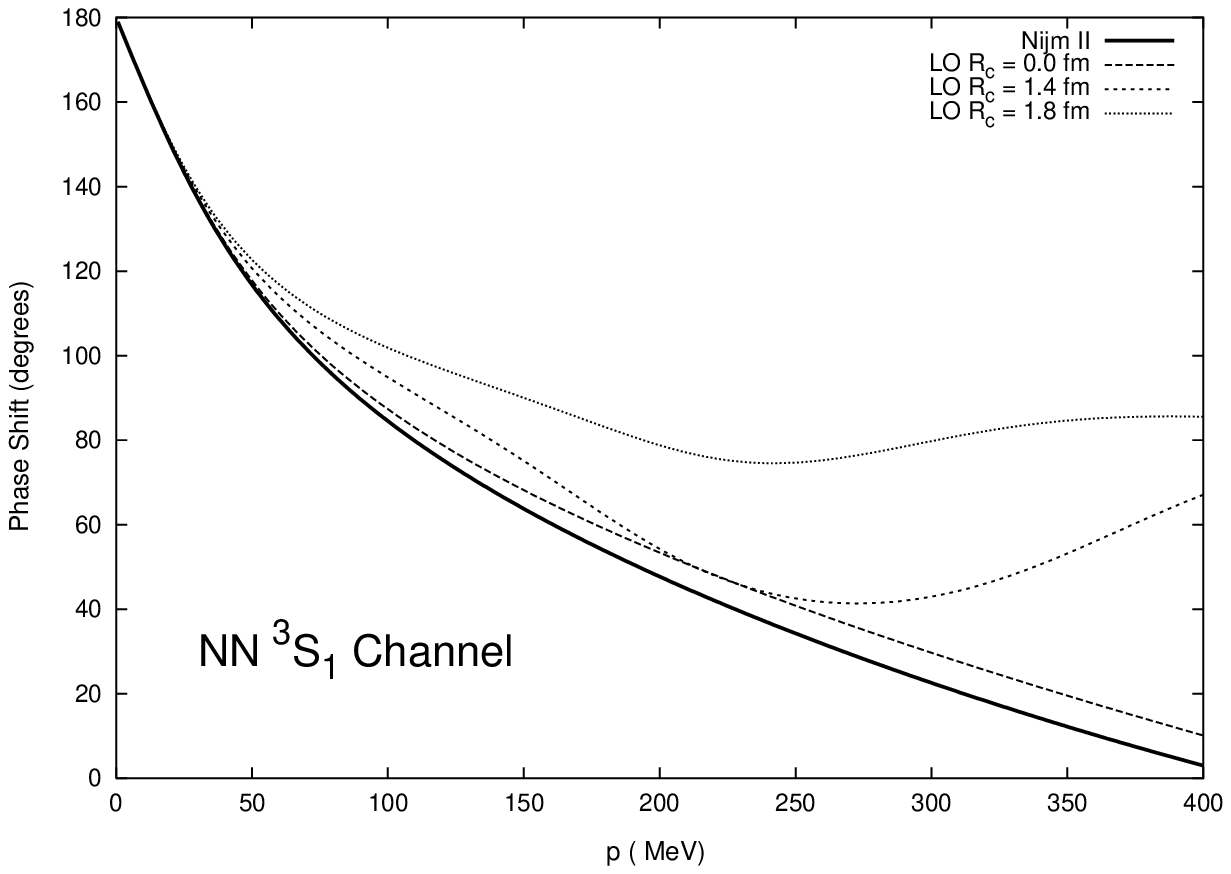,height=5.5cm,width=5.5cm} 
\epsfig{figure=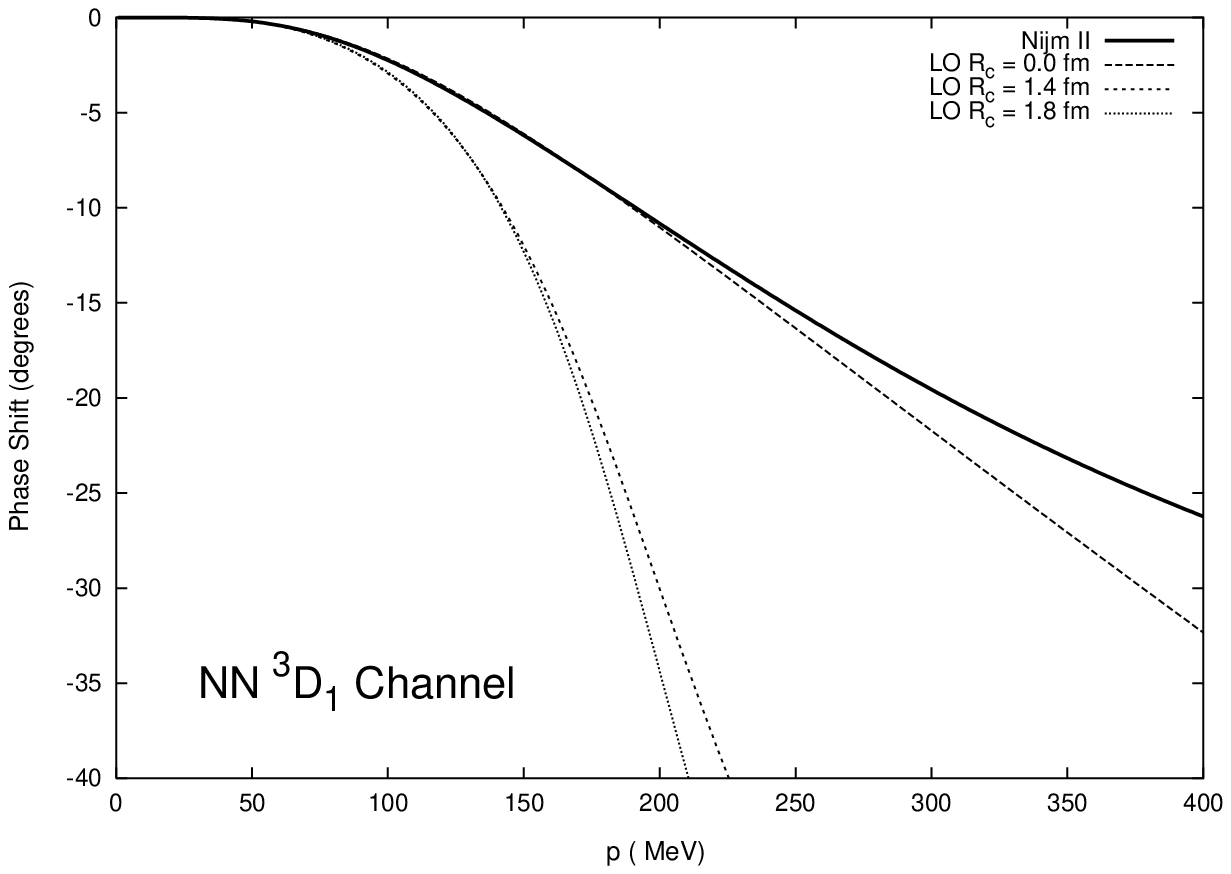,height=5.5cm,width=5.5cm} 
\epsfig{figure=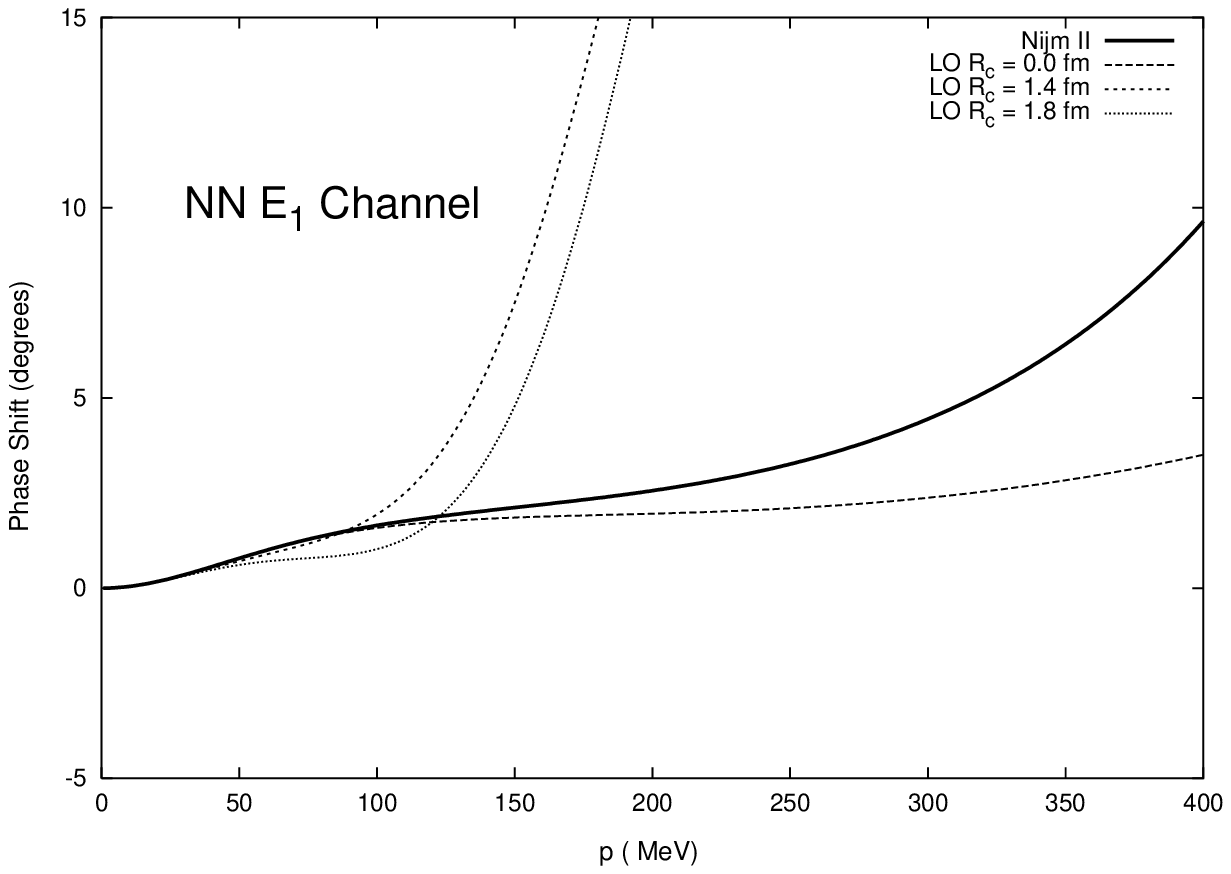,height=5.5cm,width=5.5cm} 
\end{center}
\begin{center}
\epsfig{figure=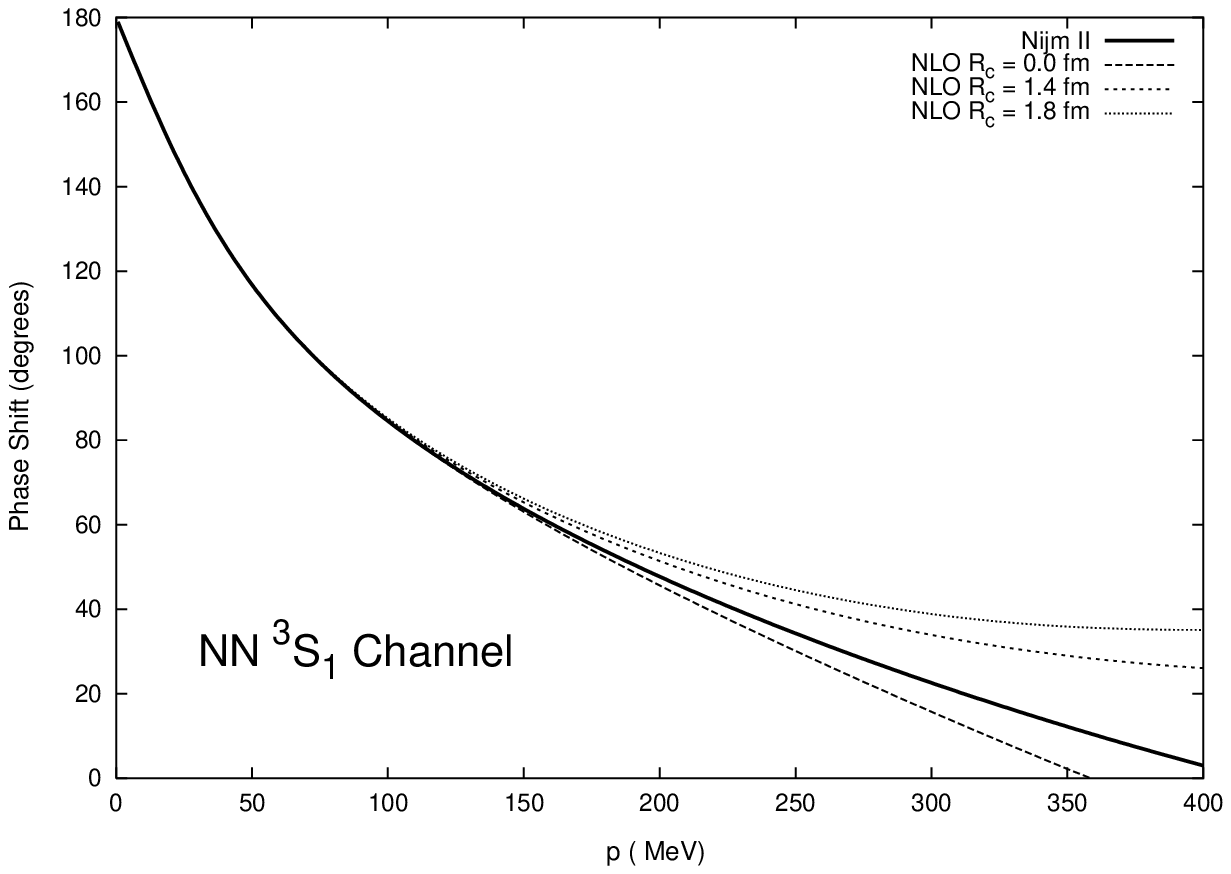,height=5.5cm,width=5.5cm} 
\epsfig{figure=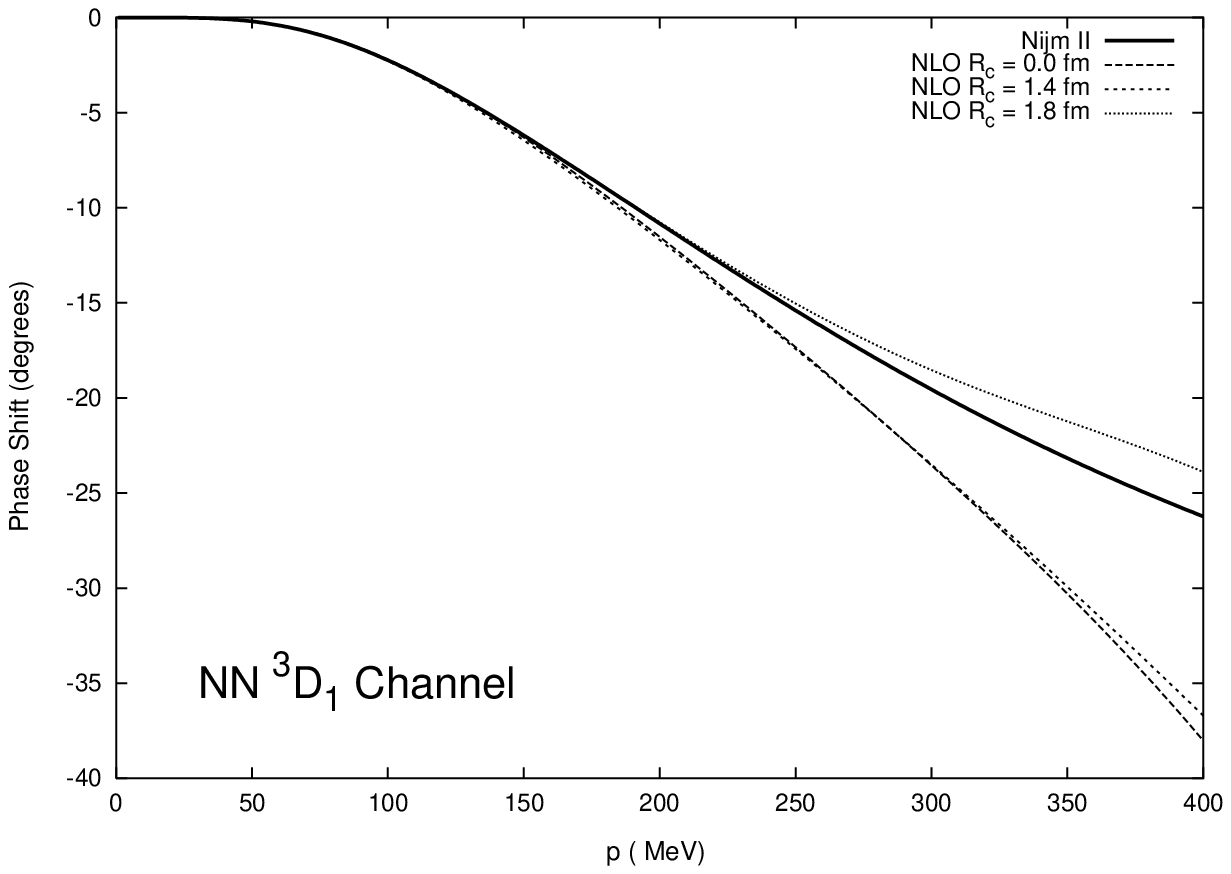,height=5.5cm,width=5.5cm} 
\epsfig{figure=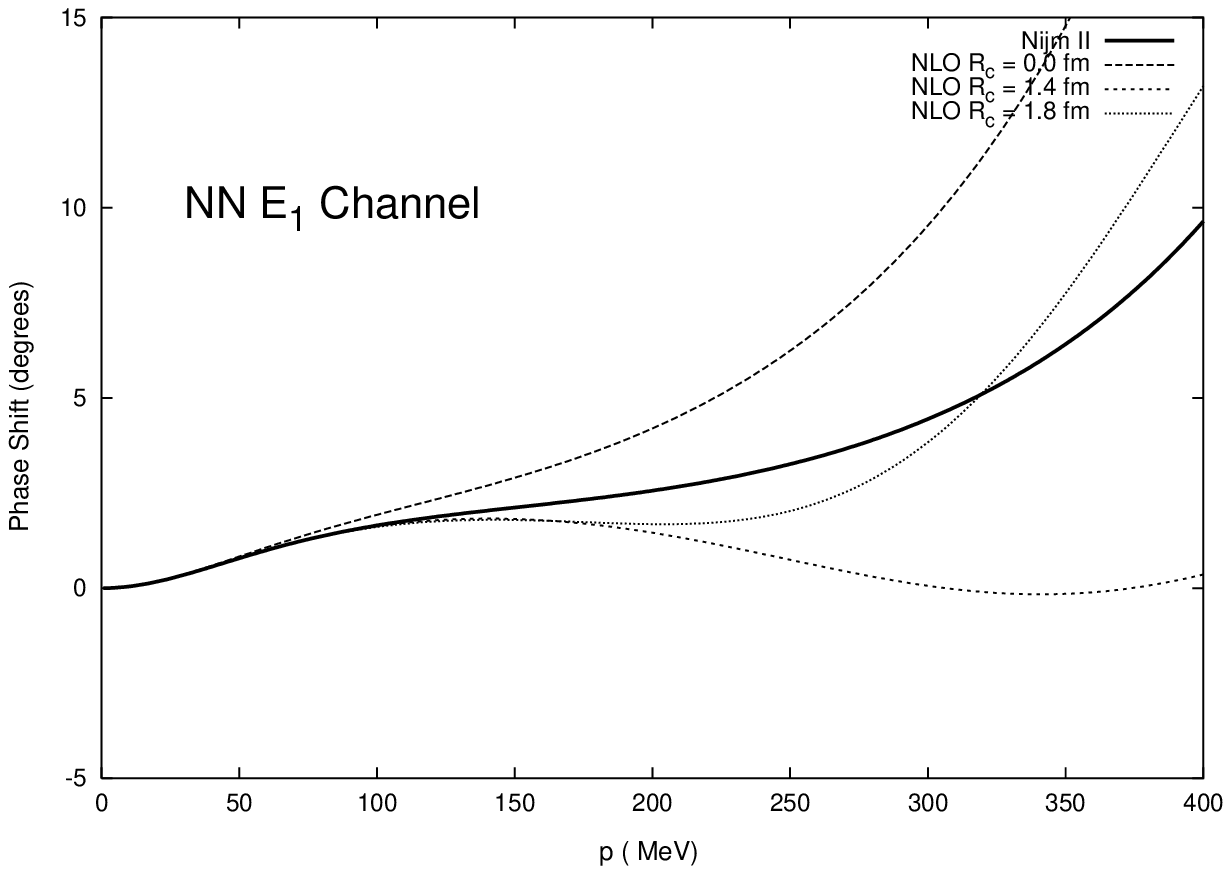,height=5.5cm,width=5.5cm} 
\end{center}
\begin{center}
\epsfig{figure=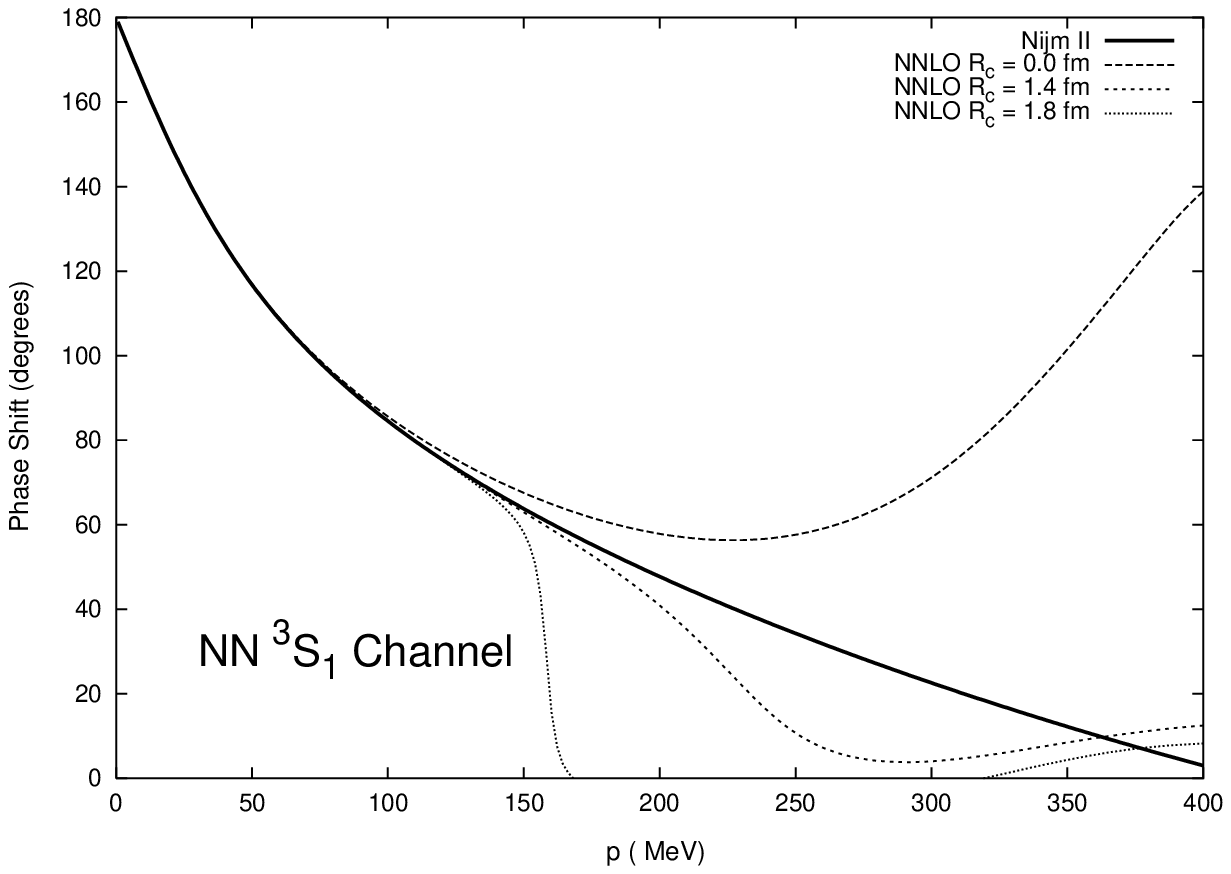,height=5.5cm,width=5.5cm}
\epsfig{figure=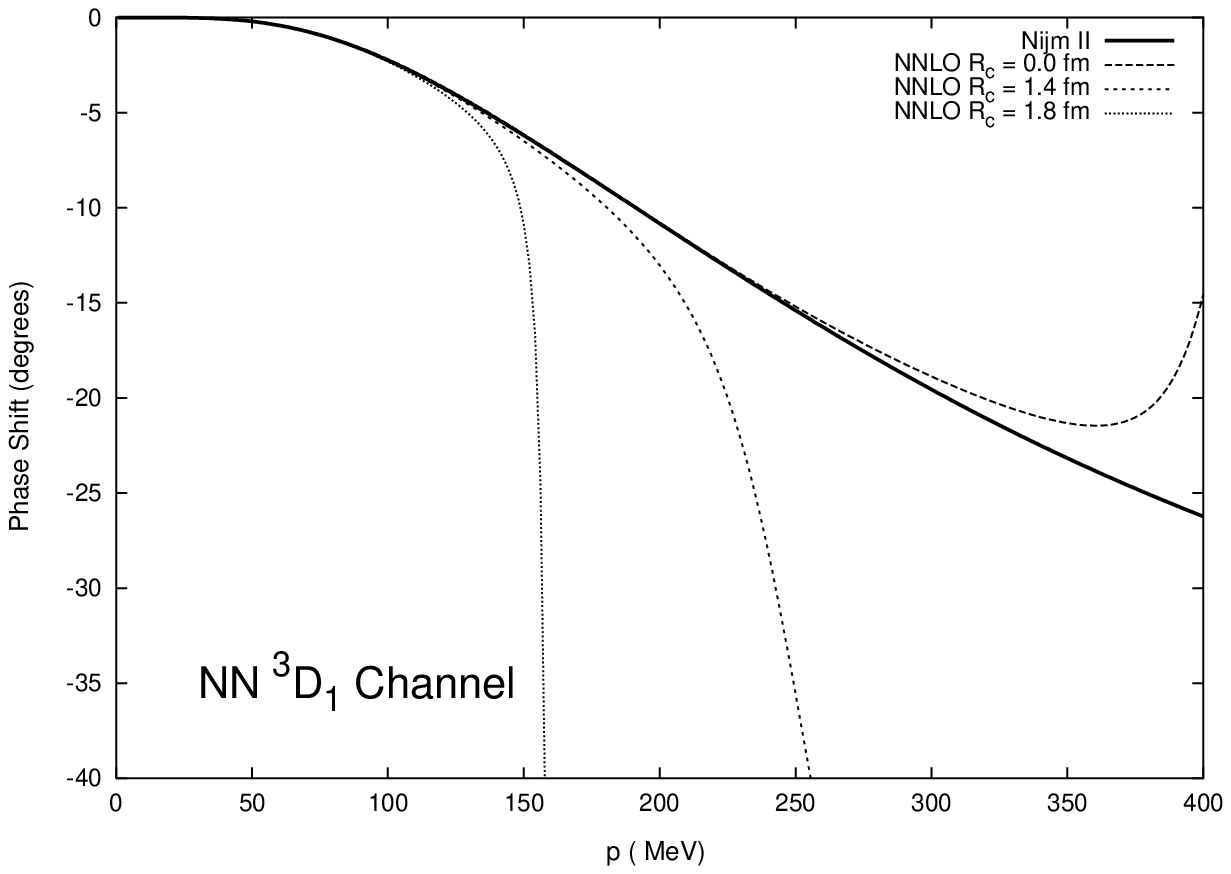,height=5.5cm,width=5.5cm}
\epsfig{figure=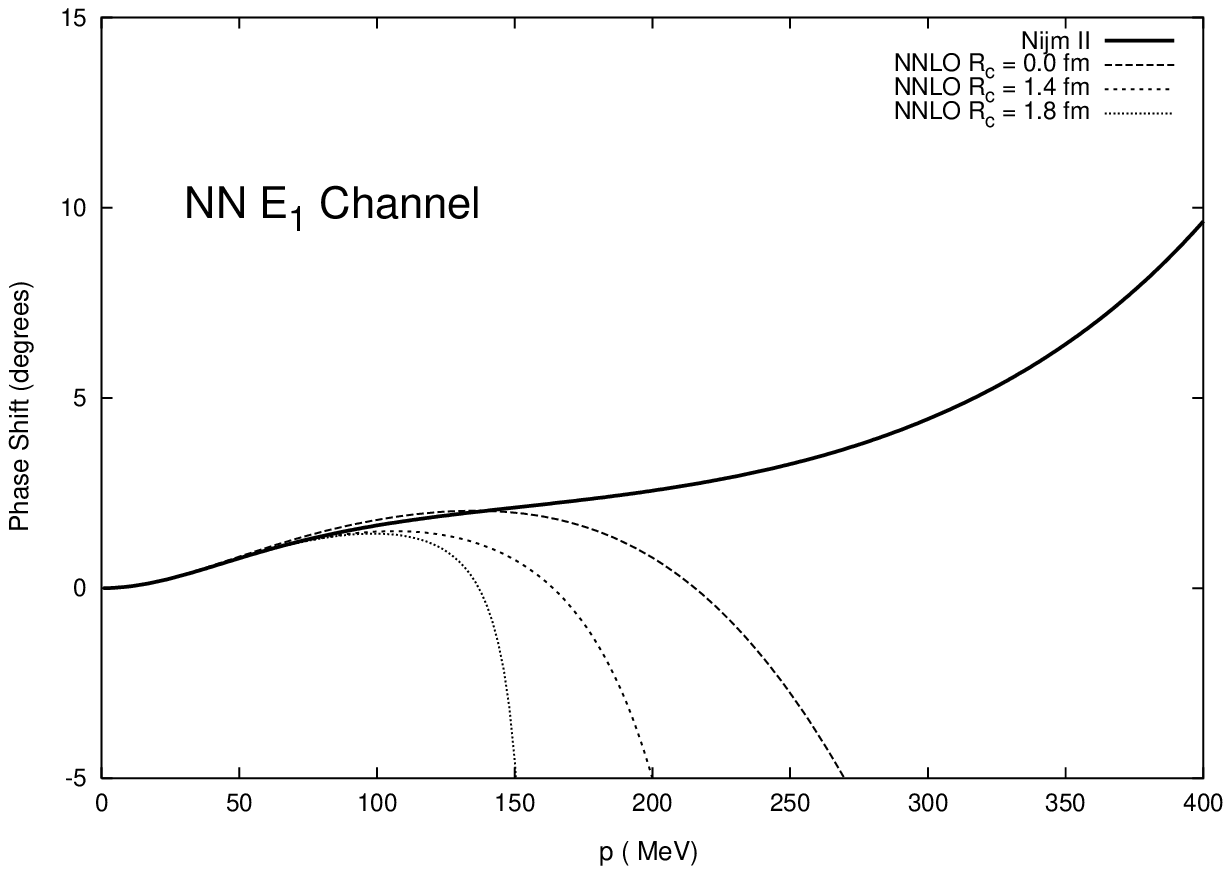,height=5.5cm,width=5.5cm}
\end{center}
\caption{The effect of having a finite short distance boundary radius
for the OPE potential on top of the effective range expansion
(ERE). We compare the theory with finite radii $R=1.4 {\rm fm} $ and $
R=1.8 {\rm fm} $ with the renormalized theory $R=0$. Top panel :
$^3S_1$, $^3D_1$ and $E_1$ OPE-LO. Middle panel: same but for
NLO. Bottom panel: same but NNLO. In all cases the low energy
threshold parameters coincide with those extracted from the NijmII
potential. Data are the PWA from Ref.~\cite{Stoks:1993tb}.}
\label{fig:cut-off}
\end{figure*}

Finite short distance cut-off effects in the scattering phase shifts
can be seen in Fig.~\ref{fig:cut-off} for finite radii $R_S = 1.4 {\rm
fm} $ and $R_S= 1.8 {\rm fm} $ as compared to the renormalized $R_S
=0$ case, for the OPE-LO, OPE-NLO and OPE-NNLO approximations. As one
naively expects these finite effects increase for larger energies,
since they probe smaller wavelengths. A very important feature which
can be deduced from the plots is that these effects are sizable for
momenta where TPE effects should not play a decisive role $ m_\pi / 2
< k < m_\pi$ role. Thus, letting a finite short distance boundary
radius $R_S \sim 1.4 {\rm fm} $ provides a large systematic error,
already in the region where OPE dominates. Thus, it is not clear
whether TPE can bee {\it seen} in the central NN waves with a finite
cut-off distance of about $R_c=1.4 {\rm fm} $. Of course, one should
include TPE contributions in order to make a definite statement.

\subsection{Are pions perturbative ?} 

The discussion of which power counting is the appropriate one for the
NN interaction corresponds physically to the question whether or not
the pion cloud can be considered to be perturbative. It is important
to realize that within our framework we are considering OPE departures
from the effective range expansion to a given order. Thus, at
sufficiently low $k$ explicit pion effects can always be considered
perturbative. This is so regardless of the number of $k^2$ terms
included in the initial condition. Actually, the point is rather if
the low energy threshold parameters can be considered large or
small. According to our results in Fig.~(\ref{fig:phase_shifts}) it
seems that best possible agreement can be obtained when both the
scattering lengths and the effective ranges are taken to be large,
while other low energy parameters can be taken to be small.  To
properly emphasize this point we plot in Fig.~(\ref{fig:OPE-ER}) the
scaled K matrix computed including OPE and compared to the ERE to LO,
NLO and NNLO. Given this fact we expect a kind of consistent long
distance perturbation theory to work.  The details of such an
expansion will be presented elsewhere.

\subsection{Evolution of the short distance boundary condition} 

\begin{figure*}[tbc]
\begin{center}
\epsfig{figure=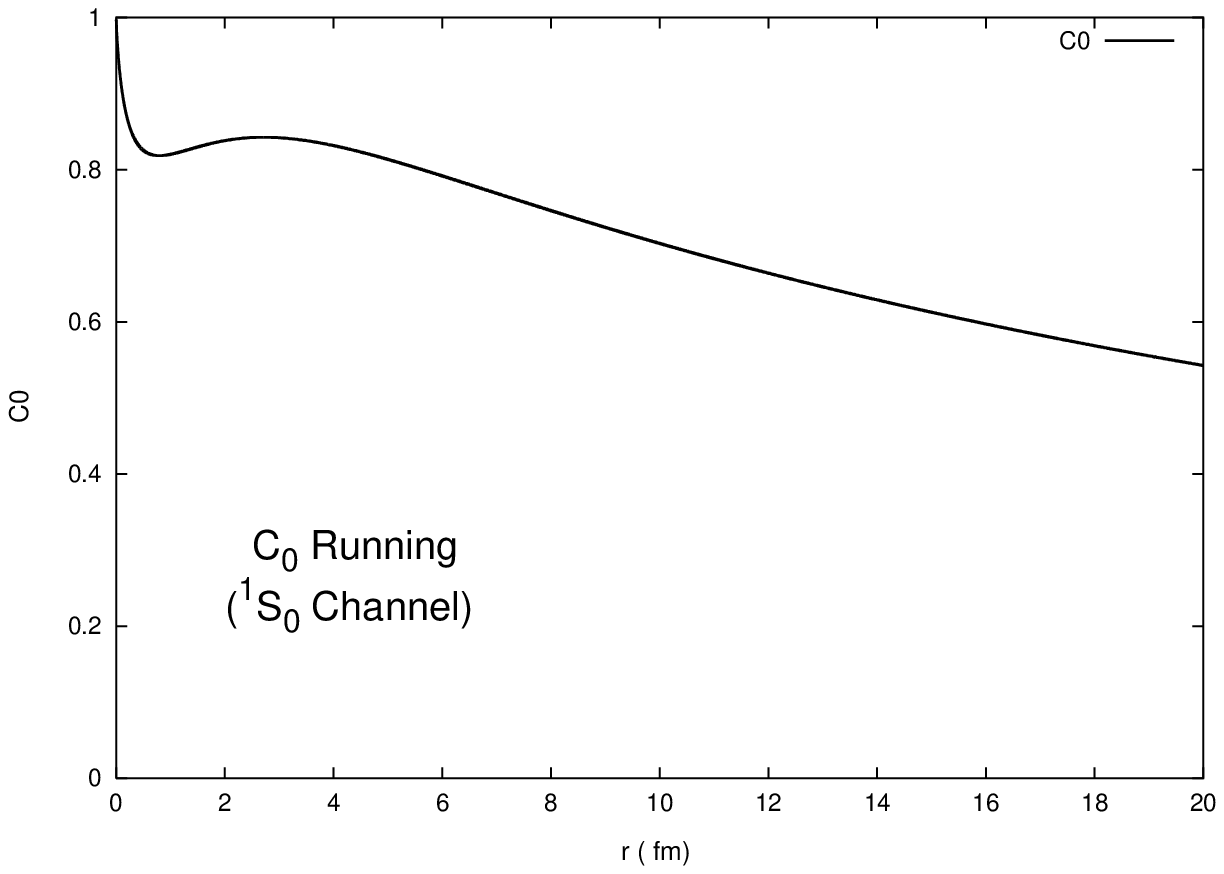,height=8cm,width=8cm}
\epsfig{figure=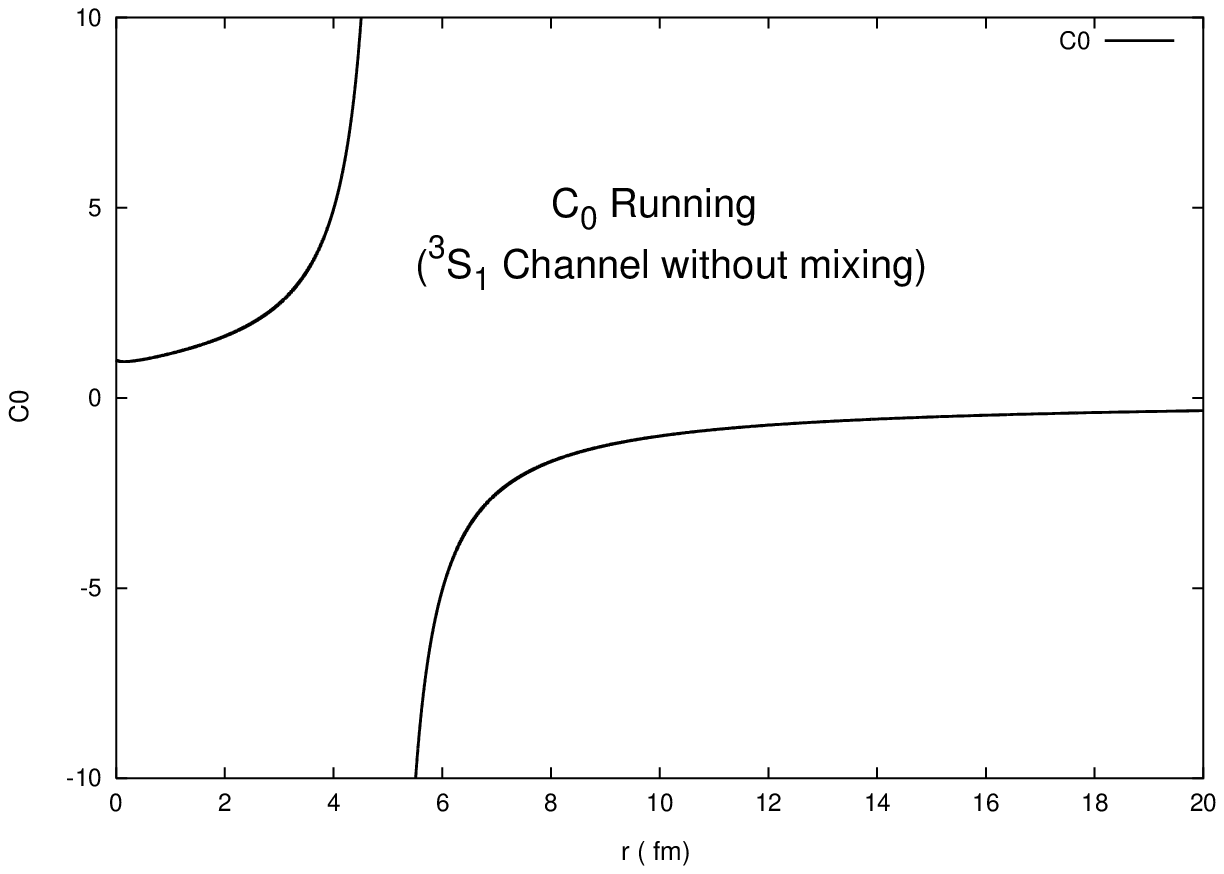,height=8cm,width=8cm}
\end{center}
\begin{center}
\epsfig{figure=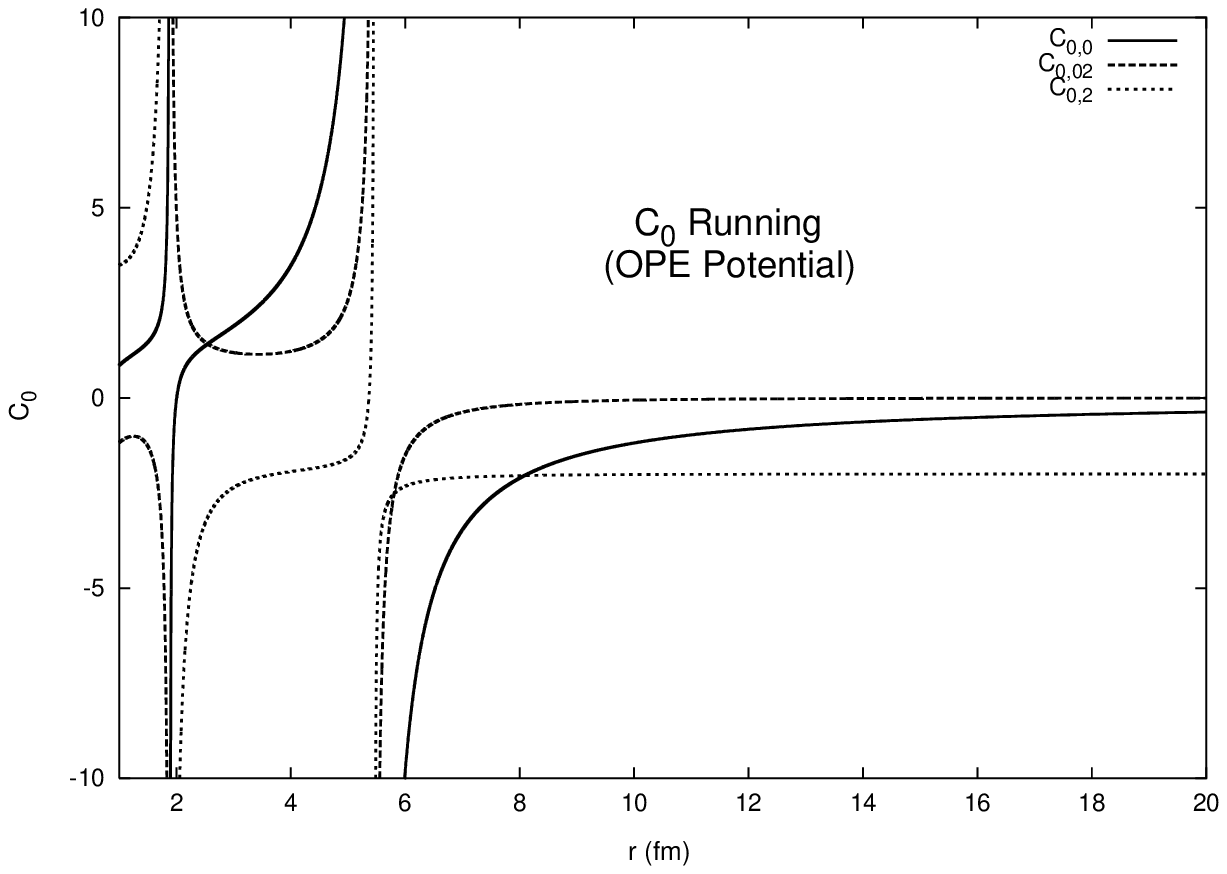,height=8cm,width=8cm}
\epsfig{figure=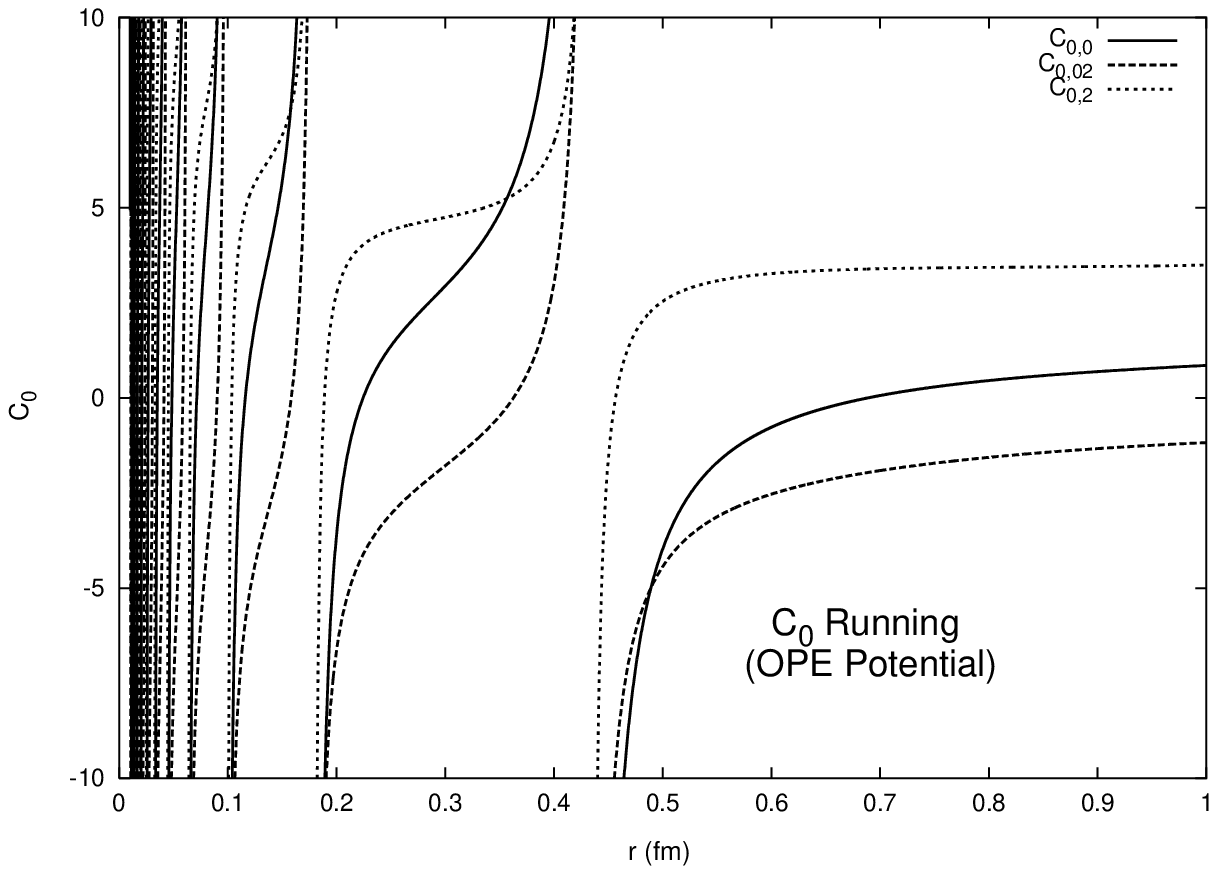,height=8cm,width=8cm}
\end{center}
\caption{Evolution of the dimensionless short distance boundary
conditions at zero energy $ {\bf C}_0 (R) = {\bf 1}-R {\bf L}_0 (R) =
{\bf 1}- R {\bf u}'_0 (R) {\bf u}_0 (R)^{-1} $ with the boundary
radius $R$ due to OPE potential. Top: Singlet $^1S_0$ channel (left)
and Triplet $^3S_1$ channel (right) without mixing using
Eq.~(\ref{eq:bc_alpha}) . Bottom: The triplet $^3S_1-^3D_1 $
channel. The coefficients $C_{ss}^0 $ $C_{sd}^0 $ and $C_{dd}^0 $ are
related to the running scattering lengths $\alpha_{00} $ ,
$\alpha_{02} $ and $\alpha_{22} $ through
Eq.~(\ref{eq:bc_alpha_coupled}). Large scale (left) and zoomed (right)
picture.}
\label{fig:bc_evol}
\end{figure*}

As we have said, the short distance singularity of the OPE potential
enforces a very precise determination of the running low energy
threshold parameters at short distances, and hence of the boundary
condition. We can directly determine this dependence by using
Eq.~(\ref{eq:bc_alpha}) and Eq.~(\ref{eq:bc_alpha_coupled}). For
simplicity and to illustrate the point we just display in
Fig.~(\ref{fig:bc_evol}) the behaviour of the boundary condition
parameters as a function of the short distances boundary radius in the
zero energy limit, both for the singlet $^1S_0$ and triplet $^3S_1$
channel without mixing and for the triplet $^3S_1-^3D_1 $ channel. The
fixed point and limit cycle behaviour obtained for the running of the
low energy threshold parameters maps into a similar behaviour for the
short distance boundary condition.  From the picture it is clear that
the standard procedure of integrating the Schr\"odinger equation
upwards from a given short distance boundary radius to infinity in
order to fit the low energy parameters would require a very high
precision determination of a rapidly varying boundary condition in the
case of the triplet $^3S_1-^3D_1$ channel. It is clear that a
determination of the $C_0 $ constants from a fit to the phase shifts
in the low energy region would be extremely delicate in the limit $R_c
\to 0$ in practice. Instead, the present approach computes directly
the boundary condition in a power expansion of the energy at any given
radius from the physical values of the low energy
parameters. Actually, our method is equivalent to integrate the
Schr\"odinger equation from that short distance boundary radius to
infinity.  In addtiiton, the singular and attractive nature of the OPE
potential allows a WKB treatment of the short distance singularity,
and allows to eliminate the finite cut-off radius taking the limit
$R_c \to 0$. Obviously, the present framework can be extended to
reanalyze the role of TPE potentials in a non perturbative way and
completely free of finite cut-off artifacts, where the short distance
behaviour is qualitatively similar~\footnote{Unlike the OPE where one
has both an attractive and repulsive $1/r^3$ singularity (see
Sect.~\ref{sec:short}), in the TPE case one encounters attractive $
1/r^6$ singularities for coupled channels.}

\section{Conclusions } 
\label{sec:concl} 

In the present paper we have analyzed the renormalization of the OPE
interaction in the presence of contact and derivative interactions of
any order for NN scattering both for the singlet and triplet channel
states.  The basic point of our approach is to regularize the {\it
unknown} short distance physics by means of a boundary condition at a
certain boundary radius, above which the OPE potential is assumed to
work, i.e. where pions are treated explicitly. Below that scale pions
contribute implicitly to the scattering properties although always in
combination with other effects which cannot be disentangled unless a
given distance scale is specified. Actually, when the boundary radius
goes to infinity, above the pion Compton wavelength, we have a low
energy theory of contact interactions and derivatives there-off. As the
boundary radius goes below the OPE range, we have a theory where pions
are eliminated above the scale set by the boundary. This allows to
remove explicitly pion effects in the threshold parameters for the OPE
potential in an unambiguous and model independent way. The
renormalization group flow implied by our non-perturbative equations
is unique provided the OPE potential is assumed to be valid al the way
down to the origin. This is obviously not a realistic assumption but
it is absolutely necessary to go to these small distances in order to
get rid of any finite short-distance cut-off effect and properly
define the OPE contributions to the scattering observables.  
This result fully complies to the spirit of an effective
field theory, the terms in a low momentum
expansion of the amplitude are shape independent while the remaining
powers depend both on the long distance OPE details (like the left branch
cut) and the shape independent low energy parameters themselves.

The short distance behaviour of threshold parameters present either a
ultraviolet fixed point structure in the $^1S_0 $ and $^3S_1
$-without-mixing channels whereas we find limit cycles for the
$^3S_1-^3D_1 $ channel due to the singular and attractive behaviour of
the OPE contribution to the tensor potential. This means that in the
latter case there is not a monotonous trend at short distances. A
direct consequence of having both ultraviolet fixed points and limit
cycles for the threshold parameters is that a delicate fine tunning of
the short distance physics is implied. In addition, for the
experimental values of the threshold parameters one obtains huge
changes for distances below $ 2 {\rm fm}$ when OPE effected are
removed. Nevertheless, we find moderate changes in the phase shifts
due to explicit pion effects. Actually, in the $^1S_0$ and
$^3S_1$-without mixing channels the effect is found to be compatible
with a perturbative treatment. In the $^3S_1-^3D_1$ channel the effect
is a bit more complicated due to the presence of ultraviolet limit
cycles triggered by the singular character of the tensor potential;
the coupled channel amplitudes are non-perturbatively renormalizable
while they become perturbatively non-renormalizable. This makes a
naive perturbative treatment slightly more subtle. One of the
advantages of having a renormalizable theory is that non-perturbative
equations make sense, and any perturbative treatment should arise as a
controllable approximation to the full equations. As we have pointed
out along the paper, this is probably an advantage of using coordinate
space methods and a boundary condition renormalization versus momentum
space methods.

Taking into account all the nice features of the present calculation,
in particular, getting a handle on the finite cut-off corrections, the
results presented in this paper are very satisfactory suggesting
several improvements. Explicit Two Pion Exchange contributions are
expected to contribute significantly at about $1.5-2 {\rm fm} $ at the
level of the potential, so our results for the evolution of the
threshold parameters should not be considered realistic below that
scale, or equivalently above CM momenta of about $100-150 {\rm MeV}$,
as it seems to be the case. In addition, our description should be
enlarged to include higher partial waves. For peripheral waves one
expects perturbative methods to work since there is a strong
centrifugal suppression of the wave function at the origin, and
perturbative renormalization methods can be applied. For those the
present approach does not have much to say.  Low partial waves,
however, are particularly interesting since a re-summation of pion
exchanges seems crucial to understand the data.  Work along these
lines will be presented elsewhere~\cite{Pavon03}.

\begin{acknowledgments}

We thank J. Nieves for discussions. This work is supported in part by
funds provided by the Spanish DGI with grant no. BMF2002-03218, Junta
de Andaluc\'{\i}a grant no. FM-225 and EURIDICE grant number
HPRN-CT-2003-00311.

\end{acknowledgments}

\appendix

\section{Another Derivation of the variable S-matrix}
\label{sec:app1} 

In order to deduce a variable $S$-matrix equation, we determine first
the infinitesimal change of the $S$ matrix under a general deformation
of the potential $\U (r) \to \U(r) + \Delta \U(r) $. Using
Schr\"odinger's equation (\ref{eq:sch_cp}) and the standard Lagrange's
identity adapted to this particular case, we get after integration
\begin{eqnarray}
\left[ \u (r)^\dagger \Delta \u'(r) - \u'(r)^\dagger \Delta \u(r)
\right] \Big|_0^\infty = \int_0^\infty dr
\u(r)^\dagger \Delta \U(r) \u(r) \nonumber \\ 
\end{eqnarray} 
which, after using the asymptotic form of the matrix wave
function, Eq.~(\ref{eq:asym}), yields  
\begin{eqnarray}
2 {\rm i} k \S^\dagger \Delta \S = \int_0^\infty dr \, \u(r)^\dagger
\Delta \U(r) \u(r)
\end{eqnarray} 
In particular, for the parametric family of potentials $\bar \U (r,R)
= \theta(R-r) \U(r) $ we get
\begin{eqnarray}
2 {\rm i} k \S^\dagger (R) \S '(R)  =  \u (R)^\dagger \U(R) \u(R)
\end{eqnarray} 
and using the value of the wave function at the outer boundary 
\begin{eqnarray}
\u (R)  =  \h^{(-)} (R) -  \h^{(+)} (R)  \S (R)  
\end{eqnarray} 
we finally get the variable S-matrix equation, Eq.~(\ref{eq:vs}).   
Note, that the variation of the potential is done with a fixed
boundary condition.

\section{Difficulties in extracting the low energy parameters} 
\label{sec:app3} 

In this appendix we want to elaborate on the problems we have
encountered while fitting the NN data base of Ref.~\cite{Stoks:1993tb}
within a generalized coupled channel effective range expansion,
Eq.~(~\ref{eq:c-ere}). Unfortunately, this data base does not provide
error estimates for their phase shifts (although 8 significant
digits), nor the typical energy resolution where these data should be
trusted, so some compromise must be made.

\begin{figure*}
\begin{center}
\epsfig{figure=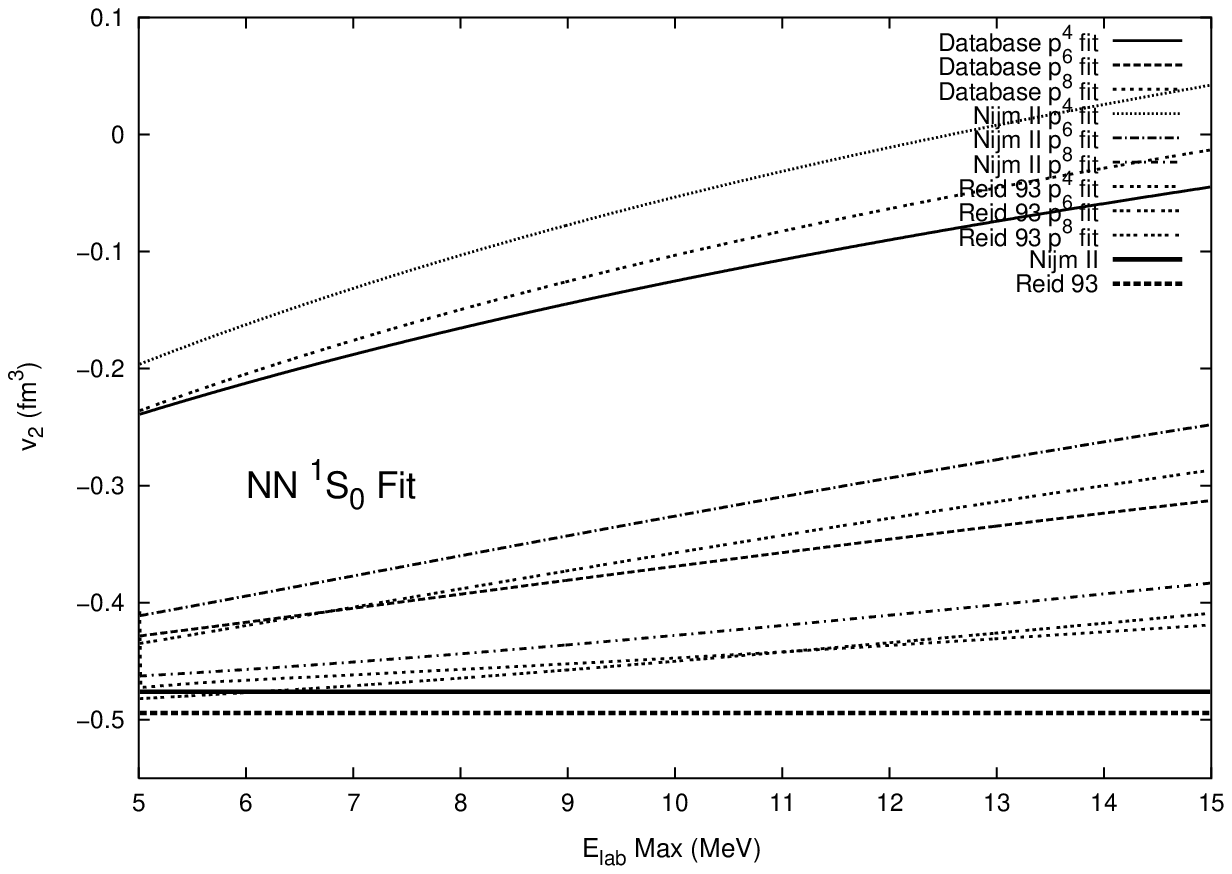,height=8cm,width=8cm} 
\epsfig{figure=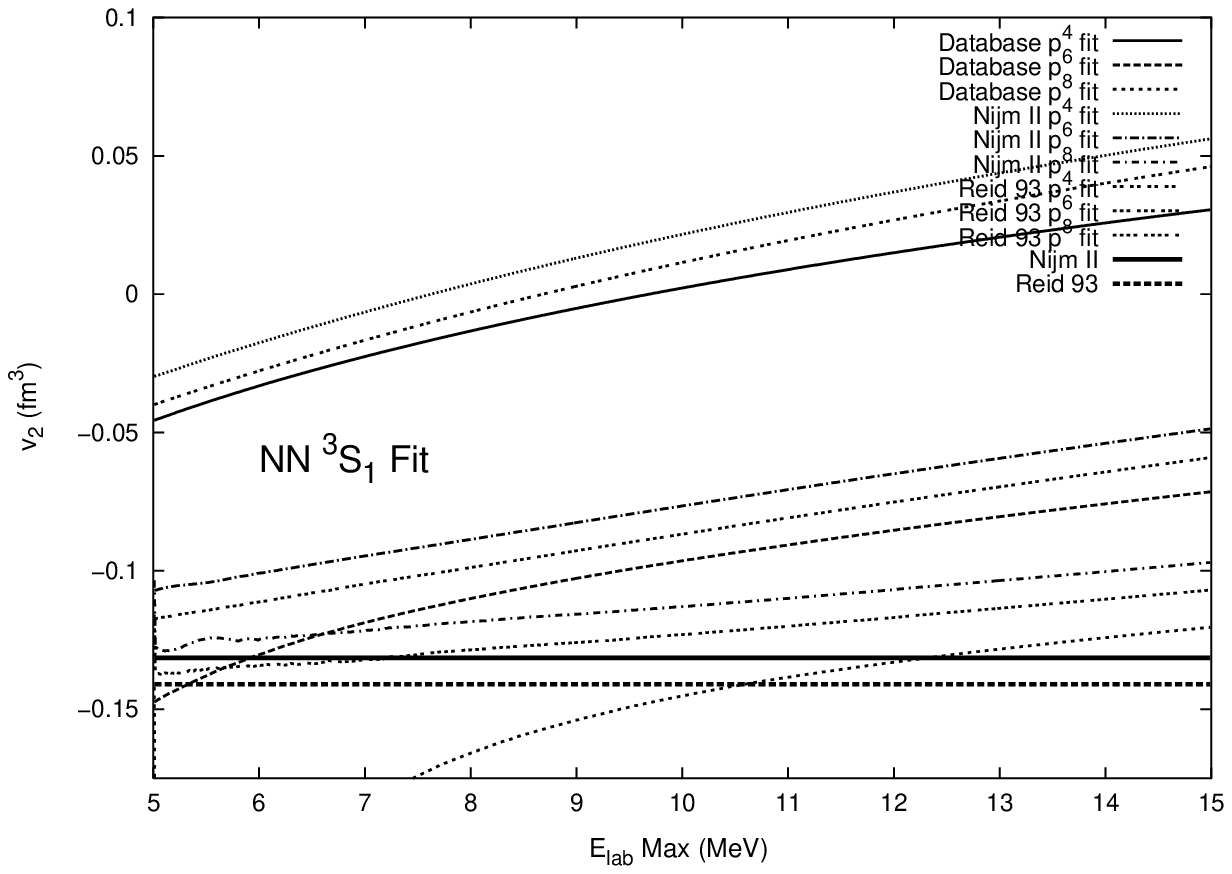,height=8cm,width=8cm} 
\end{center}
\caption{The $v_2$ parameter for the $^1S_0$ (left) and the $^3S_1$
(rught) channels determined from a fit to the low energy date of the
NN data base \cite{Stoks:1993tb} (see Eq.~(\ref{eq:chi2})) and main
text, as a function of the maximal LAB-energy considered in the fit.
$p^n$ means a fit including up to $p^n$ terms in the effective range
expansion Eq.~\ref{eq:era_fit}. ``Database'' means a fit to the
average value of the corresponding scaled $K$-matrix. ``Reid93'' and
``NijmII'' means a fit to only this data. The values we obtained by
integrating the Eqs.~(\ref{eq:valpha}),(\ref{eq:vr0}) and
(\ref{eq:vv2}) with the NijmII and Reid93 potentials in
Sect.~\ref{sec:lecs}. }
\label{fig:lecs_fit}
\end{figure*}

We use the NN-data and define the $\chi^2$ as
\begin{eqnarray}
\chi^2 = \sum_{i=1}^N \left( \frac{\hat K_{\rm ER} - \bar K_{\rm NN}}{\Delta
K_{\rm Nm} }\right)^2   \frac{M}{4p} 
\label{eq:chi2}
\end{eqnarray} 
where we take $ \Delta E_{\rm LAB } =0.01 {\rm MeV} $ , and $ \bar
K_{\rm NN} $ and $ \Delta K_{\rm NN} $ are the mean value and the
standard deviation of the six potentials listed in the NN-data
base~\cite{Stoks:1993tb}, which can be taken as independent
uncorrelated primary data.  The factor $M/(4p) $ is the Jacobian of
the transformation between the Lab-energy and the C.M. momentum, $
E_{\rm LAB} = 2 p^2 /M $, and would correspond to make an equidistant
sampling in $p$, in the limit $ \Delta E_{\rm LAB} \to 0 $ (this is
why we take a small energy spacing). This weight factor is introduced
in order to enhance the region at low momenta. On the other hand, very
low momenta, must be excluded since the resulting mean value
$K$-matrix is incompatible within the attributed errors with the
expected theoretical behaviour, Eq.~(\ref{eq:c-ere}), thus we take
$E_{\rm LAB} \ge 0.5 {\rm MeV} $ Also, the fit goes up to $E_{\rm LAB}
\le 10 {\rm MeV} $, which corresponds to a C.M. momentum about $ p =
m_\pi /2 $ where we expect the finite polynomial of the scaled
$K-matrix $ to truly represent an analytical function within the
convergence radius up to the branch cut singularity located at $ p =
\pm {\rm i} m_\pi /2 $. The form of the fitting function is 
\begin{eqnarray}
\hat K_{\rm ER}= -\beta + r_0 p^2 + v_2 p^4+ v_3 p^6 + v_4 p^8 + \dots   
\label{eq:era_fit} 
\end{eqnarray} 
In Fig.~\ref{fig:lecs_fit} we show as an illustration the $v_2$
parameter determined from a fit to the low energy region of the NN
data base \cite{Stoks:1993tb} as a function of the maximal LAB-energy
considered in the fit. As we see, instead of a plateau within some
energy window, we observe an ever changing value. We observe no
stability depending on the number of terms considered in
Eq.~(\ref{eq:era_fit}) either. For comparison we also plot the values we
obtained by integrating the Eqs.~(\ref{eq:valpha}),(\ref{eq:vr0}) and
(\ref{eq:vv2}) with the NijmII and Reid93 potentials in
Sect.~\ref{sec:lecs}, which where quite stable numerically. As we see,
the values obtained from the fit, in the chosen energy window are
hardly compatible. The deceptive features extend to other channels,
and non diagonal low energy threshold parameters such as the matrix
elements of ${\bf a} $ and ${\bf r}$.

Finally, we have also tried, with no success, other methods for the
determination of the low energy threshold parameters, like evaluation
of derivatives within several algorithms. The reason for the failure
has to do with round-off errors generated by the relatively small
number of digits provided in the NN database. 

\end{document}